\documentclass[aoas,preprint]{imsart}
\RequirePackage[OT1]{fontenc}
\RequirePackage{amsthm,amsmath}
\RequirePackage[numbers]{natbib}
\arxiv{arXiv:0000.0000}
\startlocaldefs
\numberwithin{equation}{section}
\theoremstyle{plain}

\endlocaldefs

\usepackage{amsmath}
\usepackage{multirow}
\usepackage{graphicx}
\usepackage{verbatim}
\usepackage{color}
\usepackage{rotating}
\newcommand{\bed}{\begin{definition}}
\newcommand{\eed}{\end{definition}}
\newcommand{\beq}{\begin{equation}}
\newcommand{\eeq}{\end{equation}}

\newcommand{\bitem}{\begin{itemize}}
\newcommand{\eitem}{\end{itemize}}

\newcommand{\beqn}{\begin{equation}}
\newcommand{\eeqn}{\end{equation}}
\newcommand{\balign}{\begin{align}}
\newcommand{\ealign}{\end{align}}

\newcommand{\sgn}{\mathrm{sgn}}

\usepackage{amssymb}

\begin{document}

\begin{frontmatter}
\title{Coauthorship and Citation Networks for Statisticians}
\runtitle{Coauthorship and Citation Networks}

\begin{aug}
\author{\fnms{Pengsheng} \snm{Ji}\thanksref{t1}\ead[label=e1]{psji@uga.edu}}
\and
\author{\fnms{Jiashun} \snm{Jin}\thanksref{t2}
\ead[label=e2]{jiashun@stat.cmu.edu}}

\thankstext{t2}{JJ was partially supported by NSF grant DMS-1208315.}
\runtitle{Coauthorship and Citation networks}
\runauthor{P. Ji and J. Jin}

\affiliation{University of Georgia\thanksmark{t1} and Carnegie Mellon University\thanksmark{t2}}

\address{
Pengsheng Ji\\
Department of Statistics\\
 University of Georgia\\
Athens, GA 30602\\
\printead{e1}}

\address{Jiashun Jin \\
Department of Statistics\\
Carnegie Mellon University\\
Pittsburgh, PA 15213 \\
 \printead{e2}
}
\end{aug}

\begin{abstract}
We have collected and cleaned  two network data sets: Coauthorship and Citation
networks for statisticians. The data sets are based on all research papers published in four of the top journals in statistics  from $2003$ to the first half of $2012$.  We analyze the data sets from many different perspectives, focusing on
 (a)  centrality, (b) community structures, and (c)  productivity, patterns  and trends.

For (a), we have   identified  the most prolific/collaborative/highly cited authors.  We have also identified a handful of ``hot" papers, suggesting ``Variable Selection" as one of  the ``hot" areas.

For (b),  we have identified about $15$ meaningful communities or research groups, including large-size ones such as   ``Spatial Statistics",  ``Large-Scale Multiple Testing", ``Variable Selection" as well as small-size ones such as   ``Dimensional Reduction",  ``Objective Bayes", ``Quantile Regression",  and ``Theoretical Machine Learning".

For (c), we find that over the 10-year period, both the average number of papers per author and the fraction of self citations have been decreasing, but the proportion of distant citations has been increasing. These  suggest that the statistics community has become
increasingly  more   collaborative,   competitive, and globalized.

Our findings shed light on research habits,  trends, and topological patterns of statisticians. The data sets
provide a fertile ground for future researches on or related to social networks of statisticians.
\end{abstract}



\end{frontmatter}

\section{Introduction} \label{sec:Intro}
It is frequently of interest to identify ``hot"  areas and key authors in a
scientific community, and to understand the research habits, trends, and topological patterns
of the researchers.  A better understanding of such features  is useful in many perspectives, ranging  from that of  administrations and funding agencies on  priorities  for
support,  to that of individual researchers on starting a new research topic or
 new research  collaboration.

Coauthorship and Citation networks provide  a convenient and yet appropriate approach to addressing many of these questions. On one hand,  with the boom of  online resources  (e.g., MathSciNet) and  search engines   (e.g.,  Google Scholar),  it is relatively convenient
for us to collect the Coauthorship and Citation network data of  a specific scientific community. On the other hand,  these network data provide a wide variety  of information (e.g., productivity, trends, impacts, and community structures)  that can be extracted to understand many different aspects of the scientific community.

Recent studies on such networks include but are not limited to the following:
Grossman  \cite{Grossman2002}  studied  the Coauthorship network of mathematicians; Newman \cite{Newman2001, Newman2004}  studied  the Coauthorship networks of biologists, physicists and computer scientists (see also Martin {\it et al}.   \cite{Newman2013}, which  studied  networks of physicists using a much larger data set   than that in \cite{Newman2001, Newman2004});   Ioannidis \cite{Ioannidis2008} used the Coauthorship network to help assess  the scientific impacts.

Unfortunately, as far as we know, the Coauthorship and Citation networks for {\it statisticians} have not yet been studied.  We recognize that
\begin{itemize}
\item The people who are most interested in social networks for statisticians are statisticians themselves or people with close ties to them.
It is unlikely for researchers from other  disciplines (e.g., physicists)
to devote substantial time and efforts to pay {\it specific} attention to networks for statisticians: it is the statisticians'  task  to collect and analyze such network data about themselves and of interest to themselves.
\item For many aspects of the networks,    the ``ground truth''  is  unavailable. However, as statisticians,  we have the advantage of  knowing (at least partially)  many aspects  (e.g., ``hot"  areas, community structures) of our own community.
Such ``partial ground truth''   can be very helpful in analyzing the networks and  interpreting
the results.
\end{itemize}

With substantial time and efforts, we have  collected  two {\it new} network data sets:    Coauthorship  network and Citation network for statisticians. The data sets are based on all published papers from 2003 to the first half of 2012   in four of the top statistical journals: Annals of Statistics (AoS), Biometrika,  Journal of American Statistical Association (JASA) and
Journal of Royal Statistical Society (Series B)  (JRSS-B).

The data sets provide a fertile ground for researches on social networks, especially to us statisticians, as we know the  ``partial ground truth" for many aspects of our community.
For example, we can use the data sets  to check and build network models,   to develop new methods and theory,  and to   further  understand the research habits, patterns, and topological structures of  the networks  of statisticians.  Last but not least, we can use the data sets and  the analysis  in the paper as a starting point for a more ambitious project, where we collect network data sets of this kind but  cover
many more journals in or related to statistics and span a much longer time period.

\subsection{Our findings}
\label{subsec:findings}
In this paper,   we analyze the two network data sets, and discuss each of the following three topics separately:
\begin{itemize}
\item {\it (a). Centrality}.  We identify ``hot"  areas as well as authors  that are most collaborative or  are most highly cited.
\item {\it (b). Community detection}. With possibly more sophisticated methods and analysis, we identify
meaningful  communities of statisticians.
\item {\it (c). Productivity, patterns and trends}.  We identify noticeable publication patterns of the statisticians, and how they
evolve over time.
\end{itemize}

{\bf (a). Centrality}.  Using several different centrality measures, we have identified    Peter Hall, Jianqing Fan, and Raymond Carroll as the most prolific authors,
Peter Hall, Raymond Carroll and Joseph Ibrahim as the most collaborative authors, Jianqing Fan, Hui Zou, and Peter Hall as the most   cited authors. See Table \ref{tab:keyAuthors}.

We have also identified $14$ ``hot"  papers. See  Table \ref{tab:hotPapers}.
Among these $14$ papers, $10$ are on ``Variable Selection", suggesting  ``Variable Selection" as a   ``hot" area.
Other ``hot" areas may include ``Covariance Estimation", ``Empirical Bayes", and
``Large-scale Multiple Testing".

{\bf (b). Community detection}. Intuitively,    communities in a network are groups of nodes that
have more edges within than across (note that ``community" and ``component" are very different concepts); see \cite{SCORE}  for example.    The goal of community detection is to identify such groups (i.e., clustering).

We consider the Citation network and two versions of Coauthorship networks. In each of these networks,
a node is an author.
\begin{itemize}
\item (b1). Coauthorship network (A).  In this network,  there is an (undirected) edge between two authors if and only if they have coauthored  $2$ or more papers in the range of our data sets.
\item (b2). Coauthorship network (B). This is similar to  Coauthorship network (A),  but  ``$2$ or more papers" is replaced by ``$1$ or more papers".
\item (b3). Citation network. There is a  (directed) edge from author $i$ to $j$ if author $i$ has cited $1$ or more papers  by author $j$.
\end{itemize}
While Coauthorship network (B) is defined in a more conventional way, Coauthorship network (A) is easier to analyze, and presents many meaningful research groups that are hard to  find using  Coauthorship network (B).
We now  discuss the three networks separately.

{\it (b1). Coauthorship network (A)}.   We find that  the network  is rather fragmented. It splits into many disconnected components, many of which are groups with special characteristics. The largest component is the ``High Dimensional Data Analysis (Coauthorship (A))" (HDDA-Coau-A) community (Figure \ref{fig:coauthorThreshHiDim}).  The component has $236$ nodes and is relatively large and seems to contain sub-structures;  see Section \ref{subsec:coauthorA} for more discussions.

The next two largest components are presented in Figure \ref{fig:coauthorcoauthorTheoreticalLearningAndDimReduction} and   can be interpreted as communities  of  ``Theoretical Machine Learning"   ($15$ nodes) and  ``Dimension Reduction" ($14$ nodes),  respectively.
The next $5$ components are presented in Table \ref{tab:otherGroups} and can be  interpreted as communities of
 ``Johns  Hopkins", ``Duke", ``Stanford", ``Quantile Regression", and ``Experimental Design", respectively.  These components have small sizes and there is no need for further study on
sub-structures.
\begin{table}[htb!]
\centering
\caption{A road map for $14$ communities discussed in Section \ref{subsec:findings}.    In Coauthorship Network (A),   each community is a component of the network.   In Coauthorship Network (B) and Citation Network, the communities are identified by SCORE and D-SCORE, respectively.}
\scalebox{0.87}{
\begin{tabular}{|l|l|l|l|}
\hline
Network & Communities & $\#$nodes & Visualization\\
\hline
\multirow{8}{*}{Coauthor(A)} & High-Dimensional Data Analysis (HDDA-Coau-A) &
236  & Figures~\ref{fig:coauthorThreshHiDim},3,4\\
& Theoretical Machine Learning & 15 & Figure~\ref{fig:coauthorcoauthorTheoreticalLearningAndDimReduction} \\
& Dimension Reduction &14 &  Figure~\ref{fig:coauthorcoauthorTheoreticalLearningAndDimReduction}\\
\cline{2-4}
& Johns Hopkins & 13 & \multirow{5}{*}{Table~\ref{tab:otherGroups}}\\
& Duke & 10 &\\
& Stanford & 9  &\\
& Quantile Regression & 9  &\\
& Experimental Design & 8 &\\
\hline
\multirow{3}{*}{Coauthor(B)}
& Objective Bayes & 64 & Figure~\ref{fig:coauthorBergerGraph} \\
& Biostatistics & 388 & Figure~\ref{fig:coauthorApplied}\\
& High-Dimensional Data Analysis (HDDA-Coau-B) & 1181 & Figure~\ref{fig:coauthorHiDim}\\
\hline
\multirow{3}{*}{Citation}
& Large-Scale Multiple Testing & 359 & Figure~\ref{fig:Citation-testing}\\
& Variable Selection & 1285 & Figure~\ref{fig:Citation-vs}\\
& Spatial $\&$ Semi-parametric/Non-parametric Statistics & 1010& Figure~\ref{fig:Citation-other-sub}\\
\hline
\end{tabular} \label{tab:roadmap}
}
\end{table}

{\it (b2). Coauthorship network (B)}. The network has much stronger connectivity than Coauthorship network (A), so we need
more sophisticated methods to identify communities/research groups; we propose to  use SCORE.

SCORE is a recent spectral approach to community detection for undirected networks \cite{SCORE}. Using SCORE,    we have identified three meaningful communities as follows:   ``Objective Bayes", ``Biostatistics (Coauthorship  (B))" (Biostat-Coau-B),  ``High Dimensional Data Analysis (Coauthorship (B))" (HDDA-Coau-B), presented in Figures  \ref{fig:coauthorBergerGraph}, \ref{fig:coauthorApplied}, and \ref{fig:coauthorHiDim},  respectively.

We have also investigated the network with several other community detection approaches for undirected networks:  Newman's Spectral Clustering method (NSC) \cite{NewmanSC},  Bickel and Chen's  Profile Likelihood (BCPL) method  \cite{BickelChen2009, zhu}, and Armini {\it et al}'s Profile Likelihood (APL) method \cite{Amini2013}.  Different methods have different results, but they seem to largely agree on the three communities aforementioned; see Section \ref{subsec:coauthorB} for more discussions.

{\it (b3). Citation network}. The Citation network is   directed, and  it remains largely unknown how to model such  networks and how to do community detection.
We propose D-SCORE (an adaption of SCORE for directed network)  as a new
community detection method.
Using D-SCORE, we have   identified three meaningful communities:  ``Large-Scale Multiple Testing",  ``Variable Selection" and ``Spatial and semi-parametric/nonparametric Statistics".  These communities are   presented in
Figures \ref{fig:Citation-testing}-\ref{fig:Citation-other-sub} respectively.

For convenience, we present in Table \ref{tab:roadmap} a road map for the $14$ communities we just mentioned. Note that some of these communities also have sub-communities; see  Sections \ref{sec:coauthorcomm}-\ref{sec:citationcomm} for details.

In comparison, the communities or research groups identified in  each of the three networks are connected, intertwined, but are also very  different. We discuss these in Sections \ref{subsubsec:compcoauthorA}-\ref{subsubsec:compcoauthorB}; see details therein.

{\bf (c). Productivity, patterns and trends}. We discuss the overall productivity,  coauthor patterns and trends, and citation patterns and trends.
 Our findings include but  not limited to the following.
\begin{itemize}
\item In the $10$-year period 2003-2012,  the number  of papers per author has been decreasing  (Figure \ref{fig:nPaper}).  Also, the proportion of self-citations has been decreasing  while  the proportion of distant citations has been increasing (Figure \ref{fig:LC-citations}).
These  suggest that the statistics community has become increasingly more collaborative,
 competitive, and globalized.
\item The distribution of either  the  degrees of the author-paper bipartite network or the Coauthorship  network  has  a  power-law tail (Figures  \ref{fig:LC}-\ref{fig:NCoauthor}), a phenomenon frequently found in social networks
\cite{Barabasi1999, Newman2001a}.
\end{itemize}

\subsection{Data collection and cleaning}
\label{subsec:dataset}
We have faced substantial challenges in data collection and cleaning, and it has taken us more than $6$ months to obtain high-quality data sets and prepare them in a ready-to-use format.

At first glance, it may be hard to understand why it is challenging to collect such  data:  the  data seem to be  everywhere,   very accessible and free.

This is true to some extent.
However,  when it comes to  high-volume high-quality data, the resources  become  surprisingly limited.  For example,  Google Scholar aggressively blocks any one (a person or a machine) who tries to download the data more than just a little;   when  you try to download   little by little, you will see some portion of the data are made messy and incomplete intentionally. For other online resources, we face a similar problem.

We also face other challenges:
missing paper identifiers, ambiguous author names, etc.; we explain how we have overcome these in
 Appendix II.

\subsection{Experimental design and scientific relevance}
\label{subsec:exp}
We are primarily interested in the networks for statisticians home based in USA. For this reason,
we have limited our attention to four journals (AoS, Biometrika, JASA, JRSS-B), which are regarded by many   US-based statisticians  the top
statistical journals (or leading journals in methods and theory,  except for JASA papers in the case study sector). We recognize that we may have different results when we include in our study   either journals which are the main venues for statisticians from a different country or region,   or journals which are the main venues
for statisticians with a different focus (e.g., Bioinformatics).

  We are also primarily interested in the time period when high dimensional data analysis
emerged as a new statistical area. We may have different results if we extend the study to a much longer time period.

On the other hand, it seems that the data sets we have serve well for solving our targeted scientific problems:  they provide many meaningful results in many aspects of our targeted community within the targeted time period.
They  also serve as a starting point for a more ambitious project in which we collect data from many more journals in a much longer time period.

\subsection{Disclaimers}
\label{subsec:disclaimer}
Our primary goal in the paper is to present the data sets we collect, and
to report our findings in such data sets.  It is not our intention to rank one author/paper  over the others.  We wish to clarify that ``highly cited" is not exactly the same as ``important" or ``influential".
It is not our intention either to rank one area over the other.  A ``hot" area is not exactly the same as an ``important" area  or an area that needs the most of our time and efforts.  It is not exactly an area that is exhausted (so we should not dive in)  either.

Also,   it is not our intention to label an author/paper/topic with a certain community/group/area.
A community or a research group
may contain  many authors,   and can be   hard to interpret.  For presentation, we need to assign  names to  such  communities/groups/areas, but
the  names do not always accurately reflect all the authors/papers in them.

Finally,  social networks are about ``real people", and this time,  ``us".  In order to obtain meaningful and interpretable results,  we have to use real names. We have not used any data beyond those which are publicly  available. The interest of the paper is on the statistics community {\it as a whole}, not on any individual statistician.
\subsection{Contents}
The  paper is organized as follows.  In  Section \ref{sec:centrality},  we discuss  the centrality.  In  Sections  \ref{sec:coauthorcomm}-\ref{sec:citationcomm}, we discuss community detection for the Coauthorship  network and  Citation network, respectively. Section \ref{sec:Discu} contains a brief summary and discusses the  limitations of the paper and  suggests  some future directions.  Section \ref{sec:pt} is Appendix I, where we study the productivity, patterns and trends for the statisticians' research,  and Section \ref{sec:data} is Appendix II, where we address the challenges in data collection and cleaning.

\section{Centrality}
\label{sec:centrality}
It is frequently of interest to identify the most ``important" authors or  papers, and one possible approach is to use
centrality.  There are many different  measures of centrality. In this section, we use
the degree centrality, the closeness centrality, and the betweenness centrality.
The closeness centrality is  defined as the reciprocal of the total distance to all others \cite{Sabidussi}.
The betweenness centrality measures the extent to which a node is located ``between" other pairs of nodes \cite{Freeman}.

The degree centrality is conceptually simple, but the definition varies with the types of networks.   For the author-paper bipartite network,   the centrality of an author  is the number of papers  he/she  publishes. For   Coauthorship  network, the centrality of an author is  the number of his/her coauthors.  For Citation network of {\it authors}, we are primarily interested in the in-degree, and the centrality of an author is the number of citers (i.e.,
  authors who cite his or her papers). For Citation network of {\it papers}, the centrality is the in-degree
(i.e., the number of papers which cite this paper).

 Table \ref{tab:keyAuthors} presents the key authors identified by  different measures of
centrality.  The results suggest that different measures of centrality are largely consistent with each other, which identify Raymond Carroll,  Jianqing Fan, and Peter Hall (alphabetically) as  the ``top $3$"
authors.
\begin{table}[htb!]
\centering
\caption{Top $3$ authors identified by the degree centrality (Columns $1$-$3$; corresponding networks are  the author-paper bipartite network, Coauthorship  network, and Citation network for authors),    the closeness centrality and the betweenness centrality.}
\scalebox{0.92}{
\begin{tabular}{|lllll|}
\hline
$\#$ of papers    &  $\#$ of coauthors & $\#$ of citers     & Closeness            & Betweenness \\
\hline
Peter Hall        &  Peter Hall          &  Jianqing Fan         &  Raymond Carroll   &  Raymond   Carroll    \\
Jianqing Fan       &  Raymond   Carroll   &  Hui Zou              &  Peter Hall          &  Peter Hall    \\
Raymond Carroll  &  Joseph   Ibrahim    &  Peter Hall           &  Jianqing Fan        &  Jianqing Fan    \\
\hline
\end{tabular}}
\label{tab:keyAuthors}
\end{table}

Table \ref{tab:hotPapers} presents  the ``hot"  papers identified by $3$ different measures of centrality.  For all these measures,
the ``hottest" papers seem to be in the area of variable selection.
In particular, the top $3$ most cited paper are  Zou \cite{Zou2006} (75 citations; adaptive lasso),    Meinshansen and Buhlmann \cite{Meinshausen2006} (64 citations; graphical lasso), and  Cand\`es and Tao \cite{Candes2007} (49 citations; Dantzig Selector). The three papers are all in a specific sub-area in high dimensional variable selection, where the theme  is to extend the well-known penalization methods of the lasso \cite{Donoholasso, Tibshirani}
 in various directions (these fit well with the impression of many statisticians:
in the past $10$-$20$ years, there is a noticeable wave of research papers devoted to the penalization methods).

\begin{table}[htb!]
\centering
\caption{Fourteen ``hot" papers (alphabetically) identified  by degree centrality (Column 2; for citation networks of papers), closeness centrality, and betweenness centrality.   Numbers in Column $2$-$4$ are the ranks (only shown when the rank is   smaller than $5$).}
\scalebox{0.875}{
\begin{tabular}{l|ccc}
Paper (Area)                                                &  Citations  & Closeness & Betweenness \\
\hline
Bickel $\&$ Levina (2008)  \cite{BLT} (Covariance Estimation)       &            &            & 4      \\
\hline
Candes $\&$ Tao (2007) \cite{Candes2007} (Variable Selection)       & 3          &            &       \\
\hline
Fan $\&$ Li (2004) \cite{FanLi2004}  (Variable Selection)                &            & 2          &       \\
\hline
Fan $\&$  Lv (2008) \cite{FanLv}  (Variable Selection)             &            &            & 1    \\
\hline
Fan $\&$ Peng (2004) \cite{FanPeng2004} (Variable Selection)           & 4          & 1          &       \\
\hline
Huang et al (2006) \cite{Huang2006} (Covariance  Estimation)          &            &            & 3      \\
\hline
Huang et al (2008) \cite{Huang2008} (Variable Selection)          &            &            & 5      \\
\hline
Hunter $\&$ Li (2005) \cite{HunterLi2005} (Variable Selection)           &            & 4          &       \\
\hline
Johnstone $\&$ Silverman (2005) \cite{JohnstoneSilverman2005} (Empirical Bayes) &            & 5          &       \\
\hline
Meinshausen $\&$ Buhlmann (2006) \cite{Meinshausen2006} (Variable Selection) & 2         &            &       \\
\hline
Storey (2003) \cite{Storey2003} (Multiple Testing)                &            & 3          &       \\
\hline
Zou (2006) \cite{Zou2006} (Variable Selection)          & 1          &            &       \\
\hline
Zou $\&$ Hastie (2005) \cite{ZouHastie2005} (Variable Selection)                & 5          &            &       \\
\hline
Zou $\&$  Li (2008) \cite{ZouLi2008} (Variable Selection)                    &            &            & 2          \\
\hline
\end{tabular}
\label{tab:hotPapers}
}
\end{table}

These results suggest  ``Variable Selection" as one of the ``hot" areas.  Other ``hot" areas may include ``Covariance Estimation", ``Empirical Bayes", and ``Large-Scale Multiple Testing"; see Table \ref{tab:hotPapers} for details.

For more information, note that
at   \verb+www.stat.uga.edu/~psji/+,   we have  listed the
$30$ most cited papers in the file top-cited.xlsx.  These $30$ papers account  for $16\%$ of the total number of citation counts.
 The list furthers shows  that the most highly cited papers are on the regularization methods (e.g., adaptive lasso, group lasso, etc.).

On the other hand, we must note that some important and innovative works in the  particular area of variable selection have significantly fewer citations. This includes but is not limited to the phenomenal  paper  by Efron {\it et al}. (2004)  \cite{Efron}   on  least angle regression, which
has received a lot of  attention from a broader scientific community. The paper has $4900$ citations on Google Scholar,  but is cited only $11$ times by papers in our data set (in comparison, the adaptive lasso paper \cite{Zou2006} has received $75$ citations). A similar claim can be drawn on other areas or topics.

The fact that statisticians have been very much focused on a very specific research topic and a very specific approach is an
interesting phenomenon that deserves more explanation by itself.

The centrality measures we use here are either natural choices or existing measures.  We are merely reporting what the data sets tell us,  with no intention
to rank one author or an area over the others; see Section \ref{subsec:disclaimer}.

\section{Community detection for Coauthorship networks}
\label{sec:coauthorcomm}
In this section, we study community detection for  Coauthorship networks (A) and (B). Community detection of  the Citation network is  discussed in Section \ref{sec:citationcomm}.
In Section \ref{subsec:methods}, we discuss models for general undirected networks and  recent approaches
to community detection. In Sections \ref{subsec:coauthorA}-\ref{subsec:coauthorB}, we analyze the Coauthorship network (A) and (B), respectively, using these approaches.

\subsection{Community detection methods (undirected networks)}
\label{subsec:methods}
Community detection is a  problem of major interest in  network analysis \cite{Fienberg}.  Consider an {\it undirected} and {\it connected} network ${\cal N} = (V, E)$ with $n$ nodes.
We think $V$ as the union of a few (disjoint) subsets which we
call the ``communities":
\[
V = V^{(1)} \cup V^{(2)} \ldots \cup V^{(K)},
\]
where ``$\cup$" stands for  the union of  sets and has nothing to do with networks (same below).
Intuitively, communities can be thought of as subsets of nodes where there are more edges ``within'' than ``across"communities (e.g.,  \cite{Bickel}).
Note that for simplicity, we assume the communities are non-overlapping here.
The goal of community detection is for each node $i \in V$,  to decide to which community it belongs (i.e., clustering).

There are many  community detection methods for undirected networks. In this paper, we consider
Newman's Spectral Clustering approach (NSC) \cite{NewmanSC},
Bickel and Chen's Profile Likelihood approach (BCPL) \cite{Bickel, zhu},
Armini {\it et al}.'s Pseudo Likelihood approach (APL) \cite{Amini2013}, and Jin's SCORE  \cite{SCORE}.

NSC is a spectral method, where  the  key observation is that Newman and Girvan's modularity matrix can be approximated by the leading
  eigenvectors of the  matrix \cite{NewmanSC}.   Newman introduced NSC as a  general  idea for spectral clustering, and there are several different ways for implementations.  Following
\cite{NewmanSC},  we
cluster by using the signs of the first leading eigenvectors when $K = 2$,  and by using the recursive bisections approach when $K \geq 3$.

BCPL is a penalization method proposed by Bickel and Chen \cite{Bickel} which uses greedy search to maximize the profile likelihood and works well for networks with thousands of nodes. When the network size is large, BCPL may be computationally slow. In light of this, Amini {\it et al}.  \cite{Amini2013}
propose a different Profile Likelihood approach which aims to improve the speed of BCPL. By doing so, the price it pays is to   ignore  some   dependence  structures of the data so as to simplify the likelihood and make it more tractable.

SCORE, or {\bf S}pectral {\bf C}lustering {\bf O}n {\bf R}atios of {\bf E}igenvectors,
is a recent spectral method proposed by Jin \cite{SCORE}.
Assume $K$ (number of communities) as known and let $A$ be the adjacency matrix associated with ${\cal N}$:
\begin{equation} \label{DefineA}
A(i,j) =  \left\{
\begin{array}{ll}
1, &\qquad  \mbox{if there is an edge between nodes  $i$ and $j$},   \\
0, &\qquad  \mbox{otherwise};
\end{array}
\right.
\end{equation}
note that $A$ is symmetric.
SCORE consists of the following simple steps.
\begin{itemize}
\item Let $\hat{\xi}_1, \hat{\xi}_2, \ldots, \hat{\xi}_K$ be the first $K$ (unit-norm) eigenvectors of $A$.
Obtain the $n \times (K-1)$ matrix $\hat{R}$  by
$\hat{R}(i,k) =  \hat{\xi}_{k+1}(i)/ \hat{\xi}_1(i)$,  $1 \leq i \leq n$, $1 \leq k \leq K-1$.
\item Clustering by applying the classical k-means to $\hat{R}$, assuming there are $\leq K$ communities.
\end{itemize}

{\bf Remark 1}. SCORE is motivated by the recent Degree  Corrected Block Model (DCBM, \cite{DCBM}).
In DCBM, for $n$ degree heterogeneity parameters $\{\theta(i)\}_{i = 1}^n$ and a $K \times K$ symmetric
matrix $P$,
we think $A(i,j)$, $1 \leq i < j \leq p$ as independent Bernoulli random variables
such that  $P(A(i,j) = 1) =  \theta(i) \theta(j) P_{k,\ell}$,  if $i \in V^{(k)}$ and $j \in V^{(\ell)}$, $1 \leq k, \ell \leq K$.  SCORE
recognizes that,   the parameters $\theta(i)$'s are nearly ancillary, and can be conveniently removed by taking entry-wise ratios between $\hat{\xi}_k$ and $\hat{\xi}_1$, $k = 2, \ldots, K$; see \cite{SCORE}.
Originally proposed for undirected network, SCORE is a flexible idea and can be used to analyze other types of networks.
In Section \ref{sec:citationcomm}, we extend SCORE to Directed-SCORE (D-SCORE) as an approach to community detection for {\it directed} networks, and use it to analyze the Citation network.

{\bf Remark 2}.
Note that the vectors of  predicted labels by different methods could be very different. For a pair of the predicted label vectors, we measure the similarity by the  Adjusted Rand Index (ARI)
 \cite{HubertArabie1985} and  the Variation of Information (VI)  \cite{Meila2003};
a large  ARI or a small VI suggests that two predicted label vectors are similar to each other.

\subsection{Coauthorship network (A)}
\label{subsec:coauthorA}
In this network, by definition, there is an edge between two nodes (i.e., authors) if and only if
they have coauthored $2$ or more papers (in the range of our data sets).
The network  is very much fragmented: the total of $3607$ nodes
split into  $2985 $ different components, where
$2805$ ($94\%$) of them are singletons,  $105$ ($3.5\%$) of them  are pairs, and  the average component size is $1.2$.

\begin{figure}[htb!]
\centering
\includegraphics[width = 5 in, height = 4 in]{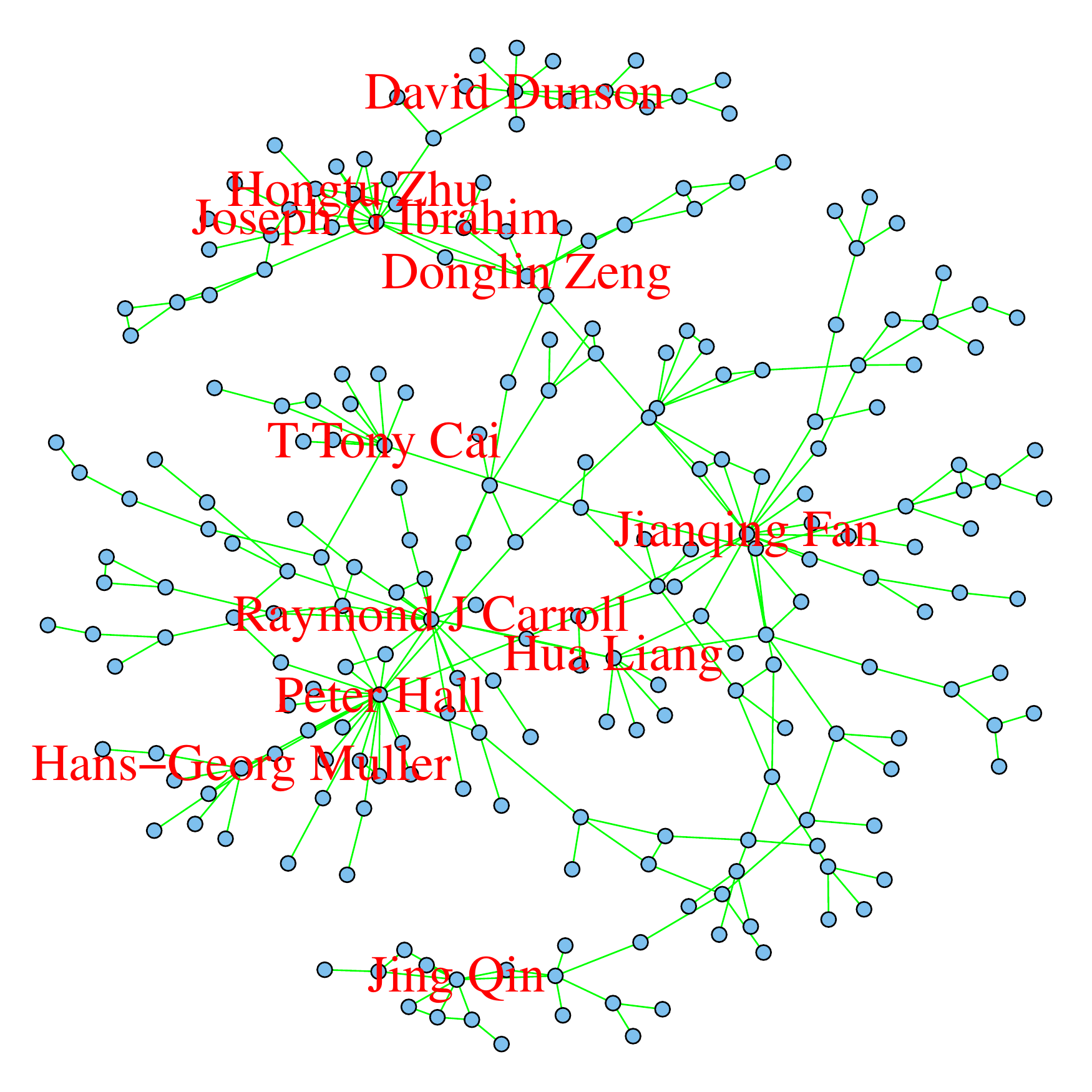}
\caption{The giant component of Coauthorship network (A). It could be interpreted as the ``High Dimensional Data Analysis (Coauthorship (A))" (HDDA-Coau-A) community. Names are only shown for $11$ nodes with a degree of $8$ or larger.  }
\label{fig:coauthorThreshHiDim}
\end{figure}

\begin{figure}[htb!]
\centering
\includegraphics[width = 1.63 in]{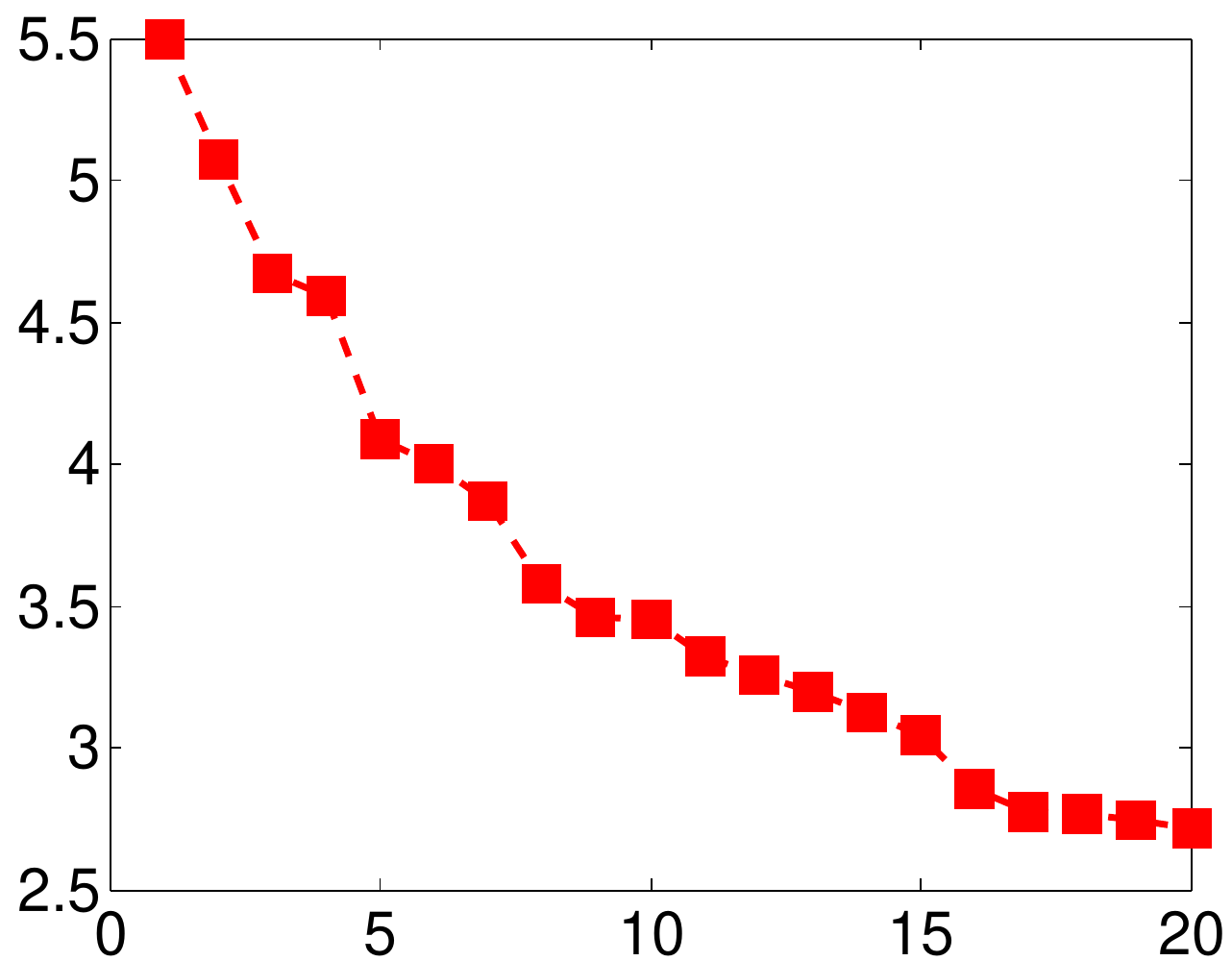}
\includegraphics[width = 1.63 in]{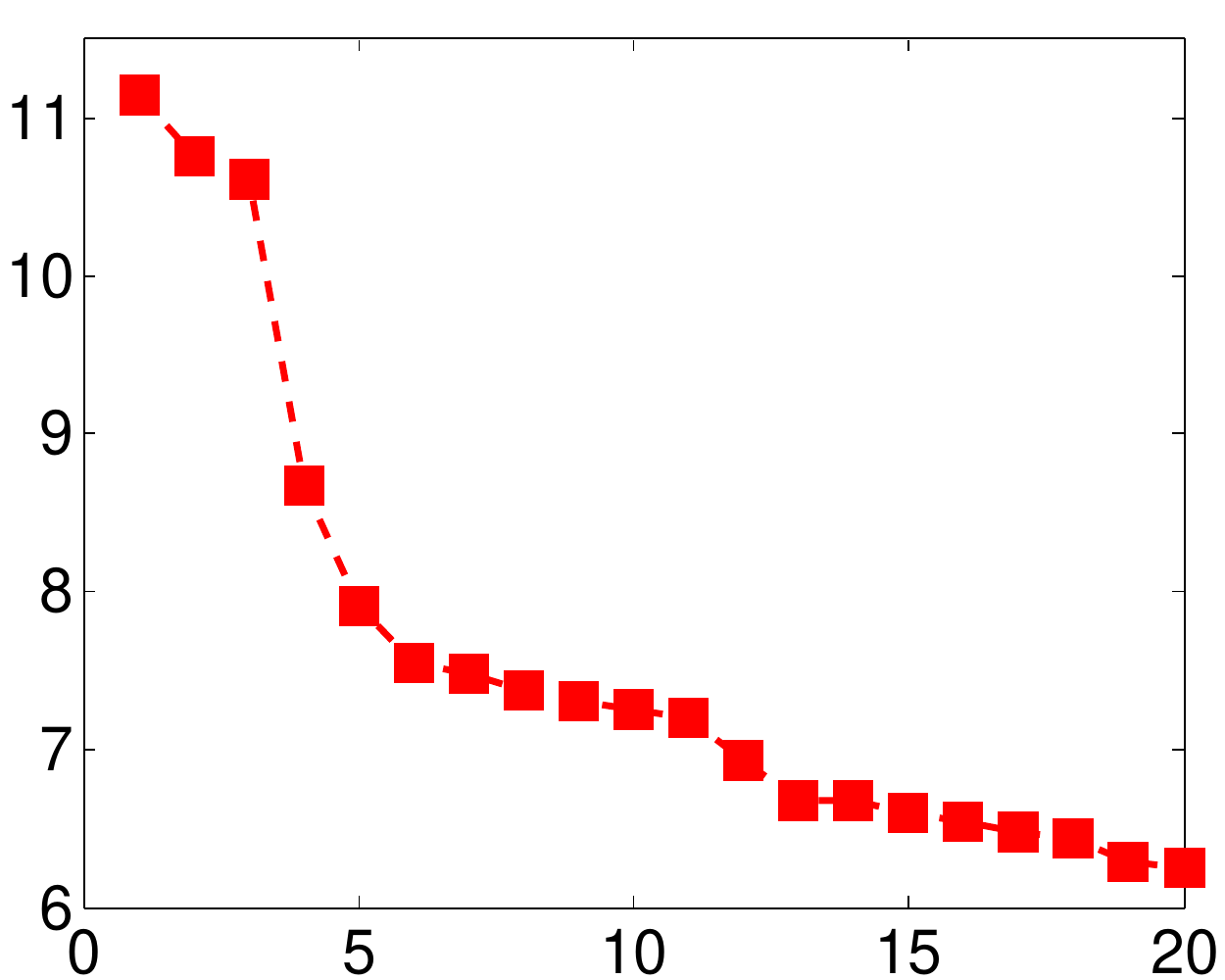}
\includegraphics[width = 1.63 in]{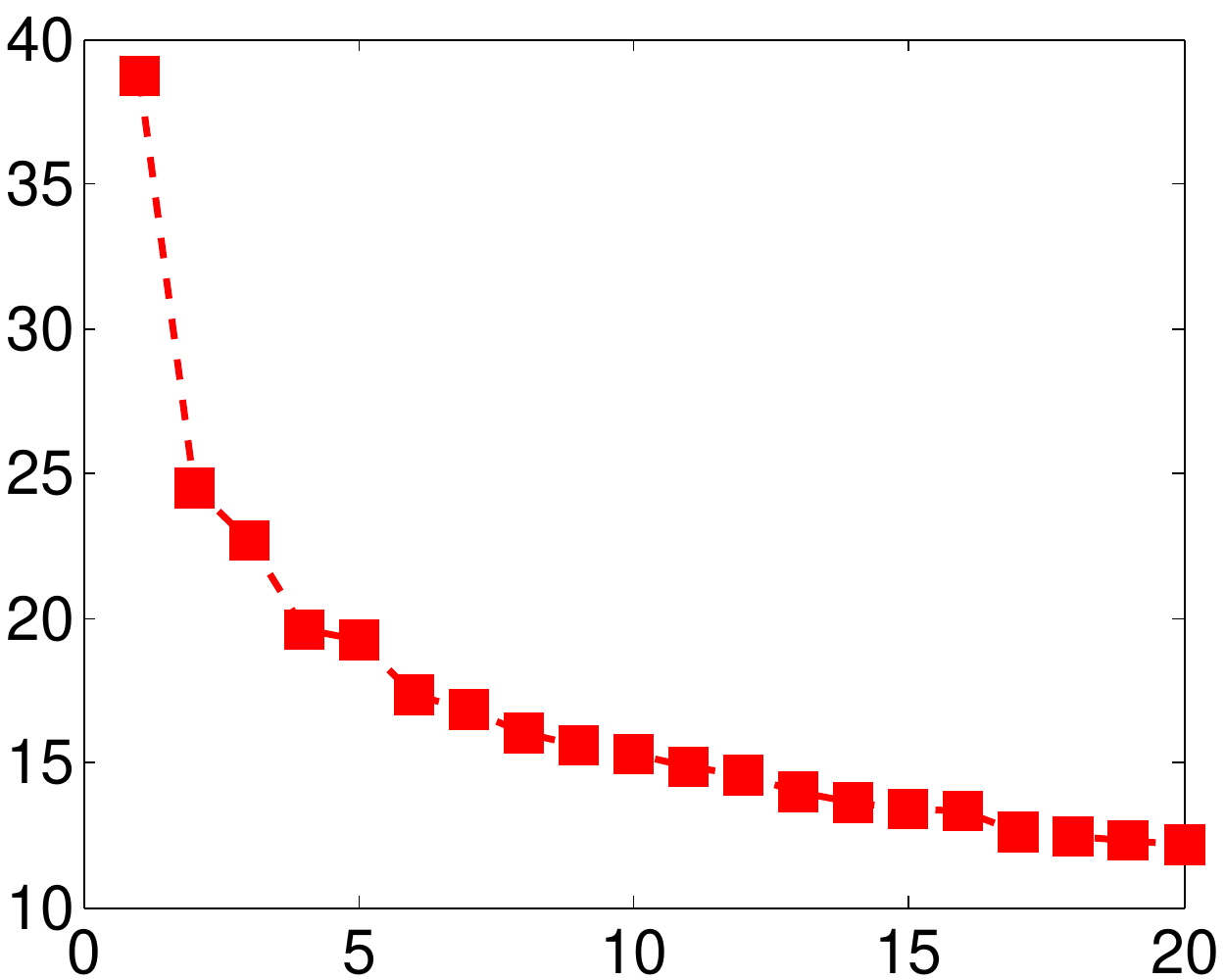}
\caption{Scree plots. From left to right:  the giant component of Coauthorship network(A),   Coauthorship network(B),  Citation network (in the last one, we display singular values instead of eigenvalues).  }
\label{fig:coauthorGiantScree}
\end{figure}

The giant component ($236$ nodes) is seen to be  the ``High Dimensional Data Analysis (Coauthorship (A))" group (HDDA-Coau-A); see  Figure \ref{fig:coauthorThreshHiDim}.
It seems that the giant component has sub-structures  (i.e., communities).
In the  left panel of Figure \ref{fig:coauthorGiantScree}, we plot the scree-plot of this group.
The elbow point of the scree-plot maybe at the $3rd$, $5th$, or $8$th largest eigenvalue, suggesting that
there may be $2$, $4$, or $7$ communities.
In light of this,  for each $K$ with $2 \leq K \leq 7$,  we run SCORE, NSC, BCPL and APL and  record the
corresponding vectors of predicted labels.  We find that for $K \geq 3$,   the results by different methods are largely inconsistent with each other:  the maximum of ARI and the minimum VI (see Remark 2 in Section \ref{subsec:methods}  for discussions on ARI and VI)  across different pairs of methods are $0.15$ and  $1.19$, respectively.

We now focus on the case of $K = 2$. In Table \ref{tab:ThreshGiantCompareK2},  we present the ARI  and VI  for each pair of the methods.  The table suggests  that: the $4$ methods split into two groups where SCORE and APL are in one of the group with an ARI of $0.72$ (between them),
and NSC and BCPL are in the other group with an ARI of $0.21$.
 The results for methods in each group are moderately consistent to each other, but those for methods in different groups are rather inconsistent. The point is confirmed by Table \ref{tab:communitySizesA}, which compares the sizes of the communities identified by the $4$ methods.

\begin{figure}[htb!]
\centering
\includegraphics[width = 5 in, height = 2.85 in]{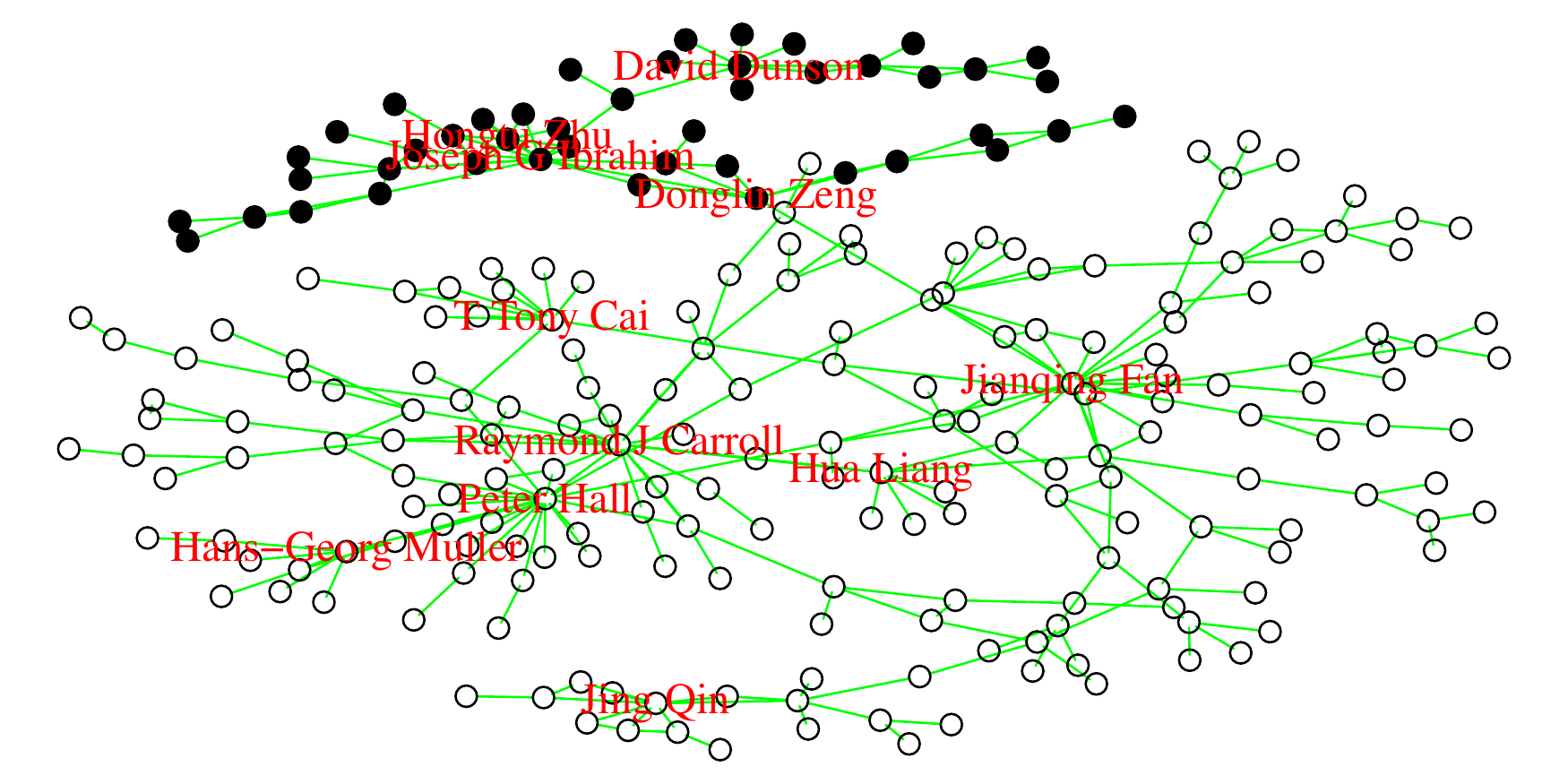}
\includegraphics[width = 5 in, height = 2.85 in]{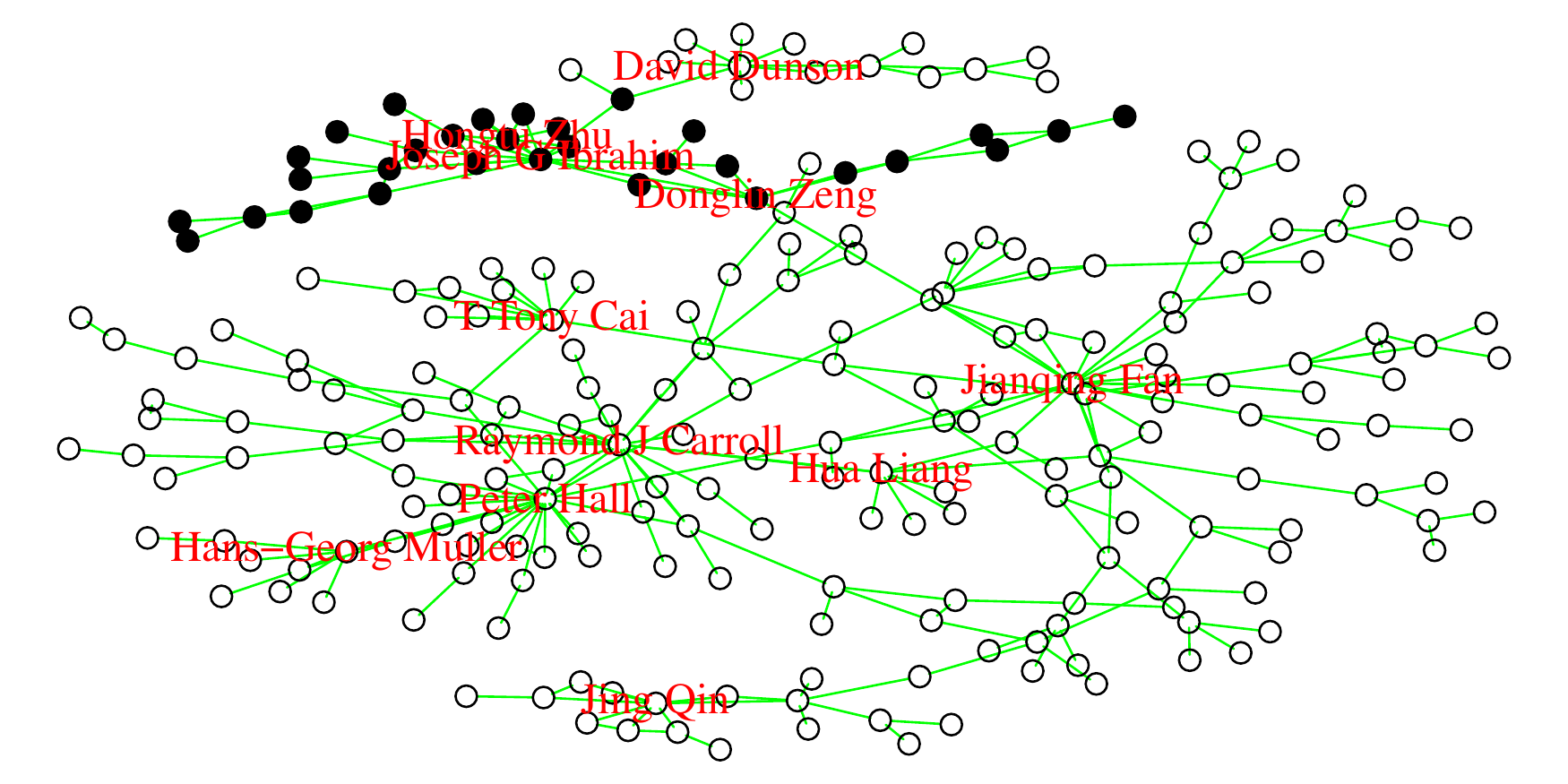}
\caption{Community detection results by SCORE (top) and APL (bottom) for the giant component of Coauthorship network (A), assuming $K = 2$.   Nodes in black (solid) dots  and white circles  represent  two different communities.}
\label{fig:coauthorThreshHiDimColor1}
\end{figure}

\begin{figure}[htb!]
\centering
\includegraphics[width = 5 in, height = 2.85 in]{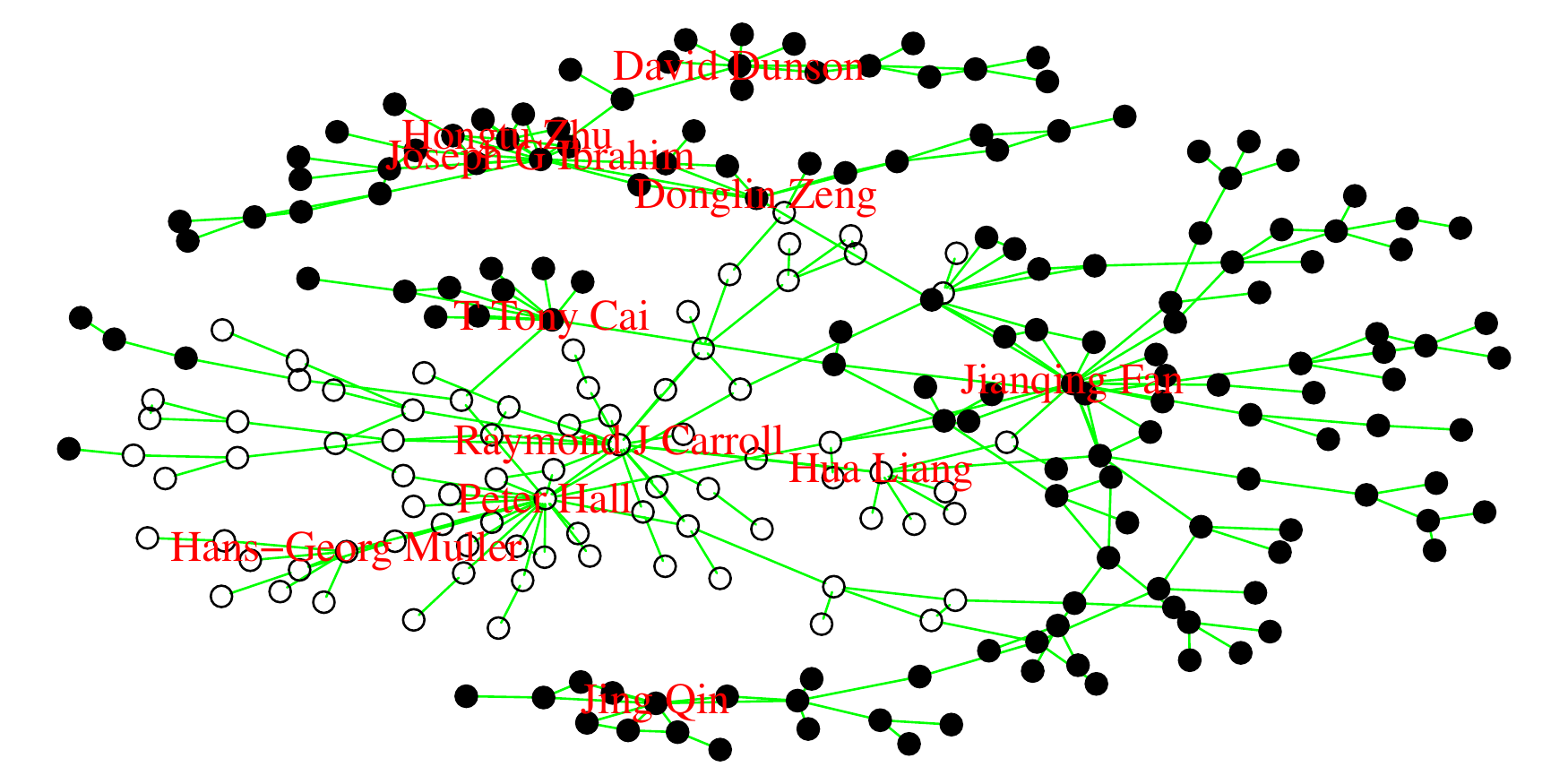}
\includegraphics[width = 5 in, height = 2.85 in]{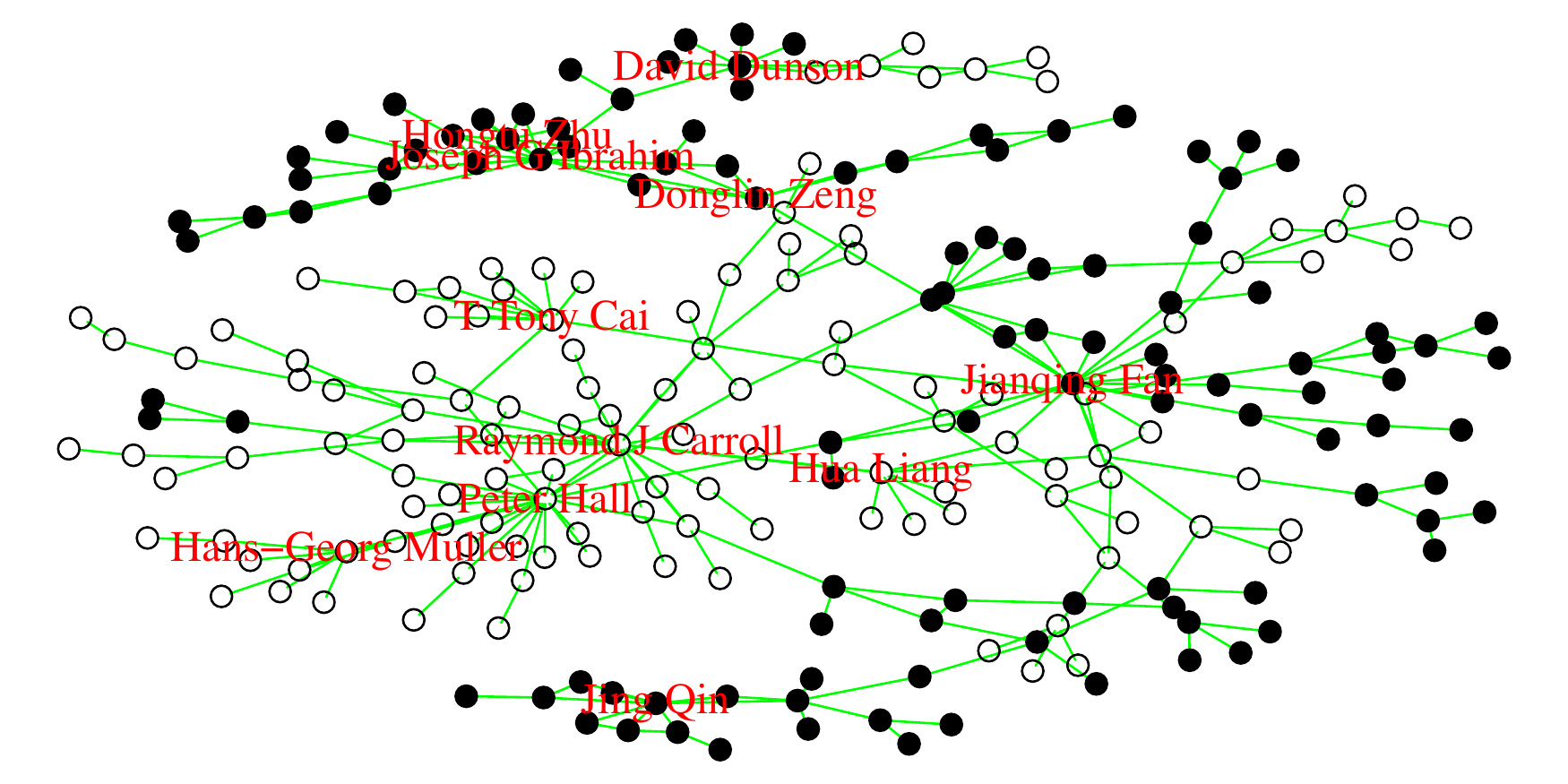}
\caption{
 Community detection results by NSC (top) and BCPL (bottom) for the giant component of Coauthorship network (A), assuming $K = 2$.   Nodes in black (solid) dots  and white circles represent two different communities.}
\label{fig:coauthorThreshHiDimColor2}
\end{figure}

In Figures \ref{fig:coauthorThreshHiDimColor1}-\ref{fig:coauthorThreshHiDimColor2}, we further compare
the community detection results by each of the $4$ methods ($K  = 2$).
In each panel, nodes are marked with either black dots or white circles,
representing two different communities.  It seems that all four methods agree that there are
two communities as follows.
\begin{itemize}
\item ``North Carolina" community. This includes a group of researchers from Duke Univ.,  Univ.  of North Carolina, North Carolina State Univ.
\item ``Carroll-Hall" community. This includes a group of researchers in non-parametric and semi-parametric statistics, functional estimation, and high dimensional data analysis.
\end{itemize}
Comparing the results by different methods, one of the major discrepancies lies in the ``Fan"  group:  SCORE and APL  cluster the  ``Fan"  group into the ``Carroll-Hall" community,
 and NSC and BCPL cluster it  into the``North Carolina" community.
A possible explanation  is that, the ``Fan"  group has strong ties
to both communities.

This may also suggest there are $3$ communities (instead of $2$) in this component. However, as mentioned before,   when we assume $K = 3$, the results by all four methods are rather inconsistent with   each other. How to obtain a more convincing explanation is an interesting  but challenging problem. We omit further discussions along this line for reasons of space.

\begin{table}[hbt!]
\centering
\caption{The Adjusted Random Index (ARI) and Variation of Information  (VI) for the vectors of  predicted community labels by four different methods for the  giant component of Coauthorship (A), assuming $K = 2$.  A large ARI/small VI suggests that the two predicted label vectors are similar to each other.}

\begin{tabular}{l|r|r|r|r}
          & SCORE & NSC & BCPL & APL \\
SCORE& 1.00/.00 & -.04/.95 & .09/1.05 & .72/.33   \\
NSC&   & 1.00/.00 & .21/1.06 & -.06/.91   \\
BCPL&   &   & 1.00/.00 & .09/.87   \\
APL &   &   &   & 1.00/.00  \\
\end{tabular}
\label{tab:ThreshGiantCompareK2}
\end{table}

\begin{table}[hbt!]
\centering
\caption{Comparison of community sizes by different methods assuming $K = 2$ for the  giant component of Coauthorship network (A).  }
\begin{tabular}{r|c|c}
& North Carolina & Carroll-Hall  \\
\hline
SCORE &  45 & 191   \\
\hline
NSC  & 155 &  81  \\
\hline
APL   &31 & 205 \\
\hline
\hline
SCORE  $\cap$  NSC  & 45 &  81 \\
\hline
SCORE $\cap$ APL    &   31 & 191 \\
\hline
NSC $\cap$ APL & 31 & 81  \\
\hline
\hline
SCORE  $\cap$  NSC $\cap$  APL  &  31 &   81 \\
\end{tabular}
\label{tab:communitySizesA}
\end{table}




Other noteworthy discrepancies are as follows:
\begin{itemize}
\item
SCORE includes the ``Dunson"  branch in the ``North Carolina" group, but APL 
clusters them into the ``Carroll-Hall" group to  which they are not directly connected. In this regard,  it seems that results by SCORE are more meaningful.
\item NSC and BCPL differ on several small branches, including the   ``Dunson" branch and two small branches connecting to  Jianqing Fan. In comparison, the results by NSC seem more meaningful.
\end{itemize}


\begin{figure}[htb!]
\centering
\includegraphics[width = 5 in, height = 2.5 in]{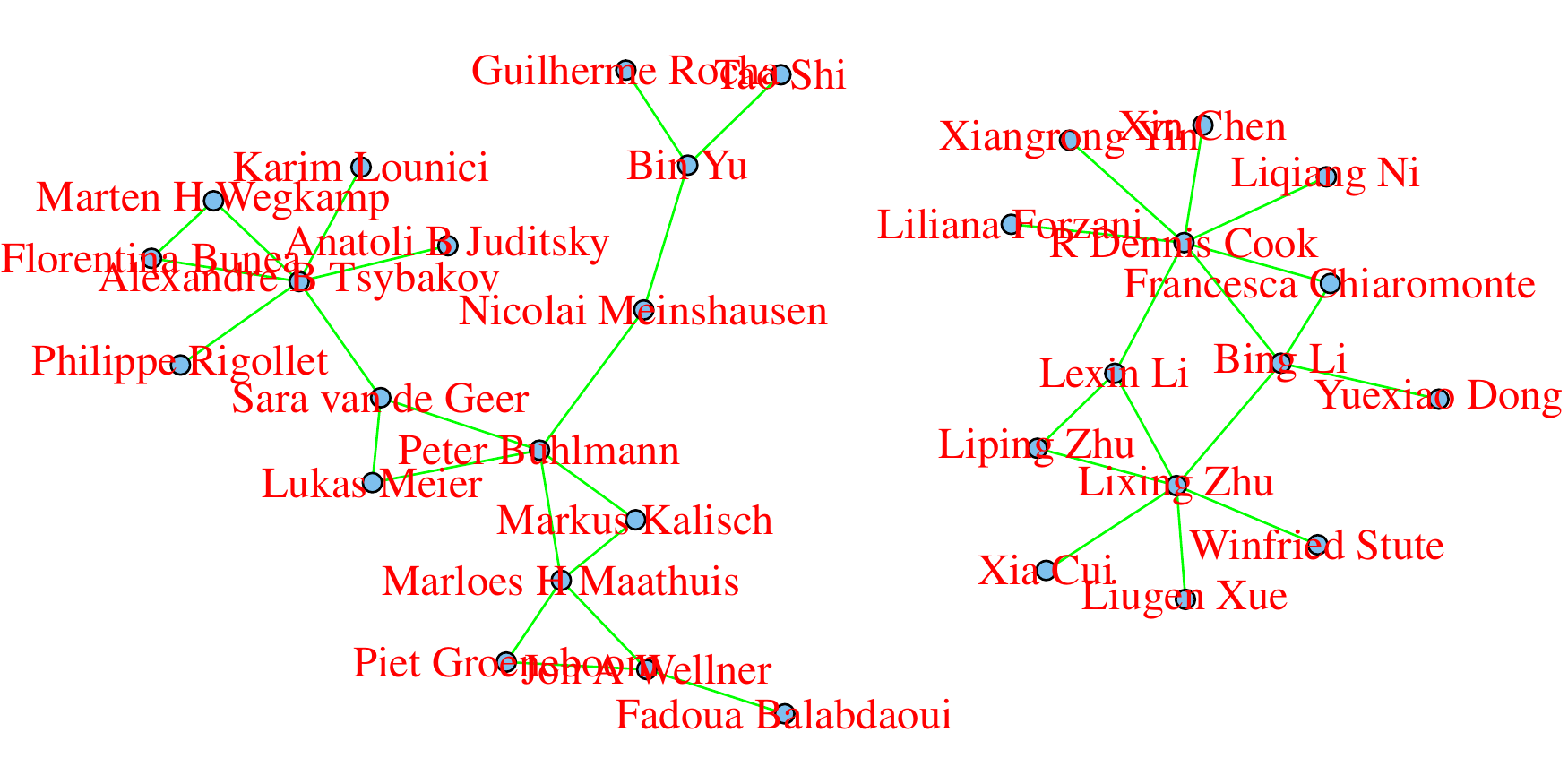}
  \caption{The second largest (left) and third largest (right) components of Coauthorship network (A). They can be possibly interpreted as the ``Theoretical Machine Learning" and ``Dimension Reduction" communities, respectively.  }
    \label{fig:coauthorcoauthorTheoreticalLearningAndDimReduction}
\end{figure}

We now move away from the giant component.
The next two largest components are the   ``Theoretical Machine Learning" group ($15$ nodes) and the ``Dimension Reduction" group ($14$ nodes); see Figure \ref{fig:coauthorcoauthorTheoreticalLearningAndDimReduction}.
The first one is a  research group who work on Machine Learning topics using sophisticated statistical theory,
 including Peter Buhlmann,  Alexandre Tsybakov, Jon Wellner, and Bin Yu.
The second one is a research group  on Dimension Reduction, including Francesca Chiaromonet,
Dennis Cook, Bing Li and their collaborators.

A conversation with Qunhua Li 
 helps to illuminate why these groups are meaningful and how they evolve  over time. In the first community, Marloes H. Maathuis obtained her Ph.D from University of Washington
(jointly supervised by Jon Wellner and Piet Groeneboom) in 2006 and then went on to work
in ETH, Switzerland, and she is possibly  the ``bridge" connecting the Seattle group and the ETH group   (Peter Buhlmann, Markus Kalische, Sara van de Geer).  Nocolai Meinshausen could be one of the  ``bridge"  nodes  between ETH and Berkeley:
he was a Ph.D student of Peter Buhlmann and then a post-doctor at Berkeley.
In the second group,
Ms. Chiaromonet obtained her Ph.D from University of Minnesota, where Dennis Cook served as the supervisor. She then went on to work in the  Statistics Department  at Pennsylvania
State University, and started to collaborate with Bing Li on Dimension Reduction. 

The next $5$ largest components in Coauthorship network (A)
 are the ``Johns Hopkins" group ($13$ nodes; including faculty at  Johns Hopkins University and their  collaborators; similar below),
``Duke" group  ($10$ nodes; including Mike West, Jonathan Stroud, Carlos Caravlaho, etc.),   ``Stanford"  group ($9$ nodes including David Siegmund, John Storey, Ryan Tibshirani, and  Nancy Zhang, etc.),    ``Quantile Regression" group ($9$ nodes; including Xuming He and his  collaborators),
and 	``Experimental Design"  group  ($8$ nodes). These groups are presented in Table  \ref{tab:otherGroups}.

\begin{table}[htb!]
\centering
\caption{Top:  the $4$-th, $5$-th, and $6$-th largest
components of Coauthorship network (A) which can be interpreted as the groups of  ``Johns Hopkins", ``Duke", and ``Stanford").
Bottom:  the $7$-th and $8$-th largest components of Coauthorship network (A) which can be interpreted as the groups of ``Quantile Regression" and ``Experimental Design".  }
\begin{tabular}{l}
 Barry Rowlingson     \\[-0.7ex]
 Brian S Caffo\\[-0.7ex]
 Chong-Zhi Di     \\[-0.7ex]
 Ciprian M Crainiceanu\\[-0.7ex]
 David Ruppert        \\[-0.7ex]
 Dobrin Marchev\\[-0.7ex]
 Galin L Jones      \\[-0.7ex]
 James P Hobert\\[-0.7ex]
 John P Buonaccorsi      \\[-0.7ex]
 John Staudenmayer\\[-0.7ex]
  Naresh M Punjabi        \\[-0.7ex]
  Peter J Diggle\\[-0.7ex]
  Sheng Luo\\[-0.7ex]
\end{tabular}
\begin{tabular}{l}
   Carlos M Carvalho \\[-0.7ex]
   Gary L Rosner\\[-0.7ex]
   Gerard Letac\\[-0.7ex]
   Helene Massam\\[-0.7ex]
  James G Scott\\[-0.7ex]
  Jonathan R Stroud\\[-0.7ex]
   Maria De Iorio\\[-0.7ex]
   Mike West\\[-0.7ex]
   Nicholas G Polson\\[-0.7ex]
  Peter Muller \\[-0.7ex]
  \\
  \\
\end{tabular}
\begin{tabular}{l}
  Armin Schwartzman\\[-0.7ex]
  Benjamin Yakir\\[-0.7ex]
  David Siegmund\\[-0.7ex]
  F Gosselin\\[-0.7ex]
  John D Storey\\[-0.7ex]
  Jonathan E Taylor\\[-0.7ex]
  Keith J Worsley\\[-0.7ex]
  Nancy Ruonan Zhang\\[-0.7ex]
  Ryan J Tibshirani\\[-0.7ex]
  \\
  \\
  \\
\end{tabular}
\begin{tabular}{l}
  Hengjian Cui\\[-0.7ex]
  Huixia Judy Wang\\[-0.7ex]
  Jianhua Hu\\[-0.7ex]
  Jianhui Zhou\\[-0.7ex]
  Valen E Johnson\\[-0.7ex]
  Wing K Fung\\[-0.7ex]
  Xuming He\\[-0.7ex]
  Yijun Zuo\\[-0.7ex]
  Zhongyi Zhu\\[-0.7ex]
\end{tabular}
\begin{tabular}{l}
  Andrey Pepelyshev\\[-0.7ex]
  Frank Bretz\\[-0.7ex]
  Holger Dette\\[-0.7ex]
  Natalie Neumeyer\\[-0.7ex]
  Stanislav Volgushev\\[-0.7ex]
  Stefanie Biedermann\\[-0.7ex]
  Tim Holland-Letz\\[-0.7ex]
  Viatcheslav B Melas\\
  \\
\end{tabular}
\label{tab:otherGroups}
\end{table}

\subsection{Coauthorship network (B)}
\label{subsec:coauthorB}
In this network,  there is an edge between nodes $i$ and $j$  if and only if they have coauthored $1$ or more papers. Compared to Coauthorship network (A), this definition is more conventional, but it also makes the network harder to analyze.

Coauthorship network (B) has a total of $3607$ nodes, where the giant component consists of  $2263$ ($63\%$ of all nodes). For analysis in this section, we focus on the giant component.
Also, for simplicity, we call the giant component the Coauthorship network (B) whenever there is no confusion.

We are primarily  interested  in  community detection.
 Figure \ref{fig:coauthorGiantScree} (middle panel) presents the scree plot associated with Coauthorship network (B),   suggesting  $3$ or more communities.
We apply all four methods: SCORE,  NSC, BCPL, and APL  assuming $K = 3$ and below are the findings.

First, in Table \ref{tab:randIndex}, we compare all $4$ methods pair-wise and tabulate the corresponding ARI and VI (see Remark 2).
Somewhat surprisingly, the results of BCPL are inconsistent with those by  all other  methods.
For example, the maximum ARI between BCPL and each of the other three methods is $.00$, and the smallest VI between BCPL and each of the other three methods is $1.29$, showing a substantial disagreement.

\begin{figure}[htb!]
\centering
\includegraphics[width=5 in, height = 2.5in]{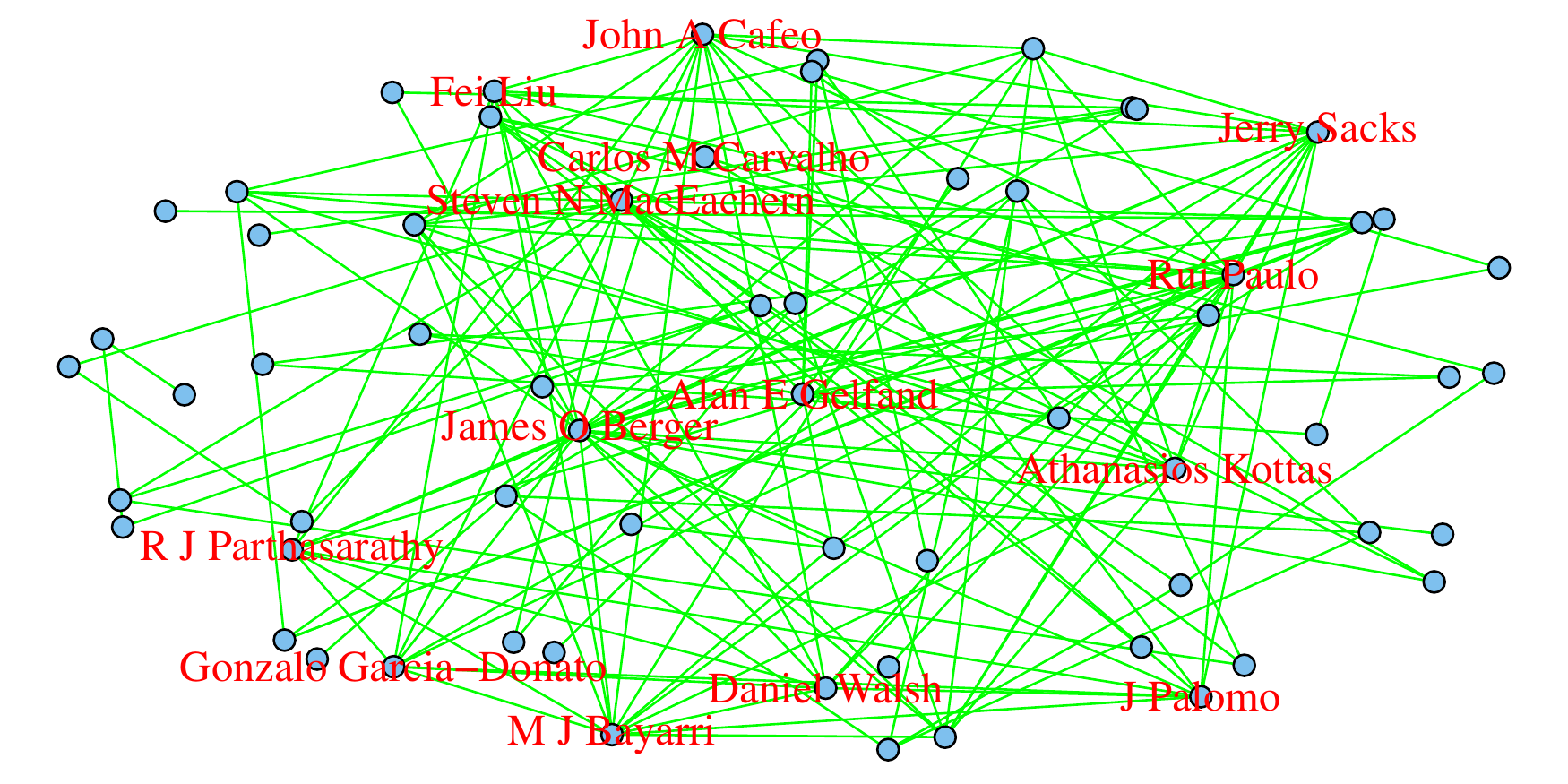}
\caption{The ``Objective Bayes" community in  Coauthorship network (B) identified by SCORE ($64$ nodes).  Only names for $14$ nodes with a degree of $9$ or larger are shown.}
\label{fig:coauthorBergerGraph}
\end{figure}

\begin{figure}[htb!]
\centering
\includegraphics[width = 5 in, height = 3.5 in]{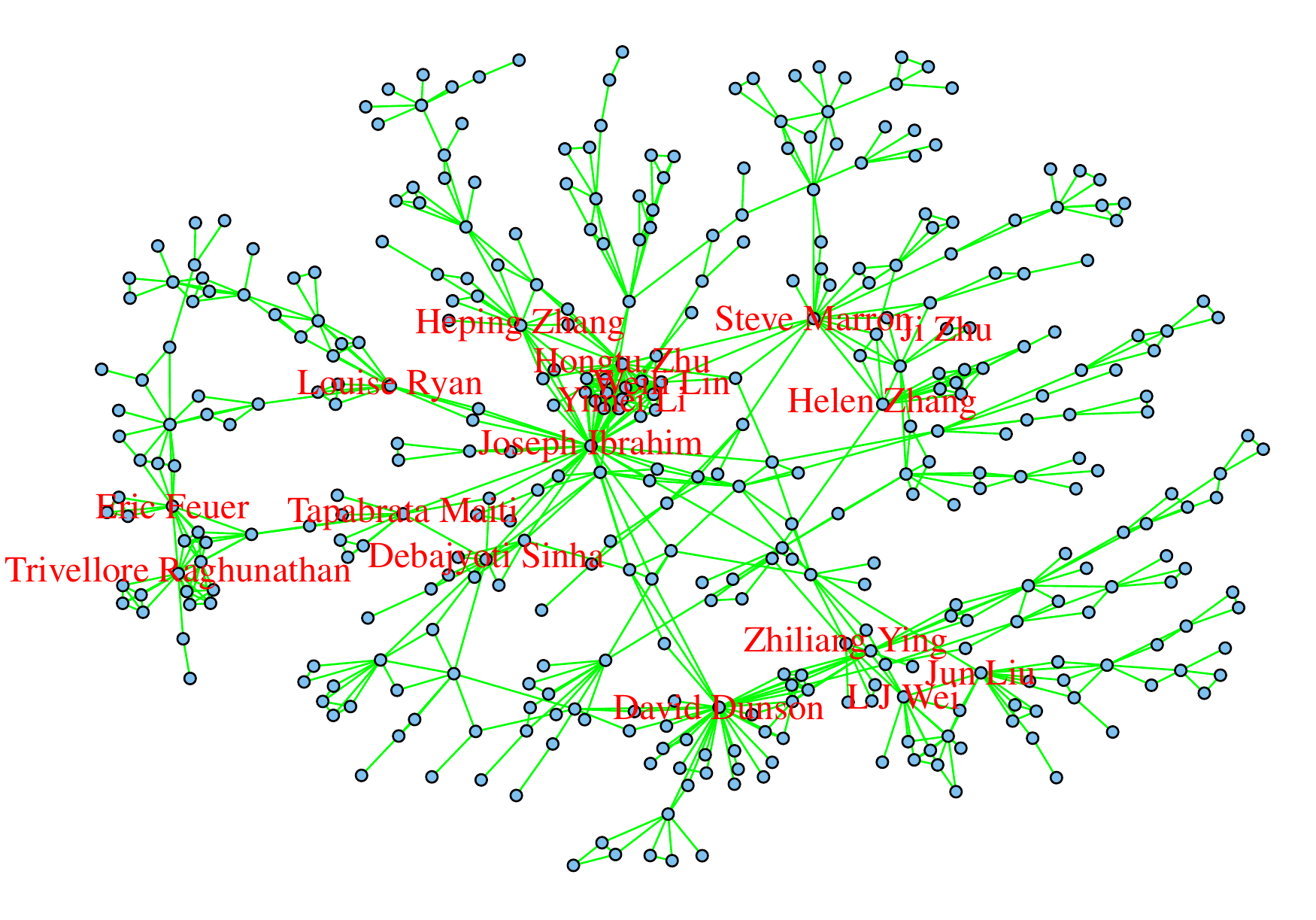}
\caption{
The ``Biostatistics" community (Biostat-Coau-B) in Coauthorship network (B)  identified by SCORE ($388$ nodes).  Only names for  $17$ nodes with a degree of $13$ or larger are shown.   A ``branch" in the figure is usually a research group in an institution or a state.}
\label{fig:coauthorApplied}
\end{figure}

\begin{figure}[htb!]
\centering
\graphicspath{ {c:\mypict~1\camera} }
\includegraphics[width = 5.25 in, height = 4.35 in]{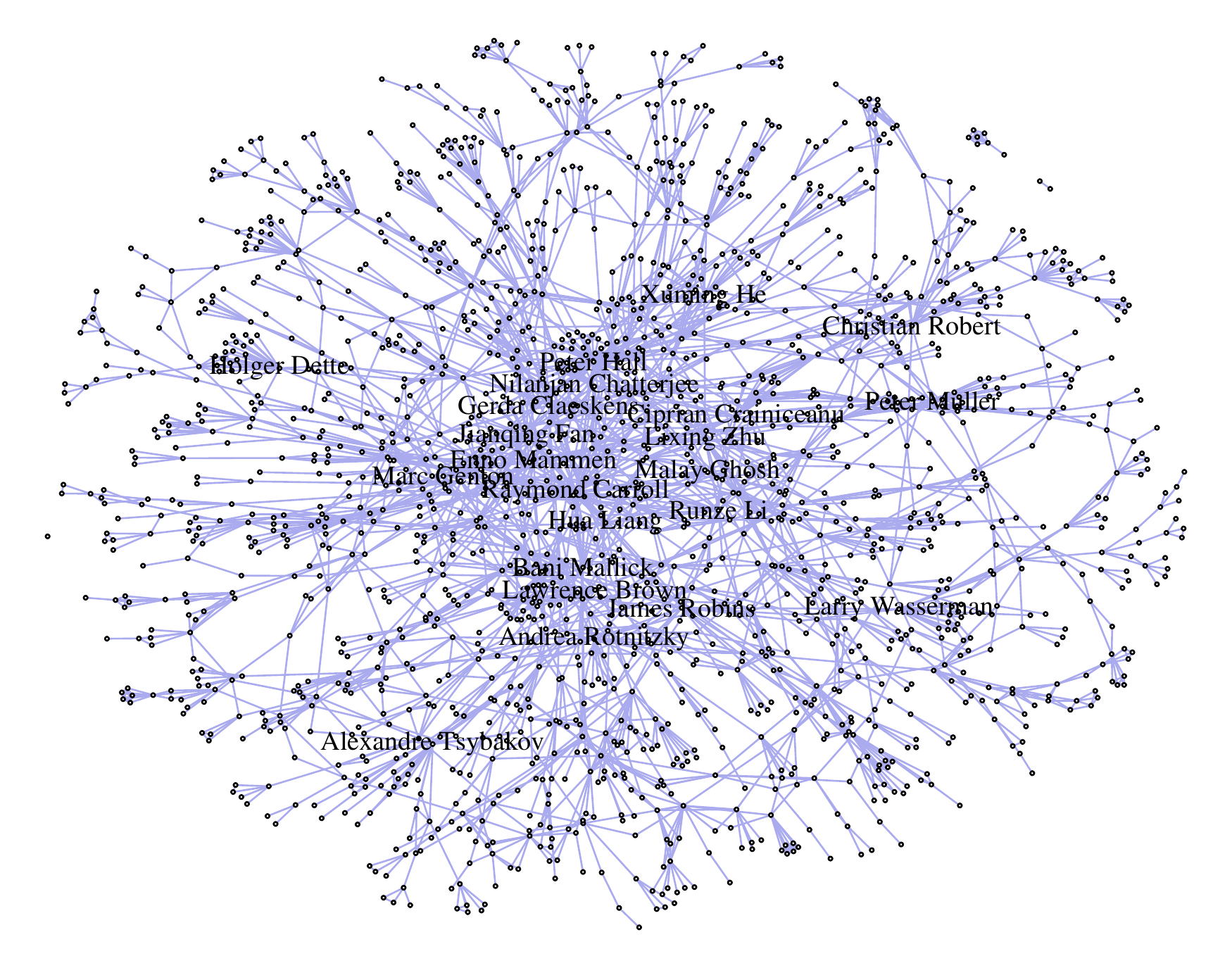}
\caption{The ``High Dimensional Data Analysis" community  (HDDA-Coau-B) in Coauthorship network (B) identified  by SCORE ($1181$ nodes).
Only names for $22$ nodes with degree of $18$ or larger are shown. }
\label{fig:coauthorHiDim}
\end{figure}

At the same time, the results by SCORE, NSC, and APL are
reasonably consistent with each other:   the ARI between   the vector of  predicted labels   by SCORE and that by NSC is $0.55$ and the ARI between  the vector of predicted labels  by NSC and that by APL is $0.41$; see Table \ref{tab:randIndex} for details.
In particular, the three methods agree on that,   the three communities each of them  identifies can be interpreted as follows (arranged in sizes  ascendingly).
\begin{itemize}
\item ``Objective Bayes" community.  This community includes a small group of researchers (group sizes are different for different methods, ranging from $20$ to $69$) including James Berger and his collaborators.
\item ``Biostatistics (Coauthorship (B))" (Biostat-Coau-B)  community.  The sizes of this community by three different methods have quite a bit variability and range from $50$ to $388$.
While it is probably not exactly right to call this community ``Biostatistics", the community consists of a number of statisticians and biostatisticians in the Research Triangle Park of  North Carolina.   It also includes many statisticians and biostatisticians  from Harvard University, University of Michigan at Ann Arbor, University of Wisconsin at Madison.
\item ``High Dimensional Data Analysis (Coauthorship (B))" (HDDA-Coau-B)  community.
The sizes of this community by three different methods range from $1811$ to $2193$.
The community includes researchers from a wide variety of research areas
in or related to high dimensional data analysis (e.g., Bioinformatics, Machine Learning).
\end{itemize}
In Figures \ref{fig:coauthorBergerGraph}-\ref{fig:coauthorHiDim}, we present these three communities (all three are identified by SCORE) respectively.

\begin{table}
\caption{The Ajusted Rand Index (ARI) and Variation of Information (VI) for the vectors of predicted community labels by
 four different methods in Coauthorship network (B), assuming $K = 3$.  A large ARI/small VI suggests that the two predicted label vectors are similar to each other.}
\begin{tabular}{l|r|r|r|r|r|r}
          & SCORE &    NSC & BCPL   & APL \\
\hline
SCORE& 1.00/.00      & .55/.51 & .00/1.65 & .19/.59   \\
NSC&      &   1.00/.00  & .00/1.46 & .41/.36   \\
BCPL &          &   & 1.00/.00 & .00/1.21   \\
APL&           &   &   & 1.00/.00
\end{tabular}
\label{tab:randIndex}
\end{table}

In Table \ref{tab:communitySizesB},  we compare the sizes of the three communities identified by each of the three methods.
There are two points worth noting.

First,  while SCORE and NSC are quite similar to each other,  there is a major  difference:  NSC clusters about 200 authors, mostly biostatisticians   from Harvard University, University of Michigan at Ann Arbor, and University of Wisconsin at Madison, into the HDDA-Coau-B community, but SCORE clusters them into the Biostat-Coau-B   community.  It seems that the results by SCORE are more meaningful.

Second,  APL behaves very differently from either SCORE or NSC.
Its estimate of the ``Objective Bayes" community is (almost) a subset of its counterpart by either SCORE or NSC,
and is much smaller in size (sizes are $20$, $64$, and $69$ for that by APL, SCORE, and NSC).
A similar claim applies to the Biostat-Coau-B community identified by each of the methods (sizes are $50$, $388$, and $169$ for that by APL, SCORE, and NSC).  This suggests that
APL may have underestimated these two communities but overestimated the HDDA-Coau-B community.

It is also interesting to  compare these results with those we obtain in Section \ref{subsec:coauthorA} for Coauthorship network (A).  Below are
three noteworthy points.

First, recall that in Figure \ref{fig:coauthorcoauthorTheoreticalLearningAndDimReduction} and Table  \ref{tab:otherGroups}, we have identified a total of $7$ different components of
Coauthorship network (A).  Among these components, the Duke component  (middle panel on top row in Table \ref{tab:otherGroups})    splits into three parts, each belongs to  the three of the communities of Coauthorship network (B)  identified by SCORE.    The other $6$ components fall into the HDDA-Coau-B
community identified by SCORE  almost completely.

Second, for the giant component of Coauthorship (A),
there is a close draw on whether we should cluster the Carroll-Hall's group and Fan's group into two communities:
SCORE and APL think that two groups belong to one community, but
NSC and BCPL do not agree with this. In Coauthorship (B), both groups are in the
HDDA-Coau-B community.  Also,  in previous studies on this giant component, BCPL and APL separate the nodes in Dunson's branch
from the North Carolina group, and cluster them into the Carroll-Hall   group. In the current study, however, the whole
North Carolina group (including Dunson's branch)  are  in the Biostat-Coau-B community.

Third,  in Coauthorship (A),  Gelfand's group is included in this $236$-node giant component, where James Berger is not a member. In  Coauthorship network (B), Gelfand's group now becomes a subset of ``Objective Baye" community where James Berger is a hub node.


\begin{table}
\centering
\caption{
Comparison of sizes of the three communities  identified by each of the  three methods in  Coauthorship network (B), assuming $K = 3$. BCPL is not  included  for comparisons for its results  are inconsistent with those by the other three methods.}
\begin{tabular}{r|c|c|c}
 &          Objective Bayes & Biostat-Coau-B  & HDDA-Coau-B \\
\hline
SCORE &    64 & 388 & 1811   \\
\hline
NSC & 69 & 163 & 2031  \\
\hline
APL &  20 & 50 &  2193  \\
\hline
\hline
SCORE $\cap$ NSC & 55 & 162 & 1807 \\
\hline
SCORE $\cap$ APL  &   20   &    50  &    1811  \\
\hline
NSC $\cap$ APL &   20   &    50  &    2032   \\
\hline
\hline
SCORE $\cap$ NSC   $\cap$ APL &      20  &   50  & 1807
\end{tabular}
\label{tab:communitySizesB}
\end{table}

\section{Community detection for Citation network}
\label{sec:citationcomm}
The Citation network is a directed network. As a result,  the study in this section is different from that in Section \ref{sec:coauthorcomm} in important ways, and provides additional insight into the structures of   statisticians' networks.
In Section \ref{subsec:citationmethod}, we discuss methods for community detection for directed networks. In Section \ref{subsec:citationcomm}, we analyze the Citation network, and compare the results with those in Section \ref{sec:coauthorcomm}.

\subsection{Community detection methods (directed networks)}
\label{subsec:citationmethod}
In the Citation network, each node is an author and there is a directed edge from node $i$ to node $j$
if and only if node $i$ has cited node $j$ at least once.
To analyze the Citation network, one usually focuses on the {\it weakly connected giant component} \cite{DIGRAPHS}.
This is the giant component of the {\it weakly connected citation network},  which is an undirected network where   there is an edge between nodes $i$ and $j$ if one has cited the other at least once.
From now on, when we say the Citation network, we mean the  weakly connected giant component of the original Citation network.

For community detection of  directed networks,  there are relatively few approaches.
In this section, we consider two methods: LNSC and  Directed-SCORE (D-SCORE).

LNSC stands for Leicht and Newman's Spectral Clustering approach proposed in  \cite{Newman2008}:  the authors extended the spectral modularity methods by  \cite{NewmanSC} for undirected networks to directed
networks, using the so-called generalized modularity \cite{Arenas2007}.
However,  it is pointed out in \cite{Kim2010} that LNSC  can not properly distinguish the directions of the edges and can not detect communities representing directionality patterns among the nodes. See
details therein.

D-SCORE  is the adaption of SCORE to directed networks.
SCORE is a community detection method for undirected networks, and
the method was  motivated by DCBM for undirected networks; see Section \ref{subsec:methods}.
Below, we first extend DCBM to directed networks, and then introduce  D-SCORE.

Let $A$ be the adjacency matrix of a directed network  ${\cal N} = (V, E)$, where
\[
A(i,j) = \left\{
\begin{array}{ll}
1, &\qquad \mbox{there is a directed edge from $i$ to $j$}, \\
0, &\qquad \mbox{otherwise},
\end{array}
\right.
\;\;\;  1 \leq i, j \leq n,
\]
and $n$ is the total number of nodes.
For DCBM of a directed network ${\cal N} = (V, E)$, similarly,  we think that all nodes  splits into $K$ different (disjoint) communities
\[
V  = V^{(1)} \cup V^{(2)} \ldots \cup V^{(K)}.
\]
Additionally, we suppose that $\{A(i,j), i \neq j, 1 \leq i, j \leq n\}$ are independent Bernoulli  with parameters $\pi_{ij}$, and that
there is a $K \times K$ non-negative matrix $P$ and two vectors with positive entries $\theta \in R^n$ and $\delta \in R^n$
such that
\[
\pi_{ij}  = \theta(i) \delta(j) P_{k,\ell}, \qquad   \mbox{if $i \in V^{(k)}$ and $j \in V^{(\ell)}$}, \qquad   1 \leq k, \ell \leq K.
\]
Here,  $\theta(i)$ models the {\it degree heterogeneity parameter} for node $i$ as a {\it citer}, and
$\delta(i)$ models the {\it degree heterogeneity parameter} for node $i$ as a {\it citee}.

This model motivates a new community detection method: D-SCORE. For detailed explanations, see the forthcoming manuscript \cite{JJZ}.
Given a directed network ${\cal N} = (V, E)$, assume ${\cal N}$ has $K$ communities.
Let $A$ be the adjacency matrix,  and let  $\hat{u}_1,  \hat{u}_2, \ldots, \hat{u}_K$ and $\hat{v}_1, \hat{v}_2, \ldots, \hat{v}_K$  be the first K  left singular vectors and the first $K$ right singular vectors of A, respectively.
Also, define two associated (undirected)  networks with the same set of nodes as follows
\begin{itemize}
\item {\it Citer network}.  There is an (undirected) edge between two distinct nodes $i$ and $j$ in $V$  if and only if both of them have cited a  node $k$ at least once,
 for some   $k \in   (V \setminus \{i,j\})$ (i.e.,   they have a common citee).
\item {\it Citee network}. There is an (undirected)  edge between two distinct node $i$ and $j$  in $V$  if and only if  each of them has been cited at least once by the same node $k \notin (V \setminus \{i,j\})$ (i.e.,  they have a common citer).
\end{itemize}
Let ${\cal N}_1$ and ${\cal N}_2$ be the giant components of the Citer network and Citee network, respectively.  Define two $n \times (K-1)$ matrices  $\hat{R}^{(l)}$
$\hat{R}^{(r)}$ by
\begin{equation} \label{DefinehatRleft}
\hat{R}^{(l)}(i,k) =
\left\{
\begin{array}{ll}
\sgn(\hat{u}_{k+1}(i) / \hat{u}_1(i)) \cdot \min\{|\frac{\hat{u}_{k+1}(i)}{\hat{u}_1(i)}|, \log(n)\},   & i \in {\cal N}_1,  \\
0, &i \notin {\cal N}_1,
\end{array}
\right.
\end{equation}
\begin{equation} \label{DefinehatRright}
\hat{R}^{(r)}(i,k) =
\left\{
\begin{array}{ll}
\sgn(\hat{v}_{k+1}(i) / \hat{v}_1(i)) \cdot \min\{|\frac{\hat{v}_{k+1}(i)}{\hat{v}_1(i)}|, \log(n)\},   & i \in {\cal N}_2,  \\
0, &i \notin {\cal N}_2.
\end{array}
\right.
\end{equation}
Note that all nodes split into four disjoint subsets:
\[
{\cal N} = ({\cal N}_1 \cap {\cal N}_2) \cup ({\cal N}_1 \setminus {\cal N}_2) \cup
({\cal N}_2 \setminus {\cal N}_1) \cup ({\cal N} \setminus ({\cal N}_1 \cup {\cal N}_2)).
\]
D-SCORE clusters nodes in each subset separately.
\begin{enumerate}
\item ({\it ${\cal N}_1 \cap {\cal N}_2$}).  Restricting the rows of $\hat{R}^{(l)}$ and $\hat{R}^{(r)}$ to the set ${\cal N}_1 \cap {\cal N}_2$ and obtaining two
matrices $\tilde{R}^{(l)}$ and $\tilde{R}^{(r)}$,
we cluster all nodes in ${\cal N}_1 \cap {\cal N}_2$ by applying the $k$-means to the matrix $[\tilde{R}^{(l)}, \tilde{R}^{(r)}]$ assuming there are $\leq K$ communities.
\item ({\it ${\cal N}_1 \setminus {\cal N}_2$}).  Note that according to the communities we identified above, the rows of $\tilde{R}^{(l)}$ partition into $\leq K$ groups. For each group,
we call the mean of the row vectors the {\it community center}.
For a node $i$ in ${\cal N}_1 \setminus  {\cal N}_2$,
if the $i$-th row of $\hat{R}^{(l)}$  is closest to the center of the $k$-th community  for some $1 \leq k \leq K$,  then we assign it to  this community.
\item ({\it ${\cal N}_2 \setminus {\cal N}_1$}).
We cluster in a similar fashion to that in the last step, but we use $(\tilde{R}^{(r)}, \hat{R}^{(r)})$ instead of  $(\tilde{R}^{(l)}, \hat{R}^{(l)})$.
\item ({\it ${\cal N} \setminus ({\cal N}_1 \cup {\cal N}_2)$}).
We say there is a weak-edge between $i$ and $j$ if there is an edge between $i$ and $j$ in the weakly connected citation network.
By $1$-$2$, all nodes in ${\cal N}_1 \cup {\cal N}_2$ partition into $\leq K$ communities.
For each node in ${\cal N} \setminus ({\cal N}_1 \cup {\cal N}_2)$,
we assign it  to the community to which it  has the largest number of weak-edges.
\end{enumerate}
For $4$, our assumption is  that  $|{\cal N} \setminus ({\cal N}_1 \cup {\cal N}_2)|$ is small, so we don't have to have a
sophisticated clustering method. For the statistical citation network data set we study in this paper,
this is true with $|{\cal N} \setminus ({\cal N}_1 \cup {\cal N}_2)| = 14$.

\begin{figure}[htb!]
\centering
\includegraphics[width=5in, height = 2 in]{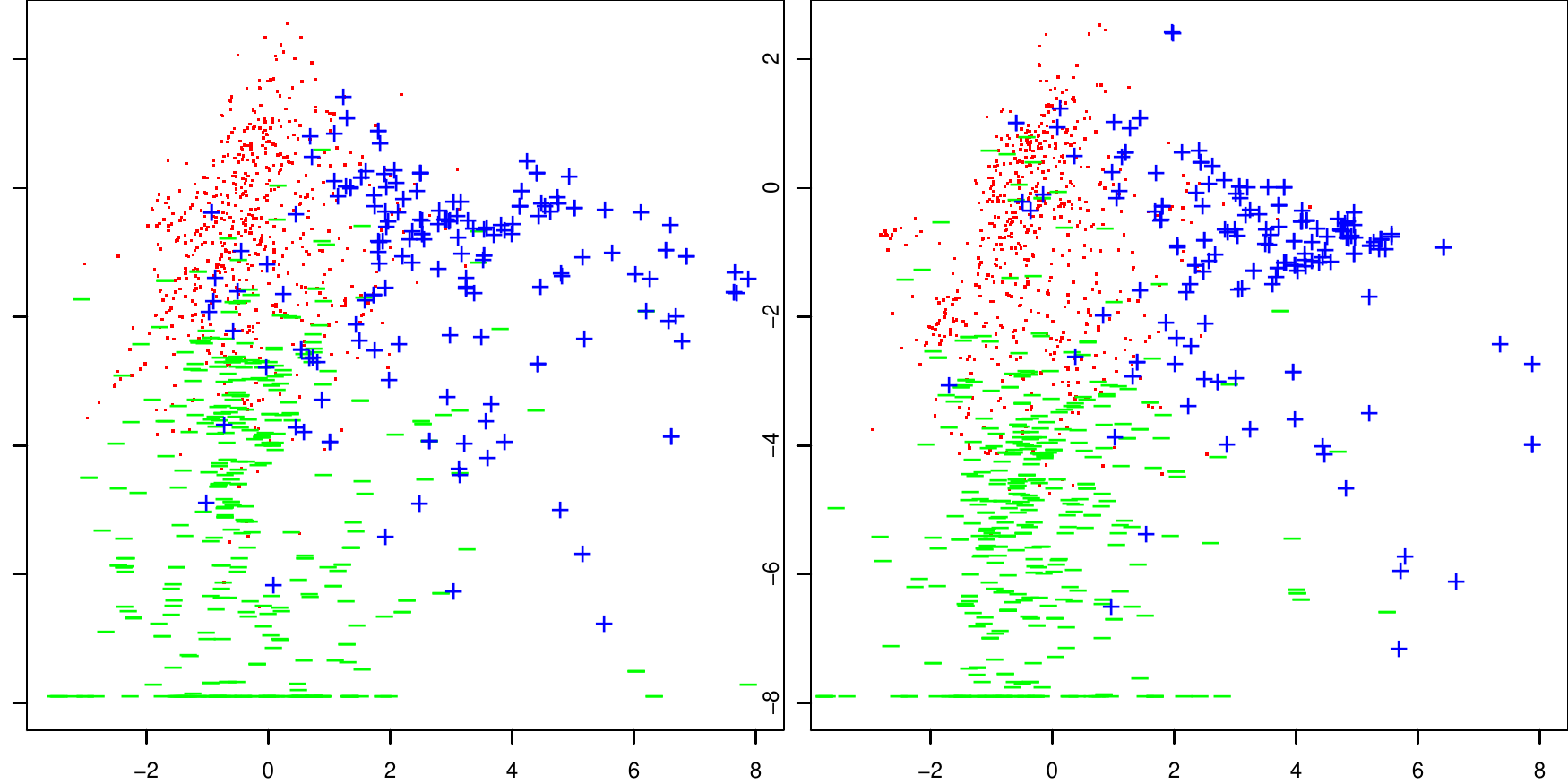}
\caption{Left: each point represents a row of the   matrix $\hat{R}^{(l)}$ (the matrix has only two columns since $K  =3$) associated with the statistical Citation network ($x$-axis: first column, $y$-axis: second column). Only rows with indices in ${\cal N}_1$ are shown. Blue pluses, green bars, and  red dots  represent $3$ different communities identified by SCORE, which can be interpreted as  ``Large-Scale Multiple testing", ``Spatial and Semi-parametric/Nonparametric Statistics" and ``Variable  Selection", Right: similar but with $(\hat{R}^{(l)}, {\cal N}_1)$  replaced by  $(\hat{R}^{(r)}, {\cal N}_2)$.}
\label{fig:Citation-clustering-result}
\end{figure}

Figure  \ref{fig:Citation-clustering-result} illustrates how D-SCORE works  using the statistical citation network
data set  with $K = 3$.  Two panels show similar clustering patterns,
suggesting that there are three communities;
see   Section \ref{subsec:citationcomm} for details.

\subsection{Citation network}
\label{subsec:citationcomm}
The original citation network data set consists of  $3607$ nodes (i.e., authors). The associated  weakly connected network has $927$ components. The giant component has $2654$ authors, accounting $74\%$ of all nodes.
All  other components have  no more than $5$ nodes.

We now restrict our attention to the weakly connected giant component  ${\cal N} = (V, E)$.
As before, let ${\cal N}_1$ and ${\cal N}_2$ be the giant components of the Citer and Citee networks associated with ${\cal N}$, respectively.
We have $|{\cal N}_1| = 2126$, $|{\cal N}_2|  =1790$,
$|{\cal N}_1 \cap {\cal N}_2| = 1276$, and $|{\cal N} \setminus ({\cal N}_1 \cup {\cal N}_2))|  = 14$.

\begin{figure}[htb!]
\centering
\includegraphics[width = 5 in, height = 4 in]{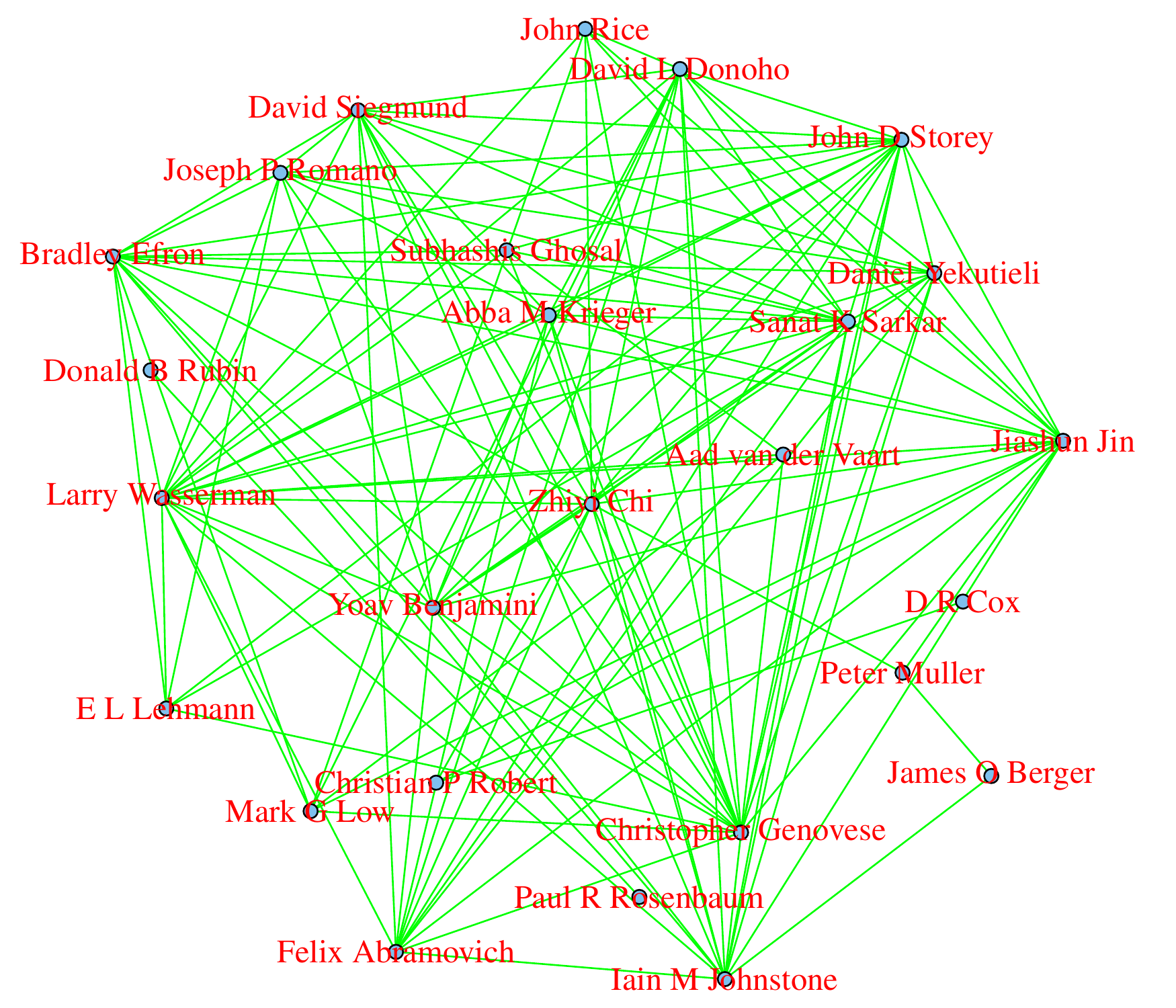}
\caption{The ``Large-Scale Multiple Testing" community identified by D-SCORE ($K = 3$) in the Citation network ($359$ nodes). Only $26$ nodes with $24$ or more citers  are shown here.}
\label{fig:Citation-testing}
\end{figure}


We are primarily interested in community detection.  In Figure \ref{fig:coauthorGiantScree} (right panel), we
present the scree plot  of $A$. Note that since $A$ is non-symmetric, we use the singular values instead of the eigenvalues in the plot.
The plot suggests that there are $K=3$ communities in ${\cal N}$.

We have applied D-SCORE and LNSC to ${\cal N}$. The results by SCORE are reported below with details.
The results of LNSC are rather inconsistent with those of SCORE, so we only discuss them briefly; see Section \ref{subsec:LNSC}.

\begin{figure}[htb!]
\centering
\includegraphics[width = 5 in, height = 4 in]{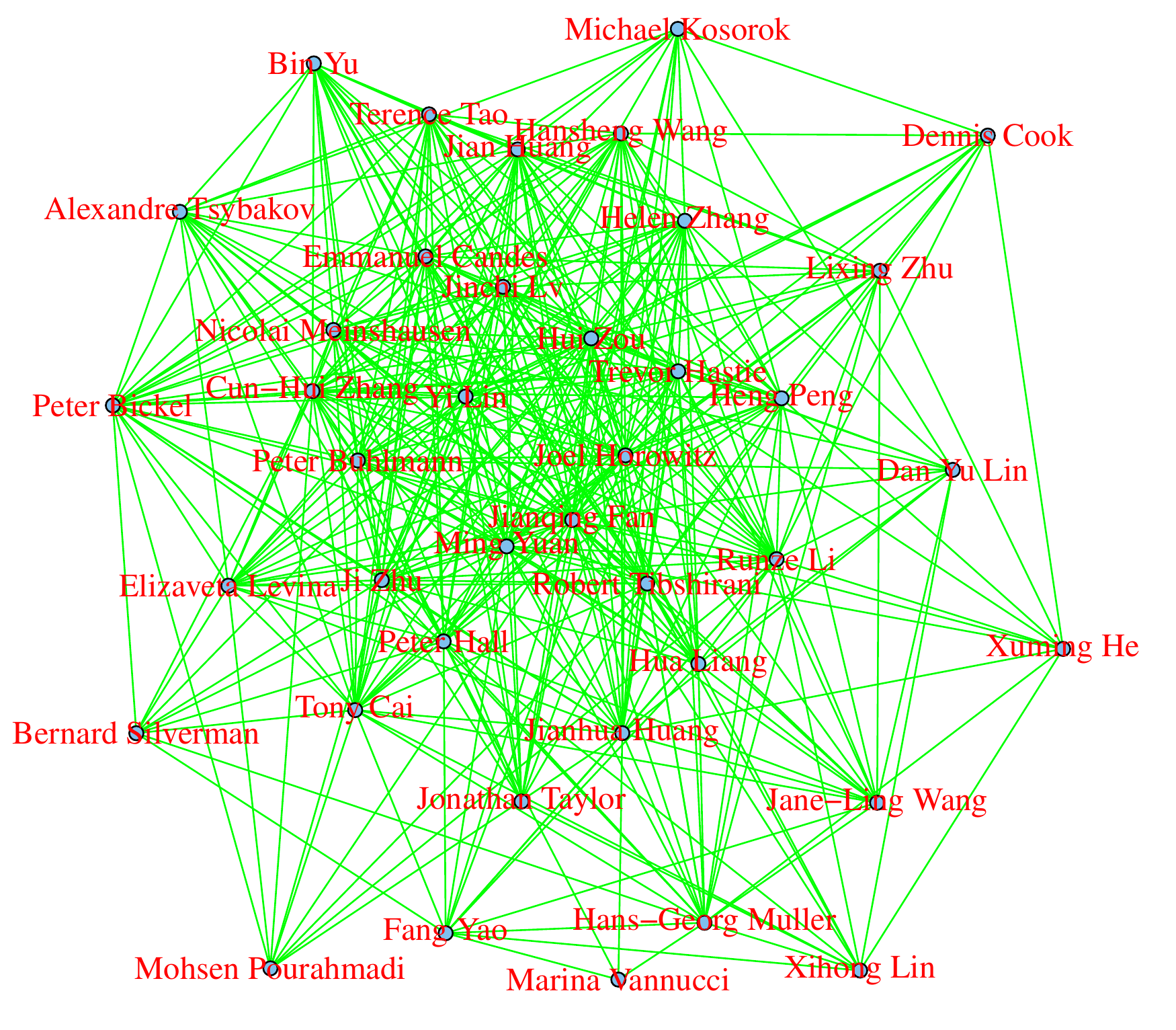}
\caption{The ``Variable Selection" community identified by D-SCORE ($K = 3$) in the Citation network ($1285$ nodes). Only  $40$ nodes with $54$ or more citers are shown here.}
\label{fig:Citation-vs}
\end{figure}

\begin{figure}[htb!]
\centering
\includegraphics[width = 4.5 in, height = 2.2 in]{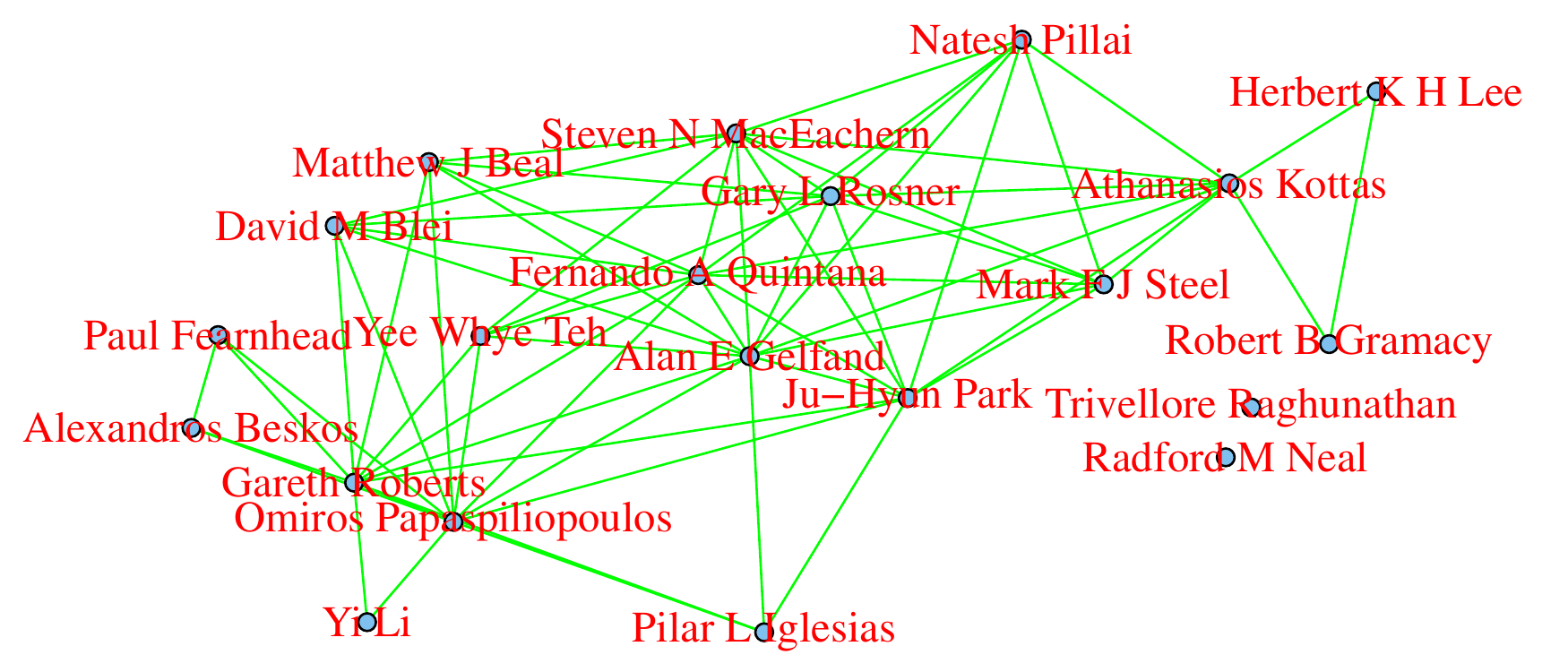}

\includegraphics[width = 4.5 in, height = 2.2 in]{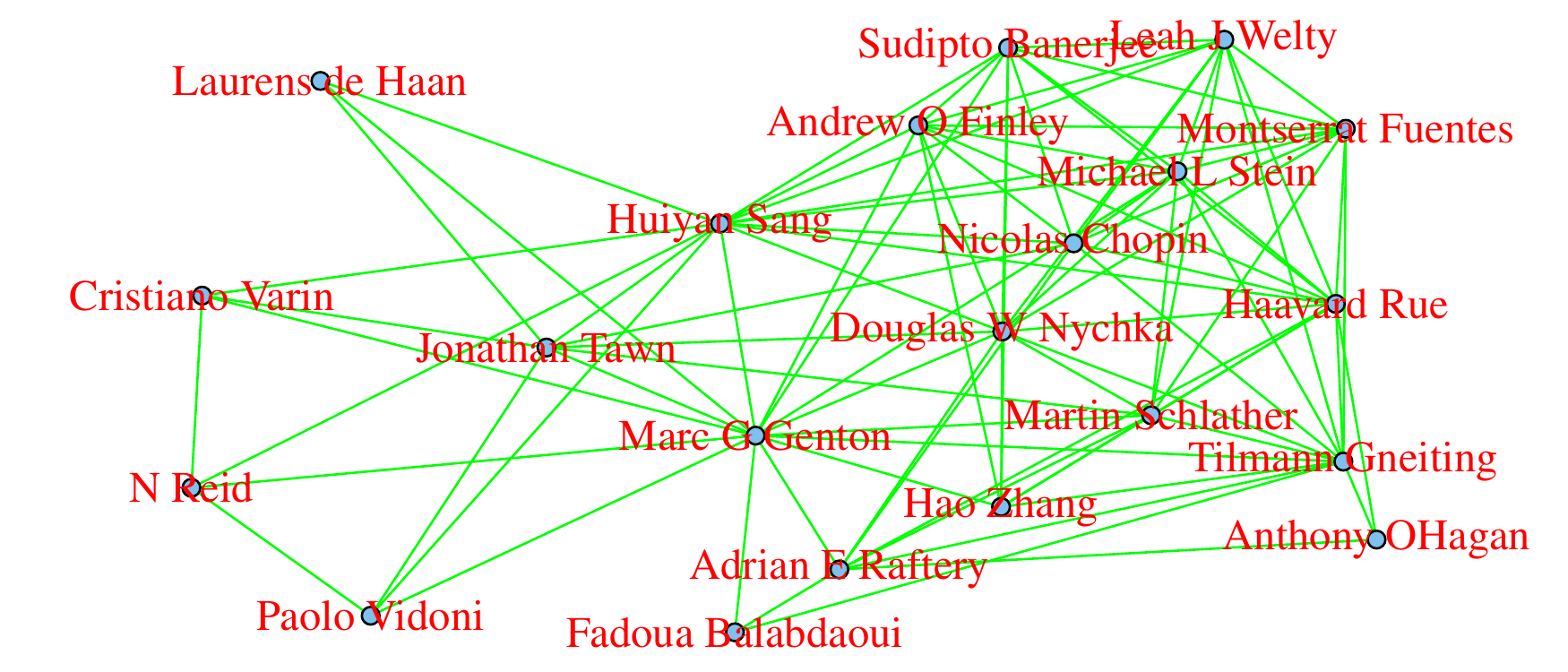}

\includegraphics[width = 4.5 in, height = 2.2 in]{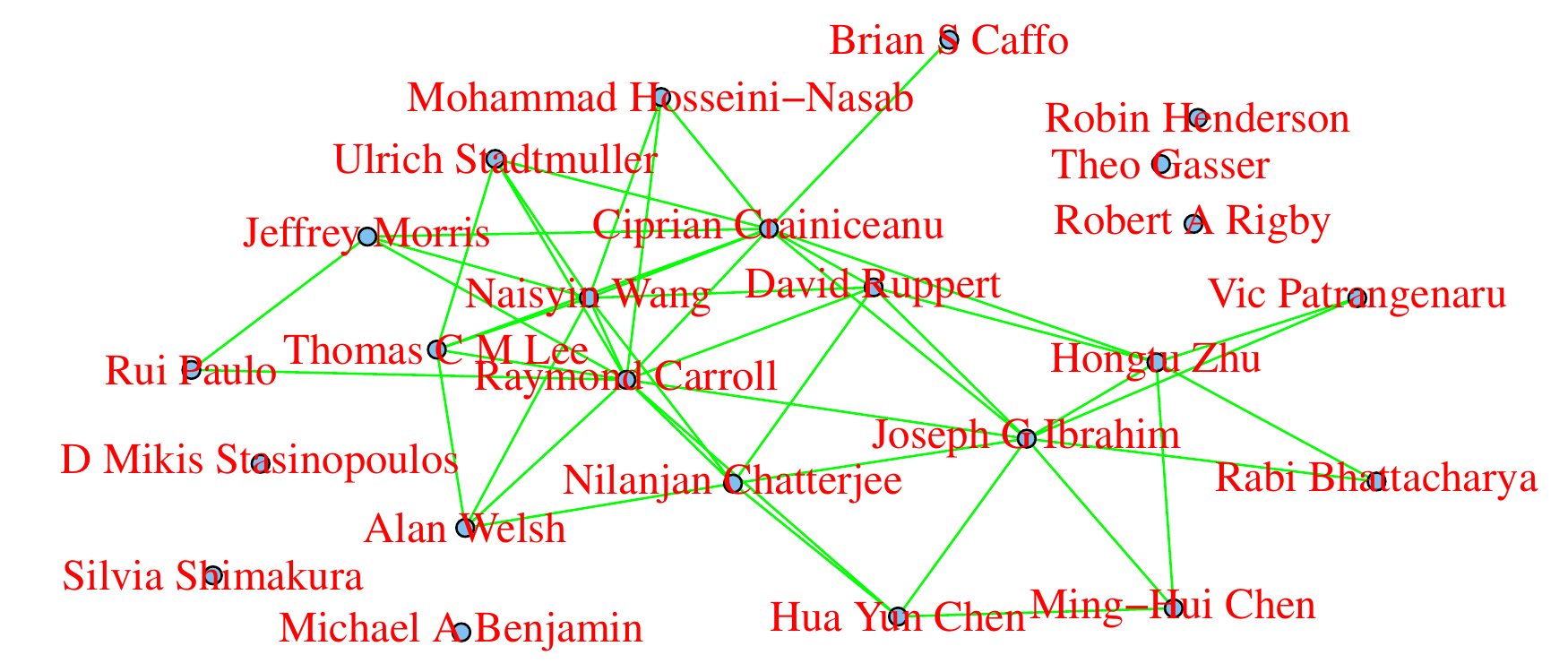}
\caption{The ``Spatial and Semi-parametric/Non-parametric Statistics"  community has sub-communities: Non-parametric Spatial (upper), Parametric Spatial (middle), Semi-parametric/Non-parametric (lower). In each, only  about 20 high-degree nodes  are shown. }
\label{fig:Citation-other-sub}
\end{figure}

 D-SCORE  identifies there communities  as follows.
\begin{itemize}
\item ``Large-Scale Multiple Testing"  community ($359$ nodes).  This consists of  researchers in
 multiple testing and  control  of  False Discovery Rate. It  includes a Bayes group (James Berger, Peter Muller), three Berkeley-Stanford  groups (Bradley Efron, David Siegmund,  John Storey;  David Donoho,  Iain Johnstone,
 Mark Low\footnote{University of Pennsylvania}, John Rice;   Erich Lehmann,  Joseph Romano),  a
 Carnegie Mellon  group (e.g., Christopher Genovese, Jiashun Jin,  Isabella Verdinelli,  Larry Wasserman),   a Causal Inference group (Donald Rubin, Paul Rosenbaum), and
a  Tel Aviv group (Felix Abramovich,  Yoav Benjamini, Abba Krieger\footnote{University of Pennsylvania}, Daniel Yekutieli), etc.
\item ``Variable Selection"   community ($1285$ nodes).  This includes   (sorted descendingly by the number of citers)
Jianqing Fan, Hui Zou, Peter Hall, Nicolai Meinshausen, Peter Buhlmann, Ming Yuan, Yi Lin, Runze Li, Peter  Bickel, Trevor   Hastie, Hans-Georg Muller, Emmanuel   Candes, Cun-Hui Zhang, Heng Peng, Jian Huang,  Tony Cai, Terence Tao, Jianhua  Huang, Alexandre  Tsybakov, Jonathan   Taylor, Xihong Lin, Jane-Ling Wang, Dan Yu Lin, Fang Yao, Jinchi Lv.
\item  ``Spatial and Semi-parametric/Nonparametric Statistics" (for short, ``Spatial Statistics") community ($1010$ nodes).
See discussions below.
\end{itemize}
The first two communities are presented in  Figures \ref{fig:Citation-testing} and \ref{fig:Citation-vs}, respectively.
The last community consists of sub-structures and is harder to interpret. To this end, we first restrict the network to this community
(i.e., ignoring all edges to/from outside) and obtain a sub-network. We than apply D-SCORE with $K = 3$ to the giant component (908 nodes) of this sub-network, and obtain three meaningful sub-communities as follows.
\begin{itemize}
\item Non-parametric spatial statistics (212 nodes), including David Blei, Alan Gelfand,  Yi Li, Steven MacEachern, Omiros Papaspiliopoulos, Trivellore Raghunathan, Gareth Roberts.
\item Parametric spatial statistics (304 nodes),  including Marc Genton, Tilmann Gneiting,  Douglas Nychka, Anthony OHagan, Adrian Raftery, Nancy Reid, Michael Stein.
\item Semi-parametric/Non-parametric statistics (392 nodes), including Raymond Carroll, Nilanjan Chatterjee, Ciprian Crainiceanu, Joseph Ibrahim, Jeffrey Morris,  David Ruppert,  Naisyin Wang, Hongtu Zhu.
\end{itemize}
These sub-communities  are presented in Figure \ref{fig:Citation-other-sub}.

\subsubsection{Comparison with Coauthorship network (A)}
\label{subsubsec:compcoauthorA}
In Section \ref{subsec:coauthorA}, we present $8$ different components of Coauthorship network (A).
In Table \ref{tab:coauthorThreshGiantVSSCORE}, we reinvestigate all these components in order to understand
their relationship with the $3$ communities identified by D-SCORE in  the Citation network.
\begin{table}[htb!]
\centering
\caption{Sizes of the intersections of the communities identified by D-SCORE ($K = 3$) in the Citation network (rows)
and the $8$ largest components of Coauthorship network (A) as presented in  Figures
\ref{fig:coauthorThreshHiDim} and \ref{fig:coauthorcoauthorTheoreticalLearningAndDimReduction} and Tables \ref{tab:otherGroups}  (columns).  ``Other": nodes outside the weakly connected giant component;     *: $9$ out of $12$ are in the ``Semi-parametric/Non-parametric"  sub-community of the ``Spatial Statistics" community.
}

\scalebox{0.93}{
\begin{tabular}{r|c|c|c|c|c|c|c|c}
            &   & Mach.    &Dim.   &Johns &  &  & Quant.  &Exp.  \\
            &  giant  &  Learn.  & Reduc. &Hopkins & Duke &Stanford & Reg.& Design \\
\hline
Spatial    &   60    &  1  &    &12* & 1  &   &    &3\\
\hline
Var. Selection     &   166   &  15 & 14 &1   & 7  & 2 &8   &2\\
\hline
Multiple Tests    &   7     &  2  &    &    & 2  & 7 &1   &3\\
\hline
Other        &   3     &    &    &    &    &&&\\
\hline
\hline
            &  236    &  18 & 14 &13  & 10 & 9 &9   &8
\end{tabular}
}
\label{tab:coauthorThreshGiantVSSCORE}
\end{table}

Among these $8$ components, the first one is the giant component, consisting of $236$ nodes. All except $3$ of these nodes fall in the $3$  communities identified by D-SCORE in the Citation network, with $60$ nodes in
``Spatial Statistics and Semi-parametric/Non-parametric statistics",  including (sorted descendingly by the number of citers; same below) Raymond Carroll, Joseph Ibrahim, Naisyin Wang, Alan Gelfand, Jeffrey  Morris, Marc  Genton, Sudipto Banerjee, Hongtu Zhu, Jeng-Min Chiou, Ju-Hyun Park, Ulrich Stadtmuller, Ming-Hui Chen, Yi Li, Nilanjan Chatterjee, Andrew  Finley,
$166$ nodes in  ``Variable Selection"  including   Jianqing Fan, Hui Zou, Peter Hall, Ming Yuan, Yi Lin, Runze Li, Trevor  Hastie, Hans-Georg Muller, Emmanuel  Candes, Cun-Hui Zhang, Heng Peng, Jian Huang,  Tony Cai, Jianhua  Huang, Xihong Lin, and $7$ nodes in ``Large-Scale Multiple Testing" including David  Donoho, Jiashun Jin, Mark  Low, Wenguang Sun, Ery Arias-Castro, Michael  Akritas, Jessie Jeng.

This is consistent with our previous claim that this $236$-node giant component contains a ``Carroll-Hall" group and a ``North Carolina" community: The ``Carroll-Hall" group has strong ties to the area of variable selection, and the ``North Carolina" group has strong ties to Biostatistics.
Raymond Carroll has close ties to both of these two groups, and it is not surprising that
SCORE assigns  him to the   ``Carroll-Hall"  group in Section \ref{subsec:coauthorA} in Coauthorship network (A) but  D-SCORE assigns him  to the ``Spatial" community in the Citation network.

For the remaining $7$ components of Coauthorship network (A),  ``Theoretical Machine   Learning", ``Dimension Reduction", ``Duke",  ``Quantile Regression" are (almost)  subsets of ``Variable Selection",  ``Stanford"  (including
John Storey, Johathan Taylor, Ryan Tibshirani) is   (almost)  a subset of ``Large-Scale Multiple Testing", and ``Johns  Hopkins" is (almost) a  subset of ``Spatial Statistics". The ``Experimental Design" group has no stronger relation to one area than to the others, so the nodes spread almost evenly to these three communities.

\subsubsection{Comparison with Coauthorship network (B)}
\label{subsubsec:compcoauthorB}
We compare the community detection results by D-SCORE for the Citation
network with those by SCORE for Coauthorship  network (B) in Section \ref{subsec:coauthorB}.
Note that for the former, we have been focused on the weakly connected giant component of the Citation network  ($2654$ nodes), and
for the latter, we have been focused on the  giant component of the Coauthorship network (B) ($2263$ nodes).
 The comparison of two sets of  results is tabulated in Table \ref{tab:intersect-citation-coauthor-B}.

Viewing the table vertically, we observe that Citation network provides additional insight into the
Coauthorship network (B), and reveals structures we have not found previously.
Below are the details.

First,  the ``Objective Bayes" community in Coauthorship network (B) contains two main parts.
The first part consists of $55\%$ of the nodes, and most of them are seen to be the researchers who have   close ties to   James Berger,  including (sorted descendingly by the number of citers; same below) Alan  Gelfand, Fernando  Quintana, Steven  MacEachern, Gary  Rosner, Rui Paulo, Herbert  Lee, Robert  Gramacy, Athanasios Kottas, Pilar  Iglesias, Daniel  Walsh, Dongchu Sun.
The second part consists of $25\%$ of the nodes, and is assigned to the ``Variable Selection" community in the Citation network by D-SCORE, including  Carlos  Carvalho, Feng Liang, Maria De Iorio, German Molina, Merlise  Clyde, Luis  Pericchi, Maria  Barbieri, Nicholas  Polson, Bala Rajaratnam, Edward  George.  For the second part, the result seems reasonable,  as many nodes in the second part (e.g., Carlos Carvalho,  Edward George, Feng Liang, Merlise Clyde) have an interest in model selection.

Second, the ``Biostatistics (Coauthorship (B))" community in Coauthorship network (B) also  has two main parts.
The first  part has $156$ nodes ($40\%$ of the total, including high-degree nodes such as    Joseph  Ibrahim, Sudipto Banerjee, Hongtu Zhu, Ju-Hyun Park, Ming-Hui Chen, Yi Li, Montserrat Fuentes, Natesh Pillai, Andrew  Finley, Amy Herring, Martin Schlather, Stuart  Lipsitz, Jonathan Tawn, Siddhartha Chib, Alexander Tsodikov.
The second part consists of $153$ nodes ($40\%$ of the total).  The high-degree nodes include
  Yi Lin, Dan Yu Lin, Ji Zhu, Helen Zhang, L J Wei, Wei Biao Wu, Donglin Zeng, Zhiliang Ying, David Dunson, Steve Marron, Anastasios  Tsiatis, Wenbin Lu, Zhezhen Jin, Xiaotong Shen, Heping Zhang, Lu Tian, Jianwen Cai, Wing Hung Wong.
The results are quite reasonable:  many nodes in the second part
(e.g., Dan Yu Lin, David Dunson,  Helen Zhang, Steve Marron, Ji Zhu, Xiaotong Shen, Yi Lin) either have works in  or  have strong ties to the area of  variable selection.

Last, the ``High Dimensional Data Analysis"  community in Coauthorship network (B) has three parts.
The first part has $459$ nodes ($25\%$), including high-degree nodes such as  Raymond  Carroll, Gareth Roberts, Naisyin Wang, Adrian Raftery, Omiros Papaspiliopoulos, David Ruppert, Tilmann Gneiting, Jeffrey Morris, Michael  Stein, Ciprian  Crainiceanu, Marc Genton, Nicolas Chopin, Alan Welsh, Anthony OHagan, Fadoua Balabdaoui, N Reid.
The second part has $840$ nodes ($46\%$), including high-degree nodes such as
  Jianqing Fan, Hui Zou, Peter Hall, Nicolai Meinshausen, Peter Buhlmann, Ming Yuan, Runze Li, Peter  Bickel, Trevor  Hastie, Hans-Georg Muller, Emmanuel  Candes, Cun-Hui Zhang, Heng Peng, Jian Huang,  Tony Cai, Terence Tao, Jianhua  Huang, Alexandre  Tsybakov, Jonathan  Taylor, Xihong Lin.
The third part has $221$ nodes ($26\%$), including high-degree nodes such as
 Iain  Johnstone, Larry Wasserman, Bradley Efron, John  Storey, Christopher Genovese, David  Donoho, Yoav Benjamini, David Siegmund, Peter Muller, Jiashun Jin, Felix Abramovich, David Cox, Daniel Yekutieli.

Respectively, the three parts are labeled as subsets of the ``Spatial and Semi-parametric/Non-parametric Statistics", ``Variable Selection", and ``Large-Scale Multiple Testing" communities in the Citation network.
This seems convincing:    (a) most of the nodes in the first part have a strong interest in spatial statistics or biostatistics (e.g., Ciprian Crainiceanu, Naisyin Wang, Raymond Carroll), (b) most of the nodes in the second part  are leaders in variable selection, and (c) most nodes in the third part are  leaders in Large-Scale Multiple Testing and in the topic of control of FDR.

Viewing the table horizontally gives similar claims but also reveals some additional insight. For example,
``Large-Scale Multiple Testing" contains  three  main parts.  One part consists of   $221$   nodes and  is a subset of the  ``High Dimensional Data Analysis" community in Coauthorship network (B). The second consists of $115$   nodes and falls outside the giant component of Coauthorship network (B).  A significant fraction  of nodes in this  part  are from Germany and have close ties to Helmut Finner, a leading researcher in Multiple Testing.  Another significant  part  (17 nodes) are researchers in Bioinformatics (e.g., Terry Speed) who do not publish  many papers in these four journals for the time period.

\begin{table}[htb!]
\centering
\caption{Sizes of the intersections of the communities identified by D-SCORE ($K = 3$) in the Citation network (rows; ``other" stands for nodes outside the weakly connected giant component) and the communities identified by SCORE in Coauthorship network (B) (columns; ``other" stands for nodes outside the giant component). *:   $14$ and $17$   are in the ``Non-parametric Spatial" and  ``Semi-parametric/Non-parametric"  sub-communities of the ``Spatial and Semi-parametric/Non-parametric Statistics" community, respectively.
 }

\begin{tabular}{r|c|c|c|c||c}
            & Obj.    Bayes  & Biostat-Coau-B & HDDA-Coau-B   &   other                       &   \\
\hline
Spatial     &   35*     &    156  &         459      &     360                           & 1010               \\
\hline
Var. Selection     &   16     &    153  &         840      &     276                          &  1285 \\
\hline
Multiple Tests    &    6      &     17  &        221     &      115                          &   359 \\
\hline
other       &    7      &     62  &        291      &            593                      &  953  \\
\hline
\hline
            &   64  &        388 &          1811     &     1344                        &      3067
\end{tabular}
\label{tab:intersect-citation-coauthor-B}
\end{table}

\subsubsection{Comparison of D-SCORE and LNSC}
\label{subsec:LNSC}
We have also applied LNSC to the Citation network, with $K = 3$. The communities are very different from those identified by D-SCORE, and maybe interpreted as follows.
\begin{itemize}
\item ``Semi-parametric and non-parametric" ($434$ nodes). We find this community hard to interpret, but it could be the community of  researchers
on semi-parametric and non-parametric models,  functional estimation, etc.. The hub nodes include
(sorted descendingly by the number of citers; same below) Peter Hall, Raymond  Carroll, Hans-Georg Muller, Xihong Lin, Fang Yao, Naisyin Wang, Marina Vannucci, David Ruppert, Gerda Claeskens, Wolfgang Hardle, Jeffrey  Morris, Enno Mammen, Ciprian  Crainiceanu, James Robins, Anastasios  Tsiatis, Catherine  Sugar, Zhezhen Jin, Alan  Welsh,  Sunil Rao, Philip  Brown.
\item ``High Dimensional Data Analysis" (HDDA-Cita-LNSC) ($614$ nodes).  The second one can be  interpreted as the ``High Dimensional Data Analysis" community, where the high-degree nodes include
 (sorted descendingly by the number of citers) Jianqing Fan, Hui Zou, Nicolai Meinshausen, Peter Buhlmann, Ming Yuan, Yi Lin, Iain Johnstone, Runze Li, Peter Bickel, Trevor Hastie, Larry Wasserman, Emmanuel  Candes, Cun-Hui Zhang, Heng Peng, Bradley Efron, John  Storey, Jian Huang,  Tony Cai, Christopher Genovese, Terence Tao.
\item ``Biostatistics" (Biostat-Cita-LNSC) ($1605$ nodes).  The community is hard to interpret and includes researchers from several different areas. For example, it includes researchers in biostatistics (e.g., Joseph Ibrahim,  L J Wei),  in
nonparametric (Bayes) methods (e.g.,  Peter Muller, David Dunson, and Nils Hjort, Fernando Quintana, Omiros Papaspiliopoulos),  and in  spatial statistics and uncertainty quantification (e.g., Mac Genton, Tilmann Gneiting, Michael Stein,  Hao Zhang).
\end{itemize}
These results are rather inconsistent to those obtained by D-SCORE:  the ARI and VI between two the vectors of predicted community labels  by LNSC and SCORE are $0.07$ and $1.68$, respectively.
Moreover, it seems that
\begin{itemize}
\item LNSC merges part of the nodes in the  ``Variable Selection" ($1285$ nodes) and ``Large-Scale Multiple Testing" ($359$ nodes) communities identified by D-SCORE into a new  HDDA-Cita-LNSC community,  but with a much smaller size ($614$ nodes).
\item The Biostat-Cita-LNSC community  ($1605$ nodes) is much larger than the ``Spatial" community identified by D-SCORE ($1010$ nodes), and hard to interpret.
\end{itemize}
Our observations here somehow agree with \cite{Kim2010} that LNSC  can not properly distinguish the directions  of the edges and can not detect communities representing directionality patterns among the nodes.

\section{Discussions}
\label{sec:Discu}
We have collected, cleaned, and analyzed two network data sets: the Coauthorship  network and Citation network for
statisticians. We investigate the network centrality and
community structures  with an array of different tools, ranging from Exploratory Data Analysis (EDA) \cite{EDA} tools   to rather sophisticated methods.   Some of these tools are relatively recent (e.g., SCORE, NSC, BCPL, APL, LNSC), and some are even new (e.g.,   D-SCORE for directed networks).
We have also presented  an array of interesting results.
For example, we identified the ``hot" authors and papers, and about $15$  meaningful communities  such as ``Spatial Statistics", ``Dimension Reduction",    ``Large-Scale Multiple Testing",    ``Objective Bayes",  ``Quantile Regression",  ``Theoretical Machine Learning", and ``Variable Selection".

The paper also has several  limitations that need further explorations.
First of all,  constrained by time and resources,
the two data sets we collect are  limited to the papers published in  four ``core" statistical journals: AoS, Biometrika,
JASA, and JRSS-B in the $10$ year period from $2003$ to $2012$.
We recognize that many statisticians not only publish
in so-called ``core" statistical journals
but also publish in a wide variety of journals of other scientific disciplines, including  but  not limited to
Nature, Science, PNAS,  IEEE journals,
journals in computer science, cosmology and astronomy,  economics and finance,  probability, and social sciences.
We also recognize  that many statisticians (even very good ones, such as  David Donoho, Steven Fienberg)  do not publish often in these journals in this specific time period.  For  these reasons, some of the results presented in this paper may be biased and they need to be interpreted with caution.

Still, the two data sets and the results we presented here serve well for our purpose of understanding
many aspects of the networks of statisticians who have USA as their home base;   see Section \ref{subsec:exp}.
They also serve as a good starting point
for a much more ambitious project on social networks for statisticians with a more ``complete" data set
for statistical publications.

Second, for  reasons of space,  we have primarily focused on data analysis in this paper, and the discussions
on models,  theory,   and methods have been kept as brief as we can.
On the other hand,  the data sets provide a fertile ground for modeling and development of methods and theory,
and there are an array of interesting problems   worthy of exploration in the near future.
For example, what could be a better model for either of the two data sets, what could be
a better measure  for centrality, and what could be a better method for community detection.
In particular, we propose D-SCORE as a new community detection method for directed network, but we only present
the idea underlying the methods, without careful analysis. We address the latter in a forthcoming paper \cite{JJZ}.
Also, sometimes,  the community detection results by different methods (e.g., SCORE, D-SCORE,  NSC,  BCPL, APL, LNSC)
are  inconsistent with each other.  When this happens, it is hard to have a conclusive comparison or  interpretation.
In light of this, it is of great interest to set up a theoretical framework and use it  to investigate the weaknesses and strengths of these methods.

Third, there are  many other interesting problems  we have not addressed here:  the issue of  mixed  membership,    link prediction, relationship between citations and recognitions (e.g., receiving an important award, elected to National Academy of Science), relationship and differences between  ``important work", ``influential work", and ``popular work".   It is of interest to explore these in the  future.

Last but not the least, coauthorship and citation networks only provide limited information for studying the research habits, trends, topological patterns, etc. of the statistical community.
There are more informative approaches  (say, using other information of the paper:  abstract, author affiliations,  key words, or even the whole paper) to studying such characteristics.  Such study is beyond the scope of the  paper, so we leave it to the future.

\section{Appendix I: Productivity, patterns and trends}
\label{sec:pt}
In this section, we report our findings on   three interconnected aspects: productivity, coauthor patterns and trends,
citation patterns and trends.

\subsection{Productivity}
\label{subsec:productivity}
\begin{figure}[htb!]
 \centering
 \includegraphics[width = 2.45 in, height = 1.6 in]{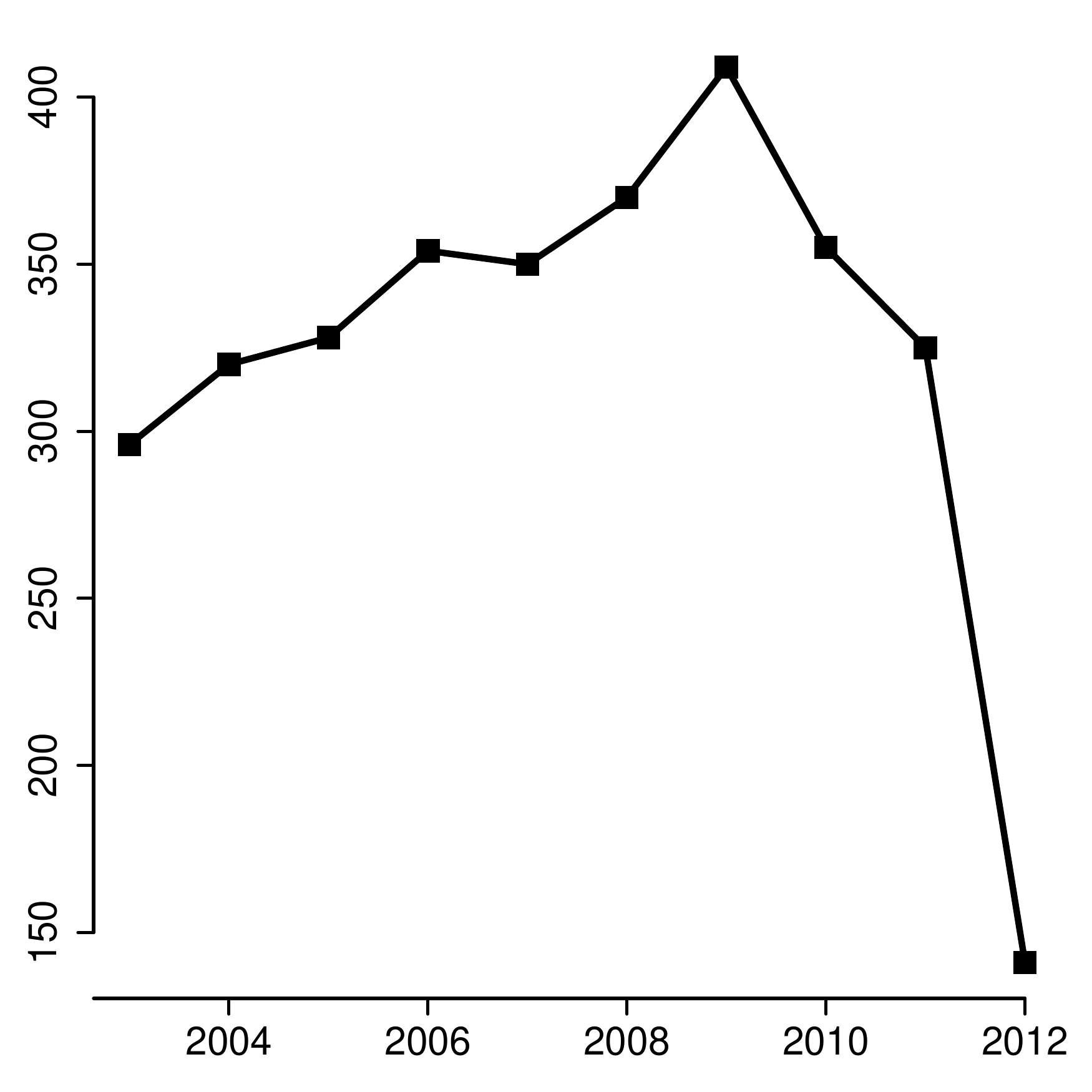}
 \includegraphics[width = 2.45 in, height = 1.6 in]{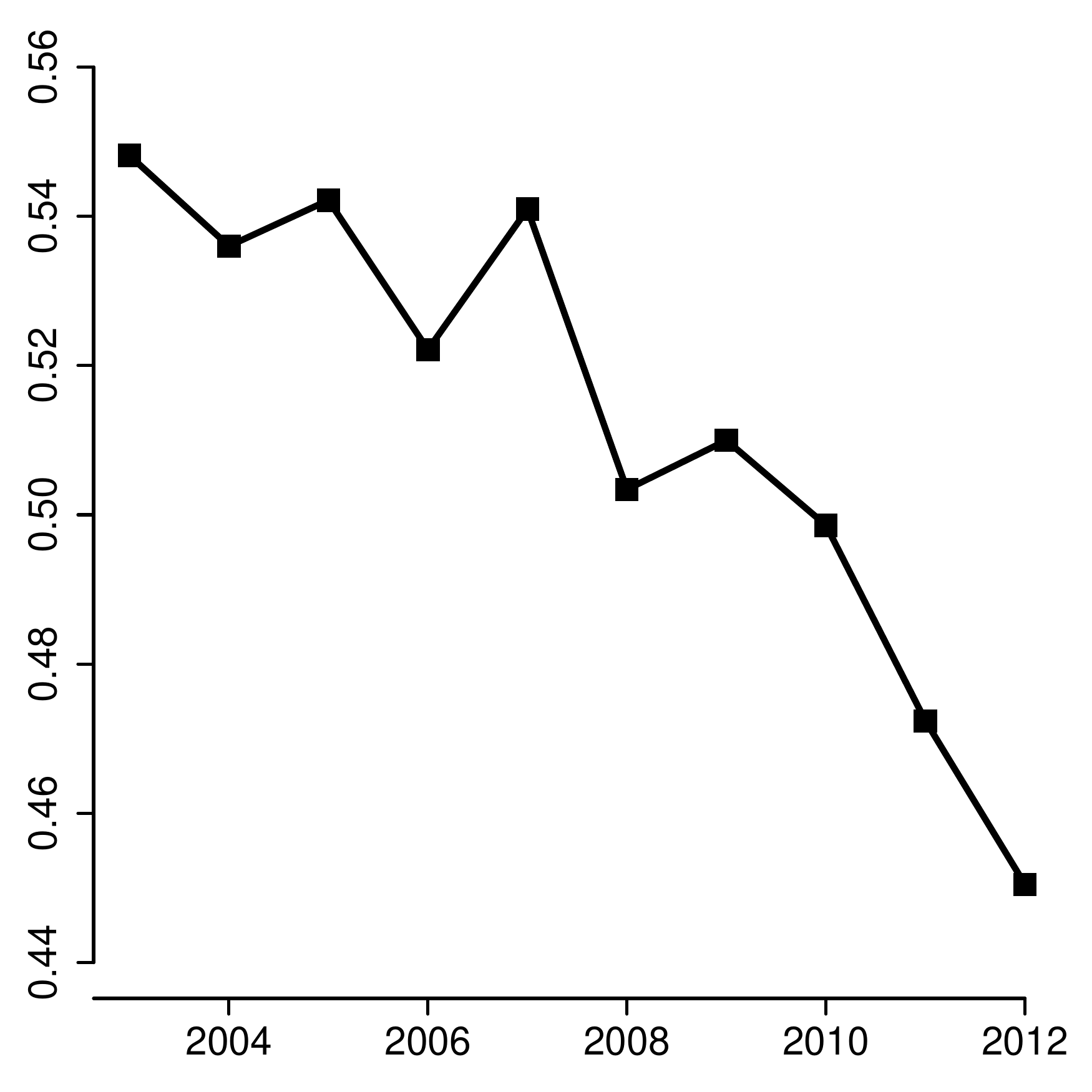}
 \caption{Left: total number of papers published each year from $2002$ to $2012$ (for the year 2012, we have only data for the first half).  Right:   the ratios between the number of papers published in each year and  the number of authors who has published in the same year. }
 \label{fig:nPaper}
 \end{figure}

Overall, there are $3248$ papers and $3607$ authors in the data set, suggesting an average of $0.90$ paper per author. It is of interest  to investigate how the productivity evolves  over the years.
In Figure \ref{fig:nPaper},  we present the total  number of papers published  in each year (left panel) and the average number of papers per author in each year (right panel), i.e., the ratio of the total number of papers published that year over the total number of authors who published at least once that year (it seems the result is inconsistent to that of an overall mean of
$.90$,   but this is due to that authors in different years largely overlap with each other).
It is interesting to note that over the $10$-year period,
the number of papers published each year has been increasing,  but  the average number of papers per author has been decreasing (drop about $18\%$ in ten years).
Possible explanations include:
\begin{itemize}
\item  {\it More collaborative}.  Collaboration between authors has been increasing.
\item  {\it More competitive}. Statistics has become a more competitive area, and there are more people who enter the area than who leave the area. Also, it  becomes increasingly more difficult to publish in these $4$ journals (which are viewed by many as
top journals in statistics).
\end{itemize}
Note that it could also be the case that  the productivity does not change much, but statisticians are publishing in a wider range of journals, and more younger ones have started making substantial contributions to the field.

We also present the distribution of the numbers of papers per author.
For any $K$-author paper,  $K \geq 1$, we have two different
ways to count each coauthor's contribution to this particular paper, either divided or non-divided.
\begin{itemize}
\item {\it Non-divided}. We count every coauthor as has published one paper.
\item {\it Divided}.  We count every coauthor as has published $1/K$ paper.
\end{itemize}
Both approaches have their virtues and disadvantages.
The first way may cause substantial ``inflation" in counting, and the second way may be insufficient, especially since for many papers, there are one or more ``leading authors" who contribute
  most of the work.

\begin{figure}[htb!]
 \centering
 \includegraphics[width = 2.45 in, height = 1.6 in]{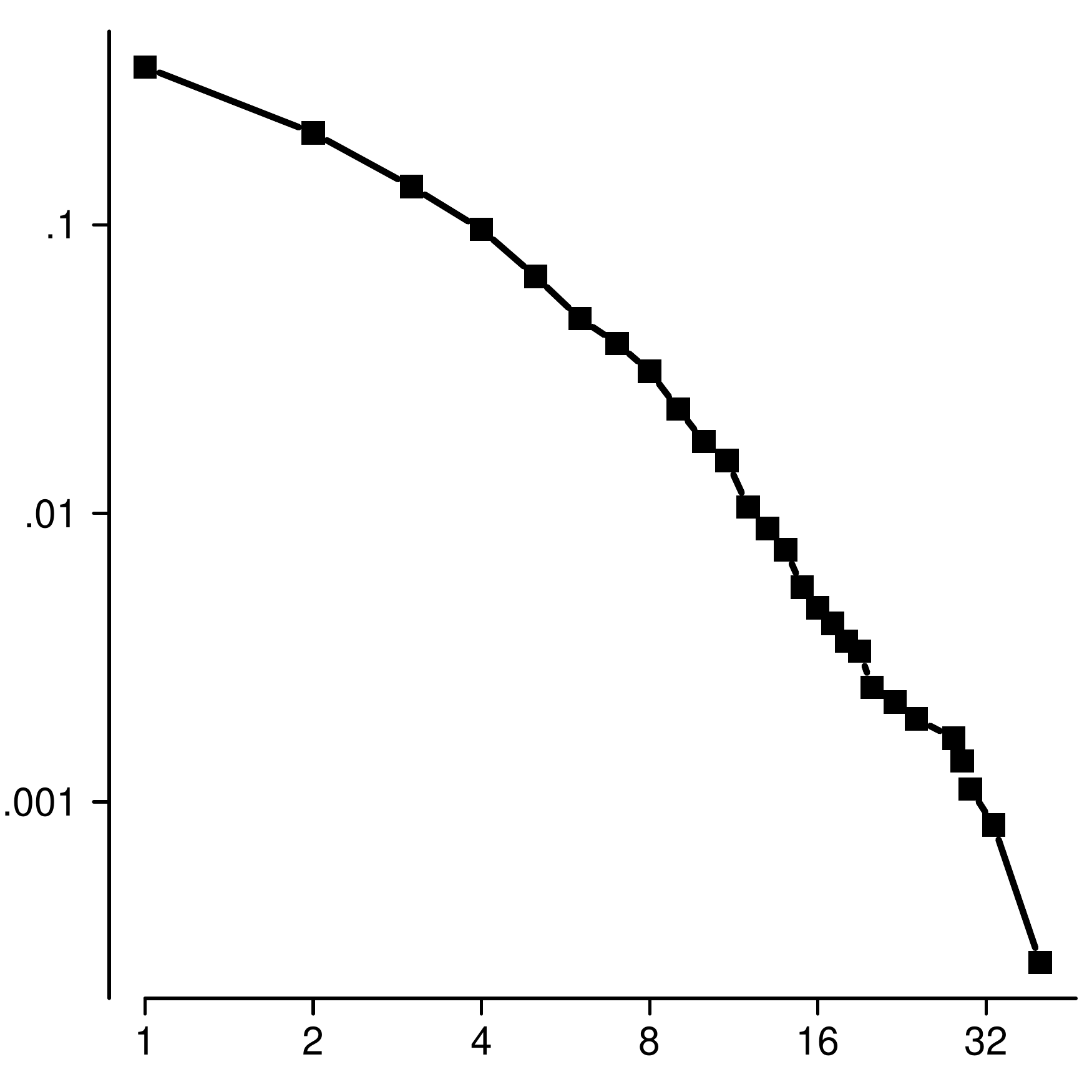}
 \includegraphics[width = 2.45 in, height = 1.6 in]{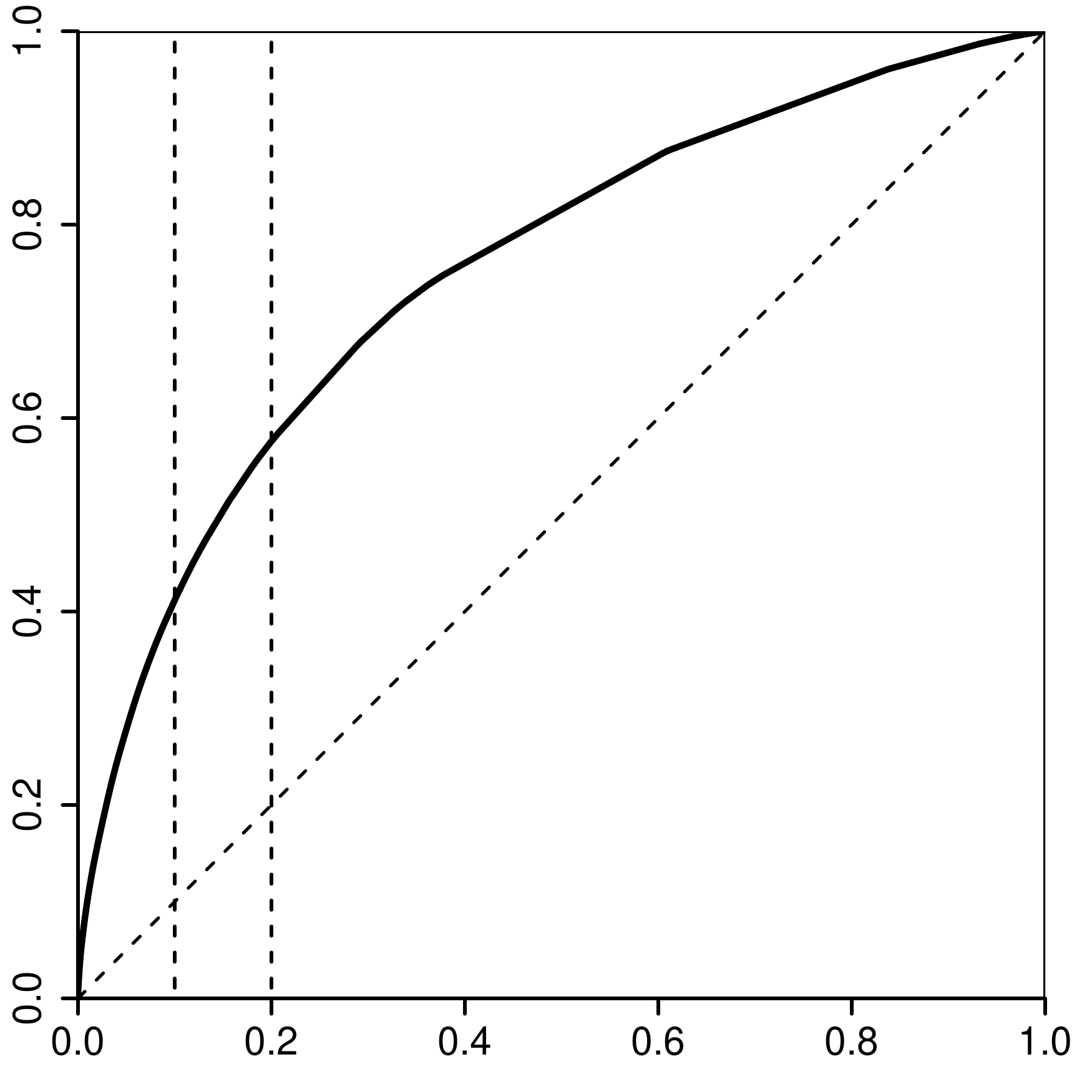}
 \caption{Left: The  proportion of authors who have written more than a certain number of papers (for a better view, both axes are evenly spaced on the logarithmic scale).  Right: The Lorenz curve for the number of papers each author with divided contributions.}
 \label{fig:LC}
 \end{figure}

Following the first approach,  we have the left panel of Figure \ref{fig:LC},
where the $x$-axis is the   number of papers, and
the $y$-axis is the proportion of authors who have written more than a certain number of papers.   Approximately, the curve looks like a straight line, especially to the right tail. This suggests that the distribution of the number of papers has a power law tail.

Following the second  approach, we present the Lorenz curve \cite{NewmanBook}
 of the number of  papers by each author (where for a $K$-author paper, each author is counted as having $1/K$ paper)  in the right panel of  Figure \ref{fig:LC}, which suggests the distribution does not have  a power law tail but is still very skewed.
   The figure shows that the top $10\%$ most prolific authors
contribute $41\%$ of the papers, and the  top $20\%$ most prolific authors  contribute $58\%$ of the papers.
Our findings are similar to that in \cite{Newman2013}  for the physics community.

The Gini coefficient  \cite{Gini} is a well-known measure of dispersion for a distribution. For our data set, the Gini coefficient for the distribution of the  number  of papers by different authors is $0.51$, which is  much smaller than the Gini coefficient of $0.70$ for that associated with the physics community \cite{Newman2013}.  This seems to suggest that the published papers are more evenly distributed among authors in the statistics community than the physics community.
 Another  possible explanation is that the data set  in \cite{Newman2013} is based on  all published papers in physics  spanning more than 100 years, while our data set is  based on  four  journals  in statistics for a 10-year period.
It is expected that in the latter,  the distribution of the  number of papers by different authors  (with divided contributions)
is less dispersed.
It is interesting to note that the Gini coefficient of the income inequality for the USA in the year of $2011$ is $0.48$, which is slightly smaller than $0.51$.

\subsection{Coauthor patterns and trends}
\label{subsec:coauthorpt}
\begin{figure}[htb!]
 \centering
 \includegraphics[width= 2.45 in, height = 1.6 in]{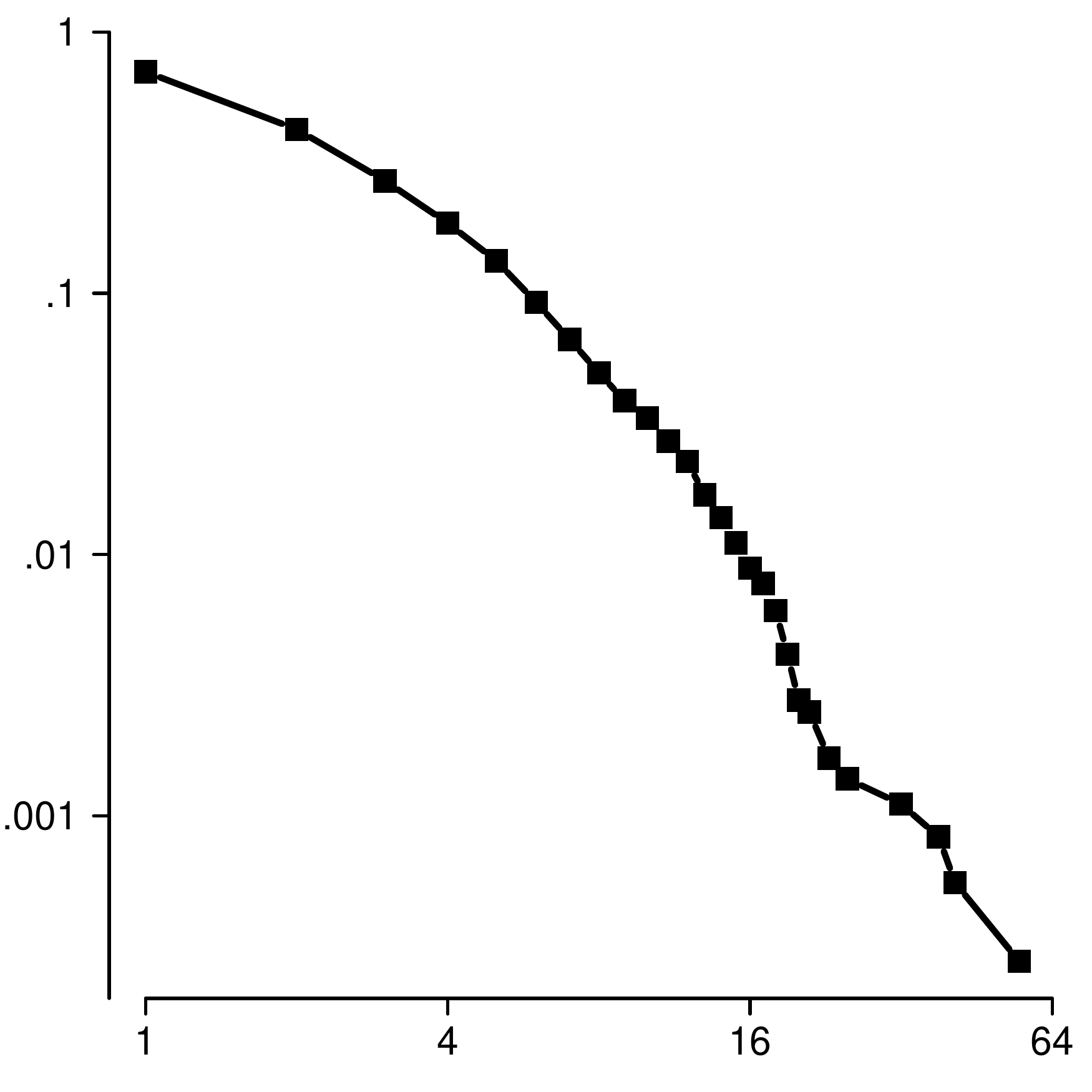}
 \includegraphics[width= 2.45 in, height = 1.6 in]{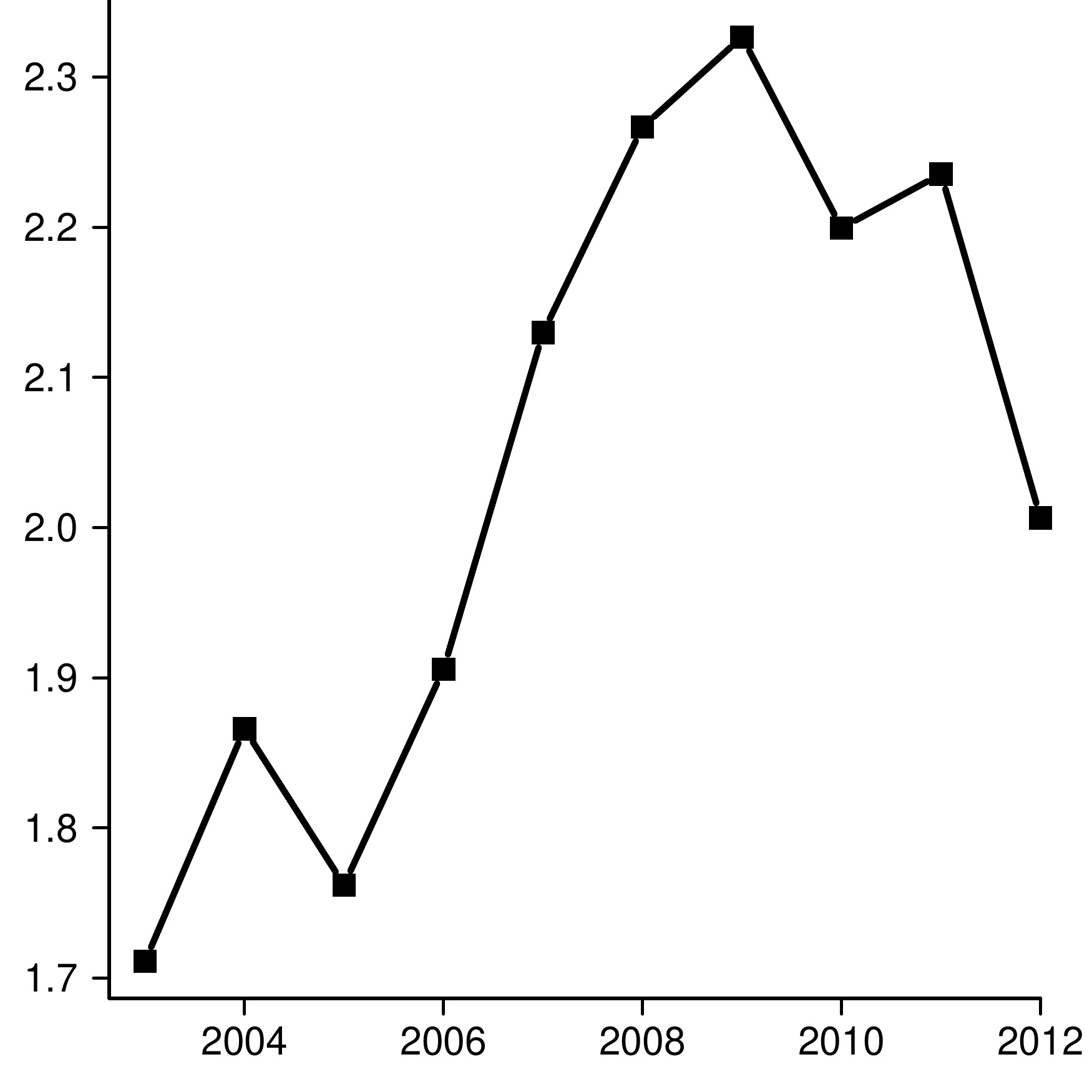}
 \caption{Left: The proportion of authors with more than a given number of coauthors (for a better view, both axes are evenly spaced on the  logarithmic scale). Right: The average number of coauthors for all authors  who has published in these journals that year. }
 \label{fig:NCoauthor}
 \end{figure}

In the coauthorship network, the degree of a node is also the number of coauthors for the node.
The degrees range from $0$ to $65$, where
Peter  Hall (65), Raymond  Carroll (55),  Joseph   Ibrahim (41), Jianqing Fan (38) and David Dunson (32)
are  the ones with the highest degrees (and so they are the most collaborative authors).
Also, $154$ authors have degree $0$, and  $913$  authors have degree $1$.
The degree distribution  is shown in Figure \ref{fig:NCoauthor} (left panel),  suggesting a power law tail.

It is of interest to investigate how the number of coauthors changes over time.
 In Figure \ref{fig:NCoauthor} (right panel), we present  the average number of coauthors in each of the $10$ years (for each year,  we consider only the authors who published in these journals).  It is seen that overall the average number of coauthors  is steadily increasing. Again, this suggests that the statistics  community   has become  increasingly more collaborative.

Many social networks are transitive  (e.g., a friend of a friend is likely to be a friend)  \cite{Wasserman1994}.    For the coauthorship network based our data sets,  the transitivity  is $0.32$, compared to
  $0.066$ for the biology community, $0.15$ for the  mathematics community, and $0.43$ for the physics community  \cite{Newman2004}.   For real-world social networks, the usual range of  transitivity is between $0.3$ and $0.6$ \cite{NewmanBook},   suggesting  that the Coauthorship network is moderately transitive.

\subsection{Citation patterns and trends}
\label{subsec:citationpt}
For the $3248$ papers  ($3607$ authors) in our data sets,  the average citation per paper is $1.76$, which is significantly lower than the Impact Factor (IF)  of these journals. Based on ISI 2010, the IFs for AoS, JRSS-B, JASA, and Biometrika are   3.84,  3.73,  3.22, and 1.94,  respectively.
This is largely due to that we  count only the citations between papers in these $4$ journals  in a $10$-year period.  Among these   papers,
(a) $1693$ ($52\%$) are not cited by any other paper in the data set,
 (b)  $1450$  ($45\%$)   do not cite any other paper in the data set,
and (c)  $778$  ($24\%$)   neither cite  nor are cited by any other papers in the data sets.

\begin{figure}[htb!]
 \centering
 \includegraphics[width = 2.45 in, height = 1.6 in]{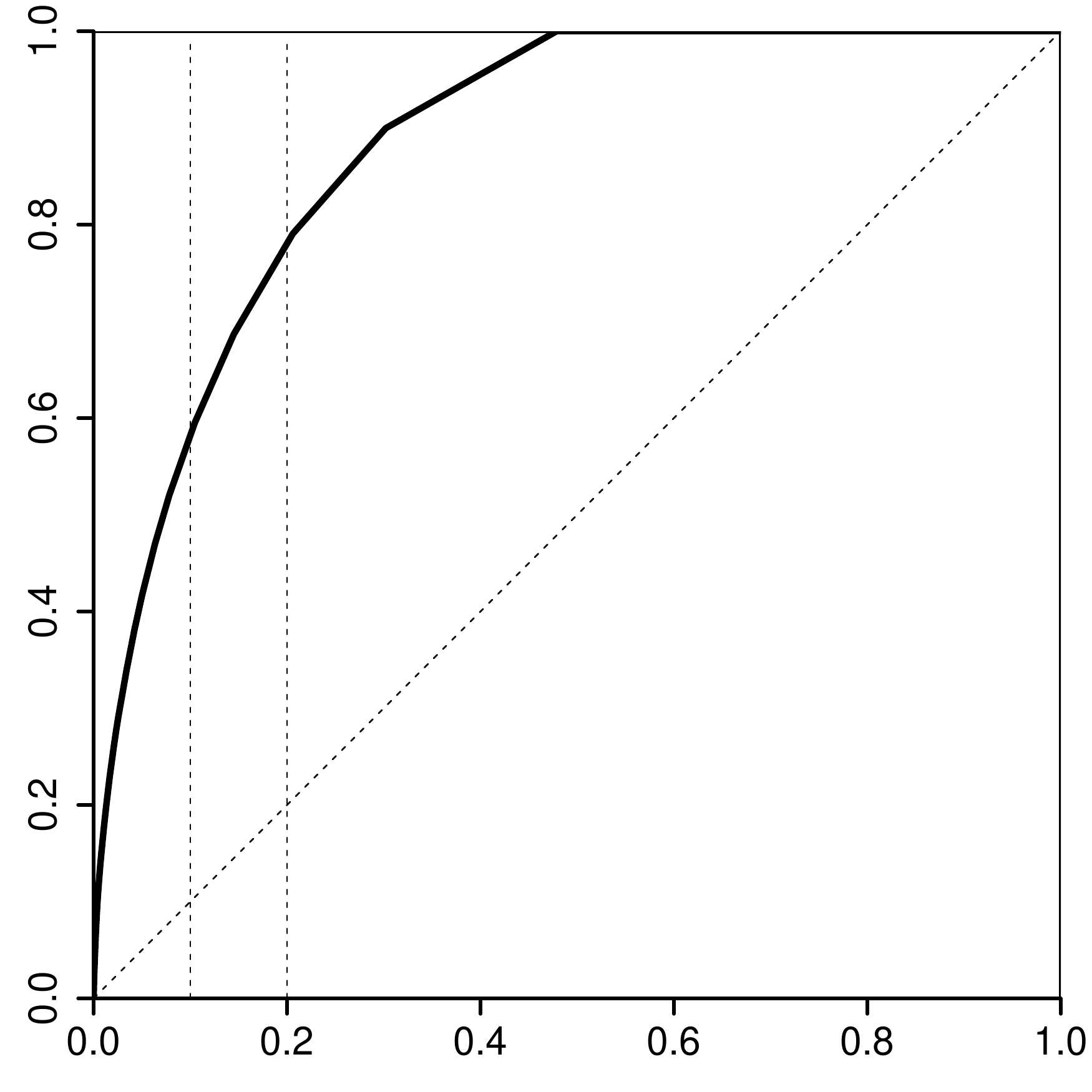}
 \includegraphics[width = 2.45 in, height = 1.6 in]{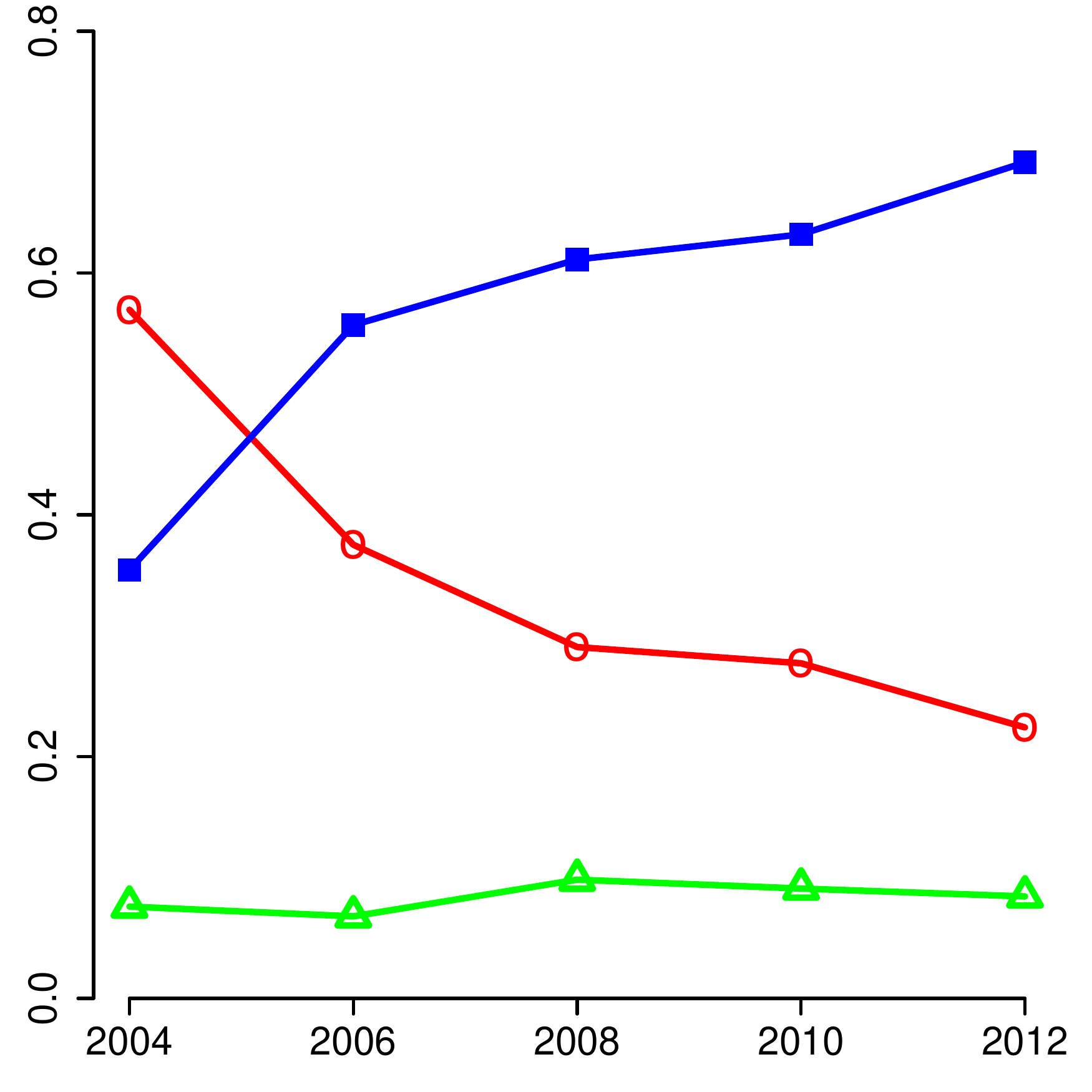}
 \caption{Left: The Lorenz curve for the number of citation received by each paper.  Right: The proportions of self-citations (red circles), coauthor citations (green triangles) and distant citations (blue rectangles) for each two-year block.}
 \label{fig:LC-citations}
 \end{figure}

The distribution of the in-degree (the number of citations received by each paper)  is highly skewed.
The top $10\%$ highly cited papers receive about $60\%$ of all citation counts, while the top $20\%$ receive about $80\%$ of all citation counts.
 The Gini coefficient is $0.77$ \cite{Gini} suggesting that the in-degree is highly dispersed.  The Lorenz curve \cite{NewmanBook}  is shown in  Figure \ref{fig:LC-citations} (left panel), confirming  that the distribution of the in-degrees is highly skewed.

We also observe some very interesting patterns. First,
the authors return a favor of citation, especially if it is from a coauthor. The proportion of (either earlier or later) reciprocation among coauthor citations is $79\%$, while that among distant citations is $25\%$.

In  Figure \ref{fig:LC-citations} (right panel), we show that over the   10-year period,    (a) the proportion of self-citations has been  slowly decreasing, (b) the proportion of citations from a coauthor remains roughly the same, and (c) the proportion of  distant citations (citations that are not from oneself or a coauthor)  has been slowly increasing. The  last item is a little unexpected, but it probably makes sense in that over the years,
the publications  have  become  increasingly more  accessible online and communications  have  become  increasingly easier and more efficient. That the blue curve and the red cross crossover with each other on the left
is probably due to the ``boundary effect":  for papers published in 2003 (say),  most the papers they have cited are probably
published earlier than 2002, which are not  included in our data sets.
Below, we   show that  the mean delay of citation is about 3 years. For this reason,  the ``boundary effect"  is probably  negligible in the later half of the time period.
 Note that   the overall proportions for self-citations, coauthor citations and distant citations are $27\%$, $9\%$, and $64\%$, respectively.

The data set also confirms a reasonable delay in citations, despite the fact that most papers appear online (such as personal website, arXiv, department archives) much earlier than the time when the paper is published.  The overall mean delay (e.g., the average difference between the years of the publication of a new paper and the papers it cites)   is $3.30$ years, and the mean delay for  self-citations, coauthor citations, and distant citations,
are  $2.81$, $3.36$ and $3.51$ years, respectively,
 suggesting the authors cite their own or their coauthors' work more quickly than that of others.

\section{Appendix II: Data collection and cleaning}
\label{sec:data}
In this section, we describe how the data were collected and preprocessed,   and how we have overcome the challenges we have faced.

We focus on all papers  published in AoS, JASA, JRSS-B, and Biometrika from  2003 to  the first  half of 2012.  For each paper in this range, we have extracted the Digital Object Identifier (DOI), title, information for the authors, abstract, keywords, journal name, volume, issue,  and page numbers,  and the DOIs of the papers in the same range that have cited this paper. The raw data set consists of about 3500 papers and 4000 authors.

Among these papers, we are only interested in those for original research, so we have removed items such as the book reviews, erratum, comments or rejoinders, etc. Usually, these items contain  signal   words such as ``Book Review",  ``Corrections" etc.  in the title. Removing such items  leaves us with a total of  $3248$ papers (about $3950$ authors) in the range of interest.

Our data collection process has three main  steps. In the first step, we identify all papers in the range of interest.
In the second step, we figure out all citations  between  the papers of interest (note that the information for {\it citation relationship between any two authors} is not directly available).
In the third step, we identify all the authors for each paper.

In the first step,  recall that the goal is to identify every paper in our range of interest, and   for each of them, to collect  the  title, author,  DOI, keywords, abstract, journal name,  etc.
In this step, we face two main challenges.

First, all popular online resources  have strict limits for high-quality high-volume downloads; we have explained this in Section \ref{subsec:dataset} with details. Eventually, we manage to overcome the challenge  by
downloading the desired data and information from Web of Science and MathSciNet little by little,  each time in the maximum amour that is allowed.   Overall, it has taken us a few months to download and combine the data from two different sources.

Second, it is hard to find a good identifier for the papers.  While  the  titles  of  the papers   could serve as unique identifiers,  they are difficult to format and compare. Also,  while many online resources have their  own paper identifiers,   they are either unavailable or unusable for our purpose. Eventually, we decide to
 use the DOI  as the identifier. The DOI has been used as a unique identifier for papers by most publishers for statistical papers since $2000$.

Using DOI as the identifier,  with substantial time and efforts, we have successfully identified all paper in the range of interest with Web of Science and MathSicNet.
One more difficulty we face here is that Web of Science does not have the DOIs of  (about) $200$ papers  and MathSciNet does not have the DOIs of (about) $100$  papers, and we have to combine these  two online sources to locate the DOI for each paper in our range of interest.

We now discuss the second step. The goal is to figure out the citation relationship between any two papers in the range  of interest.
MathSciNet does not  allow automated downloads for  such information,  but, fortunately, such information is retrievable from Web of Science, if we parse  the XML pages in R  at a small amount each time.  One issue we encounter in this step is that (as mentioned above) Web of Science misses the DOIs of about $200$ papers, and we have to deal with these  papers
with extra efforts.

Consider the last step. The goal is to uniquely identify all authors for each paper  in the range of interest.  This is the most time consuming step, and we have faced  many challenges.
First, for many papers published in Biometrika, we do not have the  first name and middle initial  for each author, and this causes problems.  For instance,  ``L. Wang" can be any one of ``Lan Wang", ``Li Wang", ``Lianming Wang", etc.  Second, the name of an author is not listed consistently in different occasions. For example,  ``Lixing Zhu" may be also listed as    ``Li Xing Zhu", ``L. X. Zhu", and ``Li-Xing Zhu".  Last but not the least, different authors may have the same name:   at least three authors (from Univ. of California at Riverside, Univ. of Michigan at Ann Arbor and Iowa State Univ., respectively)  have the same name of ``Jun Li".

Note that every service has its own internal identification  system, but, unfortunately, none of them
is willing to reveal the system to the end users.
Also, people have been trying hard to create a universal author identification system, in a similar spirit to that of using DOI as a universal identifier for each paper. Among these are ResearcherID introduced by Thomson Reuters in 2008 and Open Researcher and Contributor ID (ORCID)  introduced in 2012. However, the  use of such systems is still very limited.

Eventually, we have to solve the problem on our own. First,  roughly saying,
we have written a  program  which mostly uses the author names (e.g., first, middle, and last names;  abbreviations)
to correctly identify all except  $200$ (approximately) authors, about whom we  may have problems in identification.  We then manually identify each of these $200$ authors using additional information (e.g., affiliations, email addresses,  information on their websites).   After all such cleaning, the number of authors is reduced from about $3950$ to $3607$.


For reproducibility purpose, we have prepared the data files and a demo for readers who are interested in exploring the data sets. All these can be found at  \verb+www.stat.uga.edu/~psji/+  once the paper is accepted for
publication. In particular, the data files include the following.
\label{subsec:datafiles}
\begin{itemize}
\item \verb+4Journals.bib+:   the raw bibtex data for about 3500 items including papers, book reviews, corrections, etc
\item \verb+4Journals-cleaned.bib+:   the cleaned bibtex data for 3248 papers after removing the book reviews and corrections and clustering the author names
\item \verb+author-cluster.txt+:      the final clustering rules for the author names
\item \verb+author-cluster-man.txt+:     the manually defined clustering rules for the author names
\item \verb+author-list.txt+:    the list of the 3607 authors after disambiguation
\item \verb+author-paper-adjacency.txt+:    3607x3248 bipartite adjacency matrix
\item \verb+coauthor-adjacency.txt+:   the 3607x3607 coauthor adjacency matrix
\item \verb+citation-adjacency.txt+:    the 3607x3607 citation adjacency matrix 
\end{itemize}

\section*{Acknowledgements}
JJ thanks David Donoho and Jianqing Fan; the paper was inspired
by a lunch conversation with them in 2011 on H-index.
The authors thank Stephen Fienberg, Qunhua Li, Douglas Nychka, and Yunpeng Zhao for helpful pointers.


\bibliographystyle{imsart-number}
\bibliography{Jiashun}

\begin{thebibliography}{45}

\bibitem{Amini2013}
\begin{barticle}[author]
\bauthor{\bsnm{Amini},~\bfnm{Arash.}\binits{A.}},
  \bauthor{\bsnm{Chen},~\bfnm{Aiyou}\binits{A.}},
  \bauthor{\bsnm{Bickel},~\bfnm{Peter}\binits{P.}} \AND
  \bauthor{\bsnm{Levina},~\bfnm{Elizaveta}\binits{E.}}
(\byear{{2013}}).
\btitle{{Pseudo-likelihood methods for community detection in large sparse
  networks}}.
\bjournal{{Ann. Statist.}}
\bvolume{{41}}
\bpages{{2097-2122}}.
\end{barticle}
\endbibitem

\bibitem{Arenas2007}
\begin{barticle}[author]
\bauthor{\bsnm{Arenas},~\bfnm{Alex}\binits{A.}},
  \bauthor{\bsnm{Duch},~\bfnm{Jordi}\binits{J.}},
  \bauthor{\bsnm{Fernandez},~\bfnm{Alberto}\binits{A.}} \AND
  \bauthor{\bsnm{Gomez},~\bfnm{Sergio}\binits{S.}}
(\byear{{2007}}).
\btitle{{Size reduction of complex networks preserving modularity}}.
\bjournal{{New J. Phys.}}
\bvolume{{9}(6)}
\bpages{176}.
\bdoi{{10.1088/1367-2630/9/6/176}}
\end{barticle}
\endbibitem

\bibitem{DIGRAPHS}
\begin{bbook}[author]
\bauthor{\bsnm{Bang-Jensen},~\bfnm{Jorgen}\binits{J.}} \AND
  \bauthor{\bsnm{Gutin},~\bfnm{Gregory}\binits{G.}}
(\byear{2009}).
\btitle{Digraphs: Theory, Algorithms and Applications}.
\bpublisher{Springer}.
\end{bbook}
\endbibitem

\bibitem{Barabasi1999}
\begin{barticle}[author]
\bauthor{\bsnm{Barabasi},~\bfnm{Albert-Laszlo}\binits{A.-L.}} \AND
  \bauthor{\bsnm{Albert},~\bfnm{Reka}\binits{R.}}
(\byear{{1999}}).
\btitle{{Emergence of scaling in random networks}}.
\bjournal{{Science}}
\bvolume{{286}}
\bpages{{509-512}}.
\bdoi{{10.1126/science.286.5439.509}}
\end{barticle}
\endbibitem

\bibitem{BickelChen2009}
\begin{barticle}[author]
\bauthor{\bsnm{Bickel},~\bfnm{Peter}\binits{P.}} \AND
  \bauthor{\bsnm{Chen},~\bfnm{Aiyou}\binits{A.}}
(\byear{{2009}}).
\btitle{{A nonparametric view of network models and Newman-Girvan and other
  modularities}}.
\bjournal{Proc. Nat. Acad. Sci.}
\bvolume{{106}}
\bpages{{21068-21073}}.
\end{barticle}
\endbibitem

\bibitem{BLT}
\begin{barticle}[author]
\bauthor{\bsnm{Bickel},~\bfnm{Peter}\binits{P.}} \AND
  \bauthor{\bsnm{Levina},~\bfnm{Elizaveta}\binits{E.}}
(\byear{2008}).
\btitle{Regularized estimation of large covariance matrices}.
\bjournal{Ann. Statist.}
\bvolume{36}
\bpages{199--227}.
\end{barticle}
\endbibitem

\bibitem{Bickel}
\begin{barticle}[author]
\bauthor{\bsnm{Bickel},~\bfnm{Peter}\binits{P.}} \AND
  \bauthor{\bsnm{Levina},~\bfnm{Elizaveta}\binits{E.}}
(\byear{2008}).
\btitle{Covariance regularization by thresholding}.
\bjournal{Ann. Statist.}
\bvolume{36}
\bpages{2577--2604}.
\end{barticle}
\endbibitem

\bibitem{Candes2007}
\begin{barticle}[author]
\bauthor{\bsnm{Candes},~\bfnm{Emmanuel}\binits{E.}} \AND
  \bauthor{\bsnm{Tao},~\bfnm{Terrence}\binits{T.}}
(\byear{2007}).
\btitle{The Dantzig selector: statistical estimation when $p$ is much larger
  than $n$ (with discussion)}.
\bjournal{Ann. Statist.}
\bvolume{35}
\bpages{2313--2351}.
\end{barticle}
\endbibitem

\bibitem{Donoholasso}
\begin{barticle}[author]
\bauthor{\bsnm{Chen},~\bfnm{Shaobing}\binits{S.}},
  \bauthor{\bsnm{Donoho},~\bfnm{David}\binits{D.}} \AND
  \bauthor{\bsnm{Saunders},~\bfnm{Michael}\binits{M.}}
(\byear{1998}).
\btitle{Atomic decomposition by basis pursuit}.
\bjournal{SIAM J. Sci. Comput.}
\bvolume{20}
\bpages{33--61}.
\end{barticle}
\endbibitem

\bibitem{Efron}
\begin{barticle}[author]
\bauthor{\bsnm{Efron},~\bfnm{Bradley}\binits{B.}},
  \bauthor{\bsnm{Hastie},~\bfnm{Trevor}\binits{T.}},
  \bauthor{\bsnm{Johnstone},~\bfnm{Iain}\binits{I.}} \AND
  \bauthor{\bsnm{Tibshirani},~\bfnm{Robert}\binits{R.}}
(\byear{2004}).
\btitle{Least angle regression}.
\bjournal{Ann. Statist.}
\bvolume{32}
\bpages{407--499}.
\end{barticle}
\endbibitem

\bibitem{FanLi2004}
\begin{barticle}[author]
\bauthor{\bsnm{Fan},~\bfnm{Jianqing}\binits{J.}} \AND
  \bauthor{\bsnm{Li},~\bfnm{Runze}\binits{R.}}
(\byear{2004}).
\btitle{New estimation and model selection procedures for semiparametric
  modeling in longitudinal data analysis}.
\bjournal{J. Amer. Statist. Assoc.}
\bvolume{99}
\bpages{710--723}.
\bdoi{10.1198/016214504000001060}
\bmrnumber{2090905 (2005d:62053)}
\end{barticle}
\endbibitem

\bibitem{FanLv}
\begin{barticle}[mr]
\bauthor{\bsnm{Fan},~\bfnm{Jianqing}\binits{J.}} \AND
  \bauthor{\bsnm{Lv},~\bfnm{Jinchi}\binits{J.}}
(\byear{2008}).
\btitle{Sure independence screening for ultrahigh dimensional feature space}.
\bjournal{J. Roy. Statist. Soc. B}
\bvolume{70}
\bpages{849--911}.
\bid{mr={2530322}}
\end{barticle}
\endbibitem

\bibitem{FanPeng2004}
\begin{barticle}[author]
\bauthor{\bsnm{Fan},~\bfnm{Jianqing}\binits{J.}} \AND
  \bauthor{\bsnm{Peng},~\bfnm{Heng}\binits{H.}}
(\byear{2004}).
\btitle{Nonconcave penalized likelihood with a diverging number of parameters}.
\bjournal{Ann. Statist.}
\bvolume{32}
\bpages{928--961}.
\bdoi{10.1214/009053604000000256}
\bmrnumber{2065194 (2005g:62047)}
\end{barticle}
\endbibitem

\bibitem{Freeman}
\begin{barticle}[author]
\bauthor{\bsnm{Freeman},~\bfnm{Linton}\binits{L.}},
  \bauthor{\bsnm{Borgatti},~\bfnm{Stephen}\binits{S.}} \AND
  \bauthor{\bsnm{White},~\bfnm{Douglas}\binits{D.}}
(\byear{1991}).
\btitle{Centrality in valued graphs: A measure of betweenness based on network
  flow}.
\bjournal{Soc. Networks}
\bvolume{13}
\bpages{141--154}.
\end{barticle}
\endbibitem

\bibitem{Gini}
\begin{barticle}[author]
\bauthor{\bsnm{Gini},~\bfnm{Corrado}\binits{C.}}
(\byear{1936}).
\btitle{On the measure of concentration with special reference to income and
  statistics}.
\bjournal{Colorado College Publication, General Series}
\bvolume{208}
\bpages{73-79}.
\end{barticle}
\endbibitem

\bibitem{Fienberg}
\begin{barticle}[author]
\bauthor{\bsnm{Goldenberg},~\bfnm{Anna}\binits{A.}},
  \bauthor{\bsnm{Zheng},~\bfnm{Alice}\binits{A.}},
  \bauthor{\bsnm{Fienberg},~\bfnm{Steven}\binits{S.}} \AND
  \bauthor{\bsnm{Airoldi},~\bfnm{Edoardo}\binits{E.}}
(\byear{2009}).
\btitle{A survey of statistical network models}.
\bjournal{Foundations and Trends in machine learning}
\bvolume{2}
\bpages{129-233}.
\end{barticle}
\endbibitem

\bibitem{Grossman2002}
\begin{barticle}[author]
\bauthor{\bsnm{Grossman},~\bfnm{Jerrold}\binits{J.}}
(\byear{2002}).
\btitle{The evolution of the mathematical research collaboration graph}.
\bjournal{Congressus Numerantium}
\bvolume{158}
\bpages{{201-212 }}.
\end{barticle}
\endbibitem

\bibitem{Huang2008}
\begin{barticle}[author]
\bauthor{\bsnm{Huang},~\bfnm{Jian}\binits{J.}},
  \bauthor{\bsnm{Horowitz},~\bfnm{Joel}\binits{J.}} \AND
  \bauthor{\bsnm{Ma},~\bfnm{Shuangge}\binits{S.}}
(\byear{2008}).
\btitle{Asymptotic properties of bridge estimators in sparse high-dimensional
  regression models}.
\bjournal{Ann. Statist.}
\bvolume{36}
\bpages{587--613}.
\bdoi{10.1214/009053607000000875}
\end{barticle}
\endbibitem

\bibitem{Huang2006}
\begin{barticle}[author]
\bauthor{\bsnm{Huang},~\bfnm{Jianhua}\binits{J.}},
  \bauthor{\bsnm{Liu},~\bfnm{Naiping}\binits{N.}},
  \bauthor{\bsnm{Pourahmadi},~\bfnm{Mohsen}\binits{M.}} \AND
  \bauthor{\bsnm{Liu},~\bfnm{Linxu}\binits{L.}}
(\byear{2006}).
\btitle{Covariance matrix selection and estimation via penalised normal
  likelihood}.
\bjournal{Biometrika}
\bvolume{93}
\bpages{85--98}.
\bdoi{10.1093/biomet/93.1.85}
\end{barticle}
\endbibitem

\bibitem{HubertArabie1985}
\begin{barticle}[author]
\bauthor{\bsnm{Hubert},~\bfnm{Lawrence}\binits{L.}} \AND
  \bauthor{\bsnm{Arabie},~\bfnm{Phipps}\binits{P.}}
(\byear{1985}).
\btitle{Comparing partitions}.
\bjournal{J. Classif.}
\bvolume{2}
\bpages{193-218}.
\end{barticle}
\endbibitem

\bibitem{HunterLi2005}
\begin{barticle}[author]
\bauthor{\bsnm{Hunter},~\bfnm{David}\binits{D.}} \AND
  \bauthor{\bsnm{Li},~\bfnm{Runze}\binits{R.}}
(\byear{2005}).
\btitle{Variable selection using {MM} algorithms}.
\bjournal{Ann. Statist.}
\bvolume{33}
\bpages{1617--1642}.
\bdoi{10.1214/009053605000000200}
\bmrnumber{2166557}
\end{barticle}
\endbibitem

\bibitem{Ioannidis2008}
\begin{barticle}[author]
\bauthor{\bsnm{Ioannidis},~\bfnm{John}\binits{J.}}
(\byear{{2008}}).
\btitle{{Measuring co-authorship and networking-adjusted scientific impact}}.
\bjournal{{PLOS ONE}}
\bvolume{{3}}.
\bdoi{{10.1371/journal.pone.0002778}}
\end{barticle}
\endbibitem

\bibitem{JJZ}
\begin{bmisc}[author]
\bauthor{\bsnm{Ji},~\bfnm{Pengsheng}\binits{P.}},
  \bauthor{\bsnm{Jin},~\bfnm{Jiashun}\binits{J.}} \AND
  \bauthor{\bsnm{Ke},~\bfnm{Zheng}\binits{Z.}}
(\byear{2014}).
\btitle{Joint community detection for Coauthorship and Citation networks of
  statisticians by D-SCORE. \textit{ Manuscript}}.
\end{bmisc}
\endbibitem

\bibitem{SCORE}
\begin{barticle}[author]
\bauthor{\bsnm{Jin},~\bfnm{Jiashun}\binits{J.}}
(\byear{2014}).
\btitle{Fast community detection by SCORE}.
\bjournal{Ann. Statist. To appear.}
\end{barticle}
\endbibitem

\bibitem{JohnstoneSilverman2005}
\begin{barticle}[author]
\bauthor{\bsnm{Johnstone},~\bfnm{Iain}\binits{I.}} \AND
  \bauthor{\bsnm{Silverman},~\bfnm{Bernard}\binits{B.}}
(\byear{2005}).
\btitle{Empirical {B}ayes selection of wavelet thresholds}.
\bjournal{Ann. Statist.}
\bvolume{33}
\bpages{1700--1752}.
\bdoi{10.1214/009053605000000345}
\bmrnumber{2166560}
\end{barticle}
\endbibitem

\bibitem{DCBM}
\begin{barticle}[author]
\bauthor{\bsnm{Karrer},~\bfnm{Brian}\binits{B.}} \AND
  \bauthor{\bsnm{Newman},~\bfnm{Mark}\binits{M.}}
(\byear{2011}).
\btitle{Stochastic blockmodels and community structures in network}.
\bjournal{Phys. Rev.}
\bvolume{83}
\bpages{1436--1462}.
\end{barticle}
\endbibitem

\bibitem{Kim2010}
\begin{barticle}[author]
\bauthor{\bsnm{Kim},~\bfnm{Youngdo}\binits{Y.}},
  \bauthor{\bsnm{Son},~\bfnm{Seung-Woo}\binits{S.-W.}} \AND
  \bauthor{\bsnm{Jeong},~\bfnm{Hawoong}\binits{H.}}
(\byear{{2010}}).
\btitle{{Finding communities in directed networks}}.
\bjournal{{Phys. Rev. E}}
\bvolume{{81}}
\bpages{016103}.
\bdoi{{10.1103/PhysRevE.81.016103}}
\end{barticle}
\endbibitem

\bibitem{Newman2008}
\begin{barticle}[author]
\bauthor{\bsnm{Leicht},~\bfnm{Elizabeth}\binits{E.}} \AND
  \bauthor{\bsnm{Newman},~\bfnm{Mark}\binits{M.}}
(\byear{{2008}}).
\btitle{{Community structure in directed networks}}.
\bjournal{{Phys. Rev. Lett.}}
\bvolume{{100}}
\bpages{118703}.
\bdoi{{10.1103/PhysRevLett.100.118703}}
\end{barticle}
\endbibitem

\bibitem{Newman2013}
\begin{barticle}[author]
\bauthor{\bsnm{Martin},~\bfnm{Travis}\binits{T.}},
  \bauthor{\bsnm{Ball},~\bfnm{Brian}\binits{B.}},
  \bauthor{\bsnm{Karrer},~\bfnm{Brian}\binits{B.}} \AND
  \bauthor{\bsnm{Newman},~\bfnm{Mark}\binits{M.}}
(\byear{2013}).
\btitle{Coauthorship and citation patterns in the Physical Review}.
\bjournal{Phys. Rev. E}
\bvolume{88}.
\end{barticle}
\endbibitem

\bibitem{Meila2003}
\begin{bincollection}[author]
\bauthor{\bsnm{Meila},~\bfnm{Marina}\binits{M.}}
(\byear{2003}).
\btitle{Comparing clusterings by the variation of information}.
In \bbooktitle{Learning Theory and Kernel Machines: 16th Annual Conference on
  Computational Learning Theory and 7th Kernel Workshop}
(\beditor{\bfnm{B}\binits{B.}~\bsnm{Scholkopf}} \AND
  \beditor{\bfnm{M~K}\binits{M.~K.}~\bsnm{Warmuth}}, eds.)
\bpublisher{Springer}.
\end{bincollection}
\endbibitem

\bibitem{Meinshausen2006}
\begin{barticle}[author]
\bauthor{\bsnm{Meinshausen},~\bfnm{Nicolai}\binits{N.}} \AND
  \bauthor{\bsnm{B{\"u}hlmann},~\bfnm{Peter}\binits{P.}}
(\byear{2006}).
\btitle{High-dimensional graphs and variable selection with the lasso}.
\bjournal{Ann. Statist.}
\bvolume{34}
\bpages{1436--1462}.
\end{barticle}
\endbibitem

\bibitem{Newman2001}
\begin{barticle}[author]
\bauthor{\bsnm{Newman},~\bfnm{Mark}\binits{M.}}
(\byear{{2001}}).
\btitle{{The structure of scientific collaboration networks}}.
\bjournal{Proc. Natl. Acad. Sci. USA}
\bvolume{{98}}
\bpages{{404-409}}.
\bdoi{{10.1073/pnas.021544898}}
\end{barticle}
\endbibitem

\bibitem{Newman2001a}
\begin{barticle}[author]
\bauthor{\bsnm{Newman},~\bfnm{Mark}\binits{M.}}
(\byear{{2001}}).
\btitle{{Scientific collaboration networks. I. Network construction and
  fundamental results}}.
\bjournal{Phys. Rev. E}
\bvolume{{64}}
\bpages{016131}.
\end{barticle}
\endbibitem

\bibitem{Newman2004}
\begin{barticle}[author]
\bauthor{\bsnm{Newman},~\bfnm{Mark}\binits{M.}}
(\byear{2004}).
\btitle{Coauthorship networks and patterns of scientific collaboration}.
\bjournal{Proc. Natl. Acad. Sci. USA}
\bvolume{101}
\bpages{5200-5205}.
\bdoi{10.1073/pnas.0307545100}
\end{barticle}
\endbibitem

\bibitem{NewmanSC}
\begin{barticle}[author]
\bauthor{\bsnm{Newman},~\bfnm{Mark}\binits{M.}}
(\byear{2006}).
\btitle{Modularity and community structure in networks}.
\bjournal{Proc. Natl. Acad. Sci.}
\bvolume{103}
\bpages{8577-8582}.
\end{barticle}
\endbibitem

\bibitem{NewmanBook}
\begin{bbook}[author]
\bauthor{\bsnm{Newman},~\bfnm{Mark}\binits{M.}}
(\byear{2010}).
\btitle{Networks: an introduction}.
\bpublisher{Oxford University Press}.
\end{bbook}
\endbibitem

\bibitem{Sabidussi}
\begin{barticle}[author]
\bauthor{\bsnm{Sabidussi},~\bfnm{Gert}\binits{G.}}
(\byear{1966}).
\btitle{The centrality index of a graph}.
\bjournal{Psychometrika}
\bvolume{31}
\bpages{581--683}.
\end{barticle}
\endbibitem

\bibitem{Storey2003}
\begin{barticle}[author]
\bauthor{\bsnm{Storey},~\bfnm{John}\binits{J.}}
(\byear{2003}).
\btitle{The positive false discovery rate: a {B}ayesian interpretation and the
  {$q$}-value}.
\bjournal{Ann. Statist.}
\bvolume{31}
\bpages{2013--2035}.
\bdoi{10.1214/aos/1074290335}
\bmrnumber{2036398 (2004k:62055)}
\end{barticle}
\endbibitem

\bibitem{Tibshirani}
\begin{barticle}[author]
\bauthor{\bsnm{Tibshirani},~\bfnm{Robert}\binits{R.}}
(\byear{1996}).
\btitle{Regression shrinkage and selection via the lasso}.
\bjournal{J. Roy. Statist. Soc. Ser. B}
\bvolume{58}
\bpages{267--288}.
\end{barticle}
\endbibitem

\bibitem{EDA}
\begin{bbook}[author]
\bauthor{\bsnm{Tukey},~\bfnm{John}\binits{J.}}
(\byear{1977}).
\btitle{Exploratory Data Analysis}.
\bpublisher{Addison-Wesley}.
\end{bbook}
\endbibitem

\bibitem{Wasserman1994}
\begin{bbook}[author]
\bauthor{\bsnm{Wasserman},~\bfnm{Stanley}\binits{S.}}
(\byear{1994}).
\btitle{Social network analysis: methods and applications}
\bvolume{8}.
\bpublisher{Cambridge University Press}.
\end{bbook}
\endbibitem

\bibitem{zhu}
\begin{barticle}[author]
\bauthor{\bsnm{Zhao},~\bfnm{Yunpeng}\binits{Y.}},
  \bauthor{\bsnm{Levina},~\bfnm{Elizaveta}\binits{E.}} \AND
  \bauthor{\bsnm{Zhu},~\bfnm{Ji}\binits{J.}}
(\byear{2012}).
\btitle{Consistency of community detection in networks under degree-corrected
  stochastic block models}.
\bjournal{Ann. Statist.}
\bvolume{40}
\bpages{2266-2292}.
\end{barticle}
\endbibitem

\bibitem{Zou2006}
\begin{barticle}[author]
\bauthor{\bsnm{Zou},~\bfnm{Hui}\binits{H.}}
(\byear{2006}).
\btitle{The adaptive lasso and its oracle properties}.
\bjournal{J. Amer. Statist. Assoc.}
\bvolume{101}
\bpages{1418--1429}.
\end{barticle}
\endbibitem

\bibitem{ZouHastie2005}
\begin{barticle}[author]
\bauthor{\bsnm{Zou},~\bfnm{Hui}\binits{H.}} \AND
  \bauthor{\bsnm{Hastie},~\bfnm{Trevor}\binits{T.}}
(\byear{2005}).
\btitle{Regularization and variable selection via the elastic net}.
\bjournal{J. R. Stat. Soc. Ser. B Stat. Methodol.}
\bvolume{67}
\bpages{301--320}.
\bdoi{10.1111/j.1467-9868.2005.00503.x}
\bmrnumber{2137327}
\end{barticle}
\endbibitem

\bibitem{ZouLi2008}
\begin{barticle}[author]
\bauthor{\bsnm{Zou},~\bfnm{Hui}\binits{H.}} \AND
  \bauthor{\bsnm{Li},~\bfnm{Runze}\binits{R.}}
(\byear{2008}).
\btitle{One-step sparse estimates in nonconcave penalized likelihood models}.
\bjournal{Ann. Statist.}
\bvolume{36}
\bpages{1509--1533}.
\bdoi{10.1214/009053607000000802}
\bmrnumber{2435443 (2010a:62222)}
\end{barticle}
\endbibitem

\end{thebibliography}

\end{document}